\documentclass[a4paper,11pt]{article}

\usepackage{lineno,hyperref,url} 
\usepackage{amsmath,amsfonts,amssymb,amsthm,enumitem,stmaryrd,xfrac,nicefrac} 
\numberwithin{equation}{section} 
\usepackage{graphics,graphicx,subfigure} 
\usepackage{array,booktabs,diagbox,multirow,caption}  
\usepackage[most]{tcolorbox} 
\usepackage{soul}
\usepackage{tikz}
\usetikzlibrary{arrows}
\usepackage[textwidth=17cm]{geometry}

\newcommand{\footremember}[2]{%
    \footnote{#2}
    \newcounter{#1}
    \setcounter{#1}{\value{footnote}}%
}
\newcommand{\footrecall}[1]{%
    \footnotemark[\value{#1}]%
} 

\usepackage[round]{natbib}
\newcommand{\bs}{\boldsymbol} 
\newcommand{\mb}{\mathbb} 
\newcommand{\mc}{\mathcal} 
\newcommand{\mr}{\mathrm} 
\newcommand{\pd}{\partial}
\newcommand{\wt}{\widetilde} 
\newcommand{\ol}{\overline}
\newcommand{\sub}{\scriptscriptstyle} 
\newcommand{\beq}{\begin{equation}}
\newcommand{\eeq}{\end{equation}}
\newcommand{\beqs}{\begin{subequations}}
\newcommand{\eeqs}{\end{subequations}}
\newcommand{\benum}{\begin{enumerate}[label=(\roman*)]}
\newcommand{\eenum}{\end{enumerate}}
\newcommand{\bracket}[1]{\left\langle {#1} \right\rangle}
\newcommand\Ro{\mbox{R}_{\mbox{o}}}  
\newcommand\Bu{\mbox{B}_{\mbox{u}}}  

\newcommand{\dom}{\mc{D}} 
\newcommand{\area}{\mc{A}}
\newcommand{\df}{\mr{d}} 
\newcommand{\sto}{\mr{\mb{D}}} 
\newcommand{\alf}{\frac{1}{2}} 
\newcommand{\salf}{\sfrac{1}{2}}
\newcommand{\kerl}{\breve{\sigma}} 
\newcommand{\dpt}{\df p_t^{\sigma}} 
 
\newcommand{\dbt}{\df b_t^{\sigma}} 
\newcommand{\bdot}{\bs{\cdot}} 
\newcommand{\grad}{\bs{\nabla}} 
\newcommand{\gradp}{\grad^{\perp}} 
\newcommand{\bdiv}{\grad \bdot} 
\newcommand{\pdiv}{\gradp \bdot}
 
\newcommand{\adv}{\bdot \grad} 
\newcommand{\tp}{^{\sub T}} 

\newcommand{\tr}{\mr{Tr}} 
\newcommand{\Exp}{\mb{E}} 
\newcommand{\Var}{\mr{Var}}
\newcommand{\Pb}{\mb{P}} 
\newcommand{\filt}{\mc{F}} 
\newcommand{\dt}{\df t} 
\newcommand{\ds}{\df s} 
\newcommand{\bsig}{\bs{\sigma}} 
\newcommand{\bkerl}{\breve{\bs{\sigma}}} 
\newcommand{\bxi}{\bs{\xi}} 
\newcommand{\bp}{\bs{\phi}}
\newcommand{\B}{\bs{B}} 
\newcommand{\noi}{\bsig \df \B_t}
\newcommand{\wnoi}{\bsig \df \wt{\B}_t}
\newcommand{\wwnoi}{\wt{\bsig} \df \wt{\B}_t}
\newcommand{\bI}{\bs{\mr{I}}} 
\newcommand{\x}{\bs{x}}
\newcommand{\y}{\bs{y}}

\newcommand{\X}{\bs{X}}
\newcommand{\bk}{\bs{k}}
\newcommand{\bu}{\bs{u}}
\newcommand{\bv}{\bs{v}}
\newcommand{\vs}{\bv^{\star}}
\newcommand{\us}{\bu^{\star}}
\newcommand{\ws}{w^{\star}} 
\newcommand{\ba}{\bs{a}}
\newcommand{\G}{\bs{\Gamma}}
\newcommand{\dx}{\df \x} 
\newcommand{\dy}{\df \y} 
\newcommand{\dz}{\df z} 
\newcommand{\dA}{\df \area} 
\newcommand{\gradh}{\grad_h}
\newcommand{\divh}{\gradh\bdot}
\newcommand{\advh}{\bdot\gradh}
\newcommand{\noih}{\bsig_h \df \B_t}
\newcommand{\noiz}{\sigma_z \df B_t}

\usepackage{comment,color}

\newcommand{\delm}[1]{\ifmmode\text{\sout{\ensuremath{#1}}}\else\sout{#1}\fi} 

\title{\vspace{-2.75cm} Stochastic data-driven parameterization of unresolved mesoscale eddies}

\author{%
  Long Li\footremember{inria}{Centre Inria Rennes -- Bretagne Atlantique, France}
  \and Bruno Deremble\footremember{ige}{Institut des G\'{e}osciences de l'Environnement, Grenoble, France}%
  \and No\'{e} Lahaye\footrecall{inria}%
  \and Etienne Mémin\footrecall{inria}%
  }

\date{}

\begin{document}

\maketitle

\begin{abstract}
In this work, a stochastic representation based on a physical transport principle is proposed to account for mesoscale eddy effects on the large-scale oceanic circulation. This stochastic framework arises from a decomposition of the Lagrangian velocity into a smooth-in-time component and a highly oscillating noise term. One important characteristic of this random model is that it conserves the total energy of the resolved flow for any realization. Such an energy-preserving representation is successfully implemented in a well established multi-layered quasi-geostrophic dynamical core. The empirical spatial correlation of the unresolved noise is calibrated from the eddy-resolving simulation data. In particular, a stationary correction drift can be introduced in the noise through Girsanov transformation. This non intuitive term appears to be important in reproducing on a coarse mesh the eastwards jet of the wind-driven double-gyre circulation. In addition, a projection method has been proposed to constrain the noise living along the iso-surfaces of the vertical stratification. The resulting noise enables us to improve the intrinsic low-frequency variability of the large-scale current. 
\end{abstract}
\vspace{0.5cm}

\section{Introduction}

Ocean mesoscale eddies contain a large proportion of energy and have an important impact on large-scale circulations. They are found everywhere and are particularly intensive in the western boundary currents such as the Gulf Stream and the Kuroshio. However, these eddies are generally not resolved in ocean general circulation models, since the deformation radius in the ocean is at most of the order of 100 km. In particular, the effects of the mesoscale eddies need to be properly parameterized in coarse-resolution ocean models.

The most successful parameterization of mesoscale effect is based on the eddy-induced advection scheme \citep{Gent1990, Gent1995, Griffies1998}, which consists in mimicking the impact of the baroclinic instability by flattening the isopycnal surfaces to transfer the available potential energy of the resolved flow to the eddy kinetic energy at the sub-grid scales. However, such scheme does not account for the backscattering of kinetic energy from small to large scales \citep{Bachman2019}. Classical eddy viscosity models are introduced in coarse models to mimic the mixing action of the unresolved scales. The associated energy dissipation mechanism is often represented by some functional operators \citep{Leith1971, Griffies2000, Bachman2017} that depend on the resolved flow. These approaches seem to improve the effective resolution of eddy-permitting models. Nevertheless, they still lead to excessive decreases of the resolved kinetic energy in large-scale ocean models. The lack of variability due to energy missing is particularly detrimental in ensemble forecasting and data assimilation applications \citep{Karspeck2013, Franzke2015}.

To address these shortcomings, stochastic parameterizations are becoming more and more popular and several schemes have been devised and studied for ocean circulation models. These schemes introduce energy backscattering across scales and enable to increase the internal ocean variability. For instance, \citet{Berloff2005} developed a random-forcing model based on the dynamical decomposition of the flow into a large-scale component and an eddy components; \citet{Jansen2014} proposed a stochastic forcing in the form of a negative Laplacian viscosity to inject dissipated energy back to the resolved flow; \citet{Grooms2014, Grooms2015} proposed a stochastic superparameterization scheme that includes stochastic Reynolds stress terms to backscatter kinetic energy; \citet{Mana2014, Zanna2017} proposed an eddy parameterization based on a non-Newtonian stress which depends on the partially resolved scales and their variability.

We propose here a specific stochastic representation of the unresolved flow dynamics based on the location uncertainty (LU) framework \citep{Memin2014}. This representation introduces a random unresolved flow component and relies on a stochastic transport principle. Similar approaches based on the same decomposition have also been recently proposed by \citet{Holm2015, Cotter2018a, Cotter2018b, Gugole2019}. This kind of random models enable to consider systems that are less dissipative than their deterministic counterparts. Nevertheless, the ability of such models to represent faithfully the uncertainties associated to the actual unresolved small scales highly depends on the definition of the random component and on its evolution in time. Unsurprisingly, stationarity/time-varying and homogeneity/inhomogeneity characteristics of the unresolved flow component have strong influences on the numerical results. For instance, \citet{Bauer2020} illustrates that the noise inhomogeneity induces a structuration of the large-scale flow reminiscent to the action of the vortex force associated to surface wave-induced Stokes drift; \citet{Bauer2020ocemod} shows that the introduction of inhomogeneous noise into the barotropic quasi-geostrophic (QG) model enables us to reproduce accurately, on a coarse mesh, the high order statistics of the eddy-resolving data; a stochastic shallow water model preserving the resolved total energy has been proposed by \citet{Brecht2021} and the results show that the stochastic parameterization provides a good trade-off between model error representation and ensemble spread.

In this work, we investigate the LU formulation for the baroclinic QG model \citep{Hogg2004} within an idealized double-gyre circulation configuration. In particular, we focus on the reproduction of the eastwards jet as well as the prediction of the low-frequency variability for the proposed random model on a coarse resolution grid, for cases in which the baroclinic instability is not resolved. Another, important aspect, concerns the ability to include in the noise representation a stationary drift component associated to the temporal mean of the high-resolution fluctuations. As shown in this paper such stationary drift can be elegantly introduced in the noise through Girsanov transformation. In addition, a projection method has been proposed to update in time the noise along the iso-surfaces of the stratification.

The remainder of this paper is structured as follows. Section \ref{sec:cont-mod} describes the framework of LU and presents the derived stochastic QG model. Section \ref{sec:num-mod} explains the discretization of the proposed QG model and the parameterization method for the unresolved noise. Section \ref{sec:results} discusses the numerical results with some statistical diagnosis and energetic analysis. In Section \ref{sec:conclu} we draw some conclusions and provide an outlook for future works. In the Appendices we demonstrate the energy conservation of the random system and explore the energy conversion between the ensemble-mean and the ensemble-eddy components.

\section{Continuous models}\label{sec:cont-mod}

In this section, we first review the general setting of Location Uncertainty (LU) \citep{Memin2014}, we then present the resulting stochastic Boussinesq quasi-geostrophic (QG) model for a continuously stratified ocean, finally we complete the random formulation by including a time-correlated drift into the unresolved flow component through the Girsanov transformation.

\subsection{Stochastic flow}

The evolution of Lagrangian particle trajectory ($\X_t$) under LU is described by the following stochastic differential equation (SDE):
\beq\label{eq:dX}
\df \X_t (\x) = \bv \big( \X_t (\x), t \big)\, \dt + \bsig \big( \X_t (\x), t \big)\, \df \B_t,\ \quad \X_0 (\x) = \x \in \dom,
\eeq
where $\bv$ denotes the time-smooth resolved velocity that is both spatially and temporally correlated, $\noi$ stands for the fast oscillating unresolved flow component (also called \emph{noise} in the following) that is correlated in space yet uncorrelated in time, and $\dom \subset \mb{R}^d$ ($d = $ 2 or 3) is a bounded spatial domain. 

Mathematically, $\{\B_t\}_{\sub 0\leq t \leq T}$ is an $\bI_d$-cylindrical Brownian motion \citep{DaPrato2014} on a filtered probability space $(\Omega, \filt, \{\filt_t\}_{\sub 0\leq t \leq T}, \Pb)$\footnote{$\Omega$ is a set of samples, $\filt$ is a $\sigma$-field (a collection of subsets of $\Omega$), $\{\filt_t\}_{\sub 0\leq t \leq T}$ is a filtration (a family of sub $\sigma$-fields of $\filt$ that are continuously indexed in time and ordered non-decreasingly), and $\Pb$ is a probability measure.} and takes values in the Hilbert space $H := (L^2 (\dom))^d$. Informally, one can consider its time-derivative as a space-time ``white noise" (in a distribution sense) with independent components $(B_t^i)_{i=1,\ldots,d}$. 

The spatial structure of the unresolved flow component is modeled by the correlation operator, $\bsig$, which is not necessarily deterministic. In the most  general case it is random and time-dependent with some regularity conditions. More precisely, for each $(\omega, t) \in \Omega \times [0,T]$, $\bsig (\bdot, t) [\bullet]$ is assumed to be a Hilbert-Schmidt integral operator on $H$ with a bounded matrix kernel $\bkerl = (\kerl_{ij})_{i,j=1,\ldots,d}$ such that
\beq\label{eq:sigma}
\bsig (\x, t)\, \bs{f} = \int_{\dom} \bkerl (\x, \y, t) \bs{f} (\y)\, \dy,\ \quad \bs{f} \in H,\ \quad \x \in \dom .
\eeq
The composition of $\bsig (\bdot, t) [\bullet]$ and its adjoint $\bsig^* (\bdot, t) [\bullet]$ is a  trace class operator on $H$ and admits eigenfunctions $\bxi_n (\bdot,t)$ with eigenvalues $\lambda_n (t)$ satisfying $\sum_{n \in \mb{N}} \lambda_n (t) < + \infty$. Then, the noise can be equally defined by the following spectral decomposition:
\beq\label{eq:KL}
\bsig (\x, t)\, \df \B_t = \sum_{n \in \mb{N}} \lambda_n^{1/2} (t) \bxi_n (\x, t)\, \df \beta_n (t), 
\eeq
where $\beta_n$ are independent (one-dimensional) standard Brownian motions. 

In addition, we assume that the operator-space-valued process $\{\bsig (\bdot, t) [\bullet]\}_{\sub 0\leq t \leq T}$ is stochastically integrable, i.e. $\Pb \big[ \int_0^T \sum_{n \in \mb{N}} \lambda_n (t)\, \dt < +\infty \big] = 1$. As such, the stochastic integral $\{ \int_0^t \bsig (\bdot, s)\, \df \B_s \}_{\sub 0\leq t \leq T}$ is a $H$-valued Gaussian process of zero mean and of bounded (global) variance under the probability measure $\mb{P}$:
\beq
\Exp_{\sub \Pb} \left[ \int_0^t \bsig (\bdot, s)\, \df \B_s \right] = \bs{0},\ \quad \Exp_{\sub \Pb} \left[ \Big\| \int_0^t \bsig (\bdot, s)\, \df \B_s \Big\|_{\sub H}^2 \right] < +\infty .
\eeq
Moreover, the point-wise ($\x \in \dom$) and path-wise ($\omega \in \Omega$) strength of the unresolved flow component at each instant ($t \in [0,T]$) is measured by the matrix kernel of the composite operator $\bsig \bsig^*$, and denoted by $\ba (\x, t)$, namely
\beq\label{eq:bracket}
\ba (\x, t) := \int_{\dom} \bkerl (\x, \y, t) \bkerl\tp (\y, \x, t)\, \dy 
= \sum_{n \in \mb{N}} \lambda_n (t) \big( \bxi_n \bxi_n\tp \big) (\x, t) .
\eeq
As shown in \citet{Bauer2020}, the process $\int_0^t \ba (\x, s)\, \ds$ actually corresponds to the quadratic variation \citep{DaPrato2014} of $\int_0^t \bsig (\x, s)\, \df \B_s$, and it is a continuous random process of finite variation (hence time differentiable). In the particular case of a nonrandom  $\bsig$, $\ba$ reduces then to the local variance of the noise according to the It\^{o} isometry \citep{DaPrato2014}. Physically, the symmetric non-negative tensor $\ba$ represents the (possibly random) friction coefficients of the unresolved fluid motions and the eigenvalues $\lambda_n$ have the unit of m$^2$/s. 

\subsection{Stochastic transport}

The evolution law of a random tracer $\Theta$ with extensive property (e.g. temperature, salinity, buoyancy) transported by the stochastic flow, $\Theta (\X_{t + \delta t}, t + \delta t) = \Theta (\X_t, t)$ with $\delta t$ an infinitely small time variation, is derived by \citet{Memin2014, Bauer2020} using the generalized It\^{o} formula \citep{Kunita1997}. It is described by the following stochastic partial differential equation (SPDE):
\beq\label{eq:Dt-LU}
\sto_t \Theta = \df_t \Theta + (\vs\, \dt + \noi) \adv \Theta - \alf \bdiv (\ba \grad \Theta)\, \dt = 0 ,
\eeq
in which $\sto_t$ is introduced as a stochastic transport operator and $\df_t \Theta (\x) := \Theta (\x, t+\delta t) - \Theta (\x, t)$ stands for the (forward) time-increment of $\Theta$ at a fixed point $\x \in \dom$. 

This SPDE encompasses physically meaningful terms. For instance, the third term describes the tracer's advection by the unresolved flow component. As shown in \citet{Resseguier2017gafd1, Bauer2020}, the resulting multiplicative noise $\noi \adv \Theta$ continuously backscatters random energy to the system through the quadratic variation $\salf\, (\grad \Theta) \bdot (\ba \grad \Theta)$ of the random tracer. The last term in \eqref{eq:Dt-LU} depicts tracer's diffusion due to the mixing of the unresolved scales. In particular, under specific noise definitions \citep{Memin2014}, the resulting diffusion can be connected to the functional eddy viscosity as introduced in many large-scale circulation models \citep{Smagorinsky1963, Redi1982}. 

As an additional feature of interest, there exists an \emph{effective} advection velocity $\vs$ in \eqref{eq:Dt-LU} which is defined as
\beq\label{eq:def-ISD}
\vs = \bv - \alf \bdiv \ba + \bsig^* (\bdiv \bsig) .
\eeq
This statistical eddy-induced velocity drift captures the action of inhomogeneity of the random field on the transported tracer and the possible divergence of the unresolved flow component. \citet{Bauer2020} shows that the \emph{turbophoresis} term $\salf\, \bdiv \ba$ can be interpreted as a generalization of the Stokes drift, which occurs, for example, in the Langmuir circulation \citep{Craik1976, Leibovich1980}. 

Many useful properties of the stochastic transport operator $\sto_t$ have been explored by \citet{Resseguier2017gafd1, Resseguier2020b, Li2021PhD}. In particular, if a random tracer is transported by the incompressible stochastic flow under suitable boundary conditions, then the path-wise $p$-th moment $(p \geq 1)$ of the tracer is materially and integrally invariant, namely
\beq\label{eq:p-moment}
\sto_t \left( \frac{1}{p} \Theta^p \right) = 0,\ \quad \df_t \int_{\dom} \frac{1}{p} \Theta^p\, \dx = 0 .
\eeq

\subsection{Stochastic QG model}

The derivation of the stochastic geophysical models under the LU framework follows almost exactly the same path as the deterministic derivation \citep{Vallis2017}. In particular, a \emph{stochastic Euler-Boussinesq model} has been derived by \citet{Resseguier2017gafd2, Bauer2020} for large-scale atmospheric and oceanic circulations, and reads:
\beqs\label{eqs:Boussinesq-LU}
\begin{align}
&\underline{\text{Momentum equations}} \nonumber \\
&\sto_t \bv +  f \bk \times (\bu\, \dt + \noih) = (b\, \dt + \dbt) \bk - \grad (p\, \dt + \dpt) , \label{eq:moment} \\
&\underline{\text{Continuity equations}} \nonumber \\
&\bdiv \vs = 0,\ \quad \bdiv \noi = 0, \label{eq:incomp3} \\
&\underline{\text{Thermodynamic equation}} \nonumber \\ 
&\sto_t b  + N^2 (\ws\, \dt + \noiz) = \alf \bdiv (\ba_{\bdot z} N^2)\, \dt, \label{eq:buoy}
\end{align}
\eeqs
where $\bv = (\bu, w)\tp$, $\noi = (\noih, \noiz)\tp$, $\bk = (0,0,1)\tp$, $f = f_0 + \beta y$ is the Coriolis parameter using beta-plane approximation, $p$ and $\dpt / \dt$ (in a distribution sense) are the time-smooth component and the fast oscillating noise of the pressure fluctuations rescaled by the background density $\rho_0$, $N^2 (z) = - g \pd_z \overline{\rho} (z) / \rho_0$ is the Brunt-V\"{a}is\"{a}l\"{a} frequency with $g$ the gravity value and $\overline{\rho}$ the stationary density in equilibrium that only depend on height, $b (\x, t) = - g \rho' (\x, t) / \rho_0$ is the resolved buoyancy associated with the density anomaly $\rho'$, $\dbt / \dt$ is a zero-mean noise that models the unresolved buoyancy fluctuations and $\ba_{\bdot z}$ denotes the $z$-column vector of tensor $\ba$. The term $\dbt$ can be seen as a thermodynamics noise coming from the nonlinear mapping law of state applied to the noise term of salinity and temperature transports. It is of the same nature as the noise term introduced in \citet{Brankart2013}.

From the system \eqref{eqs:Boussinesq-LU} a diverse set of approximated models under LU can be obtained through nondimensionalization and asymptotic approach with proper scaling. However, the noise introduces an additional degree of freedom that must be appropriately accounted for \citep{Brecht2021, Bauer2020ocemod, Resseguier2017gafd2, Resseguier2017gafd3}. 
The horizontal components of the quadratic variation are first scaled as $\ba_h \sim \epsilon\, U L$, where $U$ and $L$ are typical velocity and length scales, and the factor $\epsilon$ is proportional to the ratio between the eddy kinetic energy (EKE) and the mean kinetic energy (MKE) and to the ratio between the small-scale correlation time and the large-scale advection time. From the definitions \eqref{eq:sigma}, \eqref{eq:KL} and \eqref{eq:bracket}, the horizontal components of the noise can be scaled as $\noih \sim \epsilon^{1/2} L$. The greater this scaling number $\epsilon$, the larger the quadratic variation, hence the stronger the path-wise noise. As shown in \citet{Resseguier2017gafd2, Resseguier2017gafd3}, using different levels of noise in the stochastic system allows us to model different physical regimes of the large-scale flow. 

In this work, we only consider the ``moderate" horizontal uncertainty, $\epsilon = \mc{O} (1)$ or $\epsilon\, \Ro = \mc{O} (\Ro)$ with a small Rossby number $\Ro = U / (f_0 L) \ll 1$, in order to maintain the classical geostrophic balance. Besides, to keep the assumption of flat isopycnal (small variation of stratification) in the QG theory, \citet{Resseguier2017gafd2} proposed to scale the ratio between the vertical scale and the horizontal scale of the noise as
$\noiz / \noih \sim (H/L) (\Ro / \Bu)$ with the height scale $H \ll L$ and the Burger number $\Bu = (L_d / L)^2 = \mc{O} (1)$, where $L_d$ stands for the scale of deformation radius. Moreover, the unresolved pressure and buoyancy are scaled by $\dpt \sim \Phi = \epsilon^{1/2} f_0 L^2$ and $\dbt \sim \Phi / H$. All the resolved variables ($\bu, w, p, b, N^2$) are scaled as in the classical framework \citep{Vallis2017} and the variation of Coriolis parameter stays small, i.e. $\beta L < f_0$. 

Expanding the resolved variables ($\bv, p, b$) in power series of the small Rossby number and truncating the equations at zeroth and first orders, then restoring the dimensions of variables, the following QG system can be derived:
\beqs\label{eqs:QG-LU-mb}
\begin{align}
&\underline{\text{Momentum equations}} \nonumber \\
&\sto_t^h \bu + f_0 \bk \times \bu_a\, \dt + \beta y \bk \times (\bu\, \dt + \noih) = - \gradh p_a\, \dt , \label{eq:QG-LU-moment}\\
&\underline{\text{Buoyancy equation}} \nonumber \\
&\sto_t^h b + N^2 w_a\, \dt = 0 , \label{eq:QG-LU-buoy} \\
&\underline{\text{Continuity equation}} \nonumber \\
&\divh \bu_a + \pd_z w_a = 0 , \label{eq:QG-LU-cont}
\end{align}
where $\bu$ and $p$ denote the geostrophic components of the resolved velocity and pressure, $(\bu_a, w_a)\tp$ and $p_a$ denote the ageostrophic components, and $\sto_t^h$ stands for the horizontal component of the stochastic transport operator, which is defined as
\beq\label{eq:Dth-LU}
\sto_t^h \Theta = \df_t \Theta + (\us\, \dt + \noih) \advh \Theta - \alf\, \divh (\ba_h \gradh \Theta)\, \dt 
\eeq
with $\us := \bu - \salf\, \divh \ba_h$ and $\gradh = (\pd_x, \pd_y)\tp$. The geostrophic component and the noise satisfy the following equilibrium:
\begin{align}
&\underline{\text{Geostrophic balances and martingale pressure equilibrium}} \nonumber \\
&\bu = \frac{1}{f_0} \gradp_h p,\ \quad \noih = \frac{1}{f_0} \gradp_h \dpt,\ \quad \divh (\us - \bu) = 0 , \label{eq:QG-LU-geost}\\
&\underline{\text{Hydrostatic balances}} \nonumber \\
&b = \pd_z p,\ \quad \dbt = \pd_z \dpt , \label{eq:QG-LU-hydro}
\end{align}
\eeqs
where $\gradp_h = (-\pd_y, \pd_x)\tp$. For sake of conciseness of the notations, we drop the horizontal subscript $h$ in the following. Combining the above equilibrium, the thermal wind balances, $\pd_z \bu \perp \grad b$ and $\pd_z \noi \perp \grad \dbt$, can be deduced. In this work, we only focus on noise fully satisfying the martingale pressure equilibrium. It means that we model the mesoscale flow as a stochastic component given a large-scale flow. However, such random equilibrium is not necessary in a general setting and could be easily relaxed.

In order to represent the previous random system in terms of potential vorticity (PV) in the same way as in the classical framework \citep{Vallis2017}, we first take the (horizontal) curl of the first-order momentum equations \eqref{eq:QG-LU-moment}, then sum the resulting equation with the vertical derivative of the buoyancy equation \eqref{eq:QG-LU-buoy} rescaled by $f_0/N^2$, together with the equilibrium \eqref{eq:QG-LU-geost}, \eqref{eq:QG-LU-hydro} and \eqref{eq:QG-LU-cont}, to obtain finally
\beqs\label{eqs:QG-LU}
\begin{align}
&\underline{\text{Stochastic evolution of PV}} \nonumber \\
&\sto_t q = \sum_{i=1,2} - \mr{J} \big( (u^{\star}\, \dt + \sigma \df B_t)^i, u^i \big) + \alf \bdiv \big( \pd_{x_i}^{\sub \perp} \ba \grad u^i \big)\, \dt - \bdiv (\ba \grad f)\, \dt  \nonumber \\
&\:\phantom{\sto_t =}\:
-  \frac{f_0}{N^2} \pd_z (\us\, \dt + \noi) \adv b + \frac{f_0}{2 N^2} \bdiv (\pd_z \ba \grad b)\, \dt , \label{eq:dPV} \\
&\underline{\text{From PV to streamfunction}} \nonumber \\
&q = \nabla^2 \psi + \pd_z \Big( \frac{f_0}{N^2} b \Big) + \beta y,\ \quad b = f_0 \pd_z \psi \label{eq:q-to-p} \\
&\underline{\text{Incompressible constraints}} \nonumber \\
&\bu = \gradp \psi,\ \quad \bdiv \noi = \bdiv (\us - \bu) = 0 \label{eq:incomp} ,
\end{align}
\eeqs
where $q$ denotes the PV, $\psi := p / f_0$ stands for the streamfunction of the resolved flow, $\mr{J} (f, g) = \pd_x f \pd_y g - \pd_x g \pd_y f$ stands for the Jacobian operator and $\nabla^2 = \pd_{xx}^2 + \pd_{yy}^2$ is the two-dimensional (2D) Laplacian operator. Note that in the deterministic setting, the sources and sinks of PV in \eqref{eq:dPV} cancels due to the anti-symmetry of the Jacobian operator, $\mr{J} (u^i, u^i) = 0$, and to the thermal wind balance, $\pd_z \bu \perp \grad b$. However, it should be noted that the presence of sub-grid terms in a large-scale deterministic Boussinesq system leads also in general to additional terms in the RHS of the PV equation. The additional terms in the RHS of \eqref{eq:dPV} are coming exactly from the same sources of difficulty inherent to large-scale flow dynamics representations (i.e. non commutation of curl operator with the sub-grid model, breaking of the Jacobian anti-symmetry as well as possibly non commutation of coarsening processes with derivatives).

\subsection{Energy conservation and transfers}\label{sec:energy}

One important characteristic of the random system \eqref{eqs:QG-LU-mb} or \eqref{eqs:QG-LU} is that it conserves path-wise (i.e. for each realization) the total energy of the resolved geostrophic flow (under natural boundary conditions), namely
\beq\label{eq:conserve-energy}
\df_t \int_{\dom} \alf \Big( |\bu|^2 + \frac{b^2}{N^2} \Big)\, \dx = \Big( \int_{\dom} w_a b\, \dx - \int_{\dom} w_a b\, \dx \Big)\, \dt = 0 .
\eeq
The demonstration can be found in \ref{sec:proof-energy}. We remark that this conservation property is consistent with the energy conservation of the stochastic barotropic QG system \citep{Bauer2020} and of the stochastic rotating shallow-water system \citep{Brecht2021}. 

The energy conversions between the ensemble-mean and ensemble-eddy components are in addition explored in \ref{sec:conv-energy}.
Based on the decompositions $\bu = \Exp [\bu] + (\bu - \Exp [\bu]) := \ol{\bu} + \bu'$ and $b = \ol{b} + b'$, the mean kinetic energy (MKE), the eddy kinetic energy (EKE), the mean potential energy (MPE) and the eddy potential energy (EPE) are respectively defined by $\salf \langle |\ol{\bu}|^2 \rangle$, $\salf \langle \ol{|\bu'|^2} \rangle$, $\salf \langle (\ol{b}/N)^2 \rangle$ and $\salf \langle \ol{(b'/N)^2} \rangle$, where $\langle \bdot \rangle$ denotes the integration over domain for simplicity. The energy conversions of the proposed random model \eqref{eqs:QG-LU-mb} are summarized in the following diagram.

\begin{center}
\begin{tikzpicture}
\tikzset{edge/.style = {->,> = latex'}}
\node (1) at  (0,0) {MKE};
\node (2) at  (11,0) {EKE};
\node (3) at  (0,-4) {MPE};
\node (4) at  (11,-4) {EPE};
\draw[edge] (1.10) -- node [above] {$-\, \big\langle \big[ \grad \ol{\bu}, \ol{(\us)' \bu'} \big]_{\sub F} \big\rangle + \salf \big\langle \big[ \grad \ol{\bu}, \ \ol{\ba} \grad \ol{\bu} + \ol{\ba' \grad \bu'} \big]_{\sub F} \big\rangle $} (2.170);
\draw[edge] (2.190) -- node [below] {$+\, \big\langle \big[ \grad \ol{\bu}, \ol{(\us)' \bu'} \big]_{\sub F} \big\rangle - \salf \big\langle \big[ \grad \ol{\bu}, \ \ol{\ba} \grad \ol{\bu} + \ol{\ba' \grad \bu'} \big]_{\sub F} \big\rangle $} (1.350);
\draw[edge] (3.10) -- node [above] {$-\, \big\langle (\grad \ol{b} / N^2) \bdot \ol{(\us)' b'} \big\rangle + \salf \big\langle (\grad \ol{b} / N^2) \bdot (\ol{\ba} \grad \ol{b} + \ol{\ba' \grad b'}) \big\rangle$} (4.170);
\draw[edge] (4.190) -- node [below] {$+\, \big\langle (\grad \ol{b} / N^2) \bdot \ol{(\us)' b'} \big\rangle - \salf \big\langle (\grad \ol{b} / N^2) \bdot  (\ol{\ba} \grad \ol{b} + \ol{\ba' \grad b'}) \big\rangle$} (3.350);
\draw[edge] (1.260) -- node [above, rotate=90] {$-\, \big\langle \ol{b} \ol{w}_a \big\rangle$} (3.100);
\draw[edge] (3.80) -- node [below, rotate=90] {$+\, \big\langle \ol{b} \ol{w}_a \big\rangle$} (1.280);
\draw[edge] (2.260) -- node [above, rotate=90] {$-\, \big\langle \ol{b' w_a'} \big\rangle$} (4.100);
\draw[edge] (4.80) -- node [below, rotate=90] {$+\, \big\langle \ol{b' w_a'} \big\rangle$} (2.280);
\end{tikzpicture}    
\end{center}
Here, $[\bs{A}, \bs{B}]_{\sub F} = \tr (\bs{A}\tp \bs{B})$ stands for the Frobenius inner product of the tensors $\bs{A}$ and $\bs{B}$. The ensemble-mean of the path-wise energy conservation \eqref{eq:conserve-energy} can be recovered by summing all the conversion terms.

Compared to the energy budget of the classical deterministic system, the random system provides additional conversion terms between the mean and eddy components. In the general setting with $\bsig$ random, the classical geostrophic Reynolds stress \citep{Bachman2019} is modified to $\ol{(\us)' \bu'}$ due to the possibly inhomogeneous nature of the noise. This term includes an additional  term related to the Ito-Stokes drift that can be positive or negative depending on its alignment with the large-scale velocity component. The fact that $\ol{\ba}\, \dt = \Var (\noi)$ is symmetric and non-negative definite, ensures that the integral $\big\langle \big[ \grad \ol{\bu}, \ \ol{\ba} \grad \ol{\bu} \big]_{\sub F} \big\rangle$ (resp. $\big\langle \big[ \grad \ol{b}, \ \ol{\ba} \grad \ol{b} \big]_{\sub F} \big\rangle$) is always positive, hence provides a positive transfer from MKE to EKE (resp. from MPE to EPE). When the correlation tensor $\bsig$ is nonrandom, only this term remains in addition to the term related to the classical geostrophic Reynolds stress. As the stochastic system \eqref{eqs:QG-LU-mb} does not contain vertical noise component, it does not explicitly modify the conversion mechanisms between KE and PE as observed in the classical system. Nevertheless, the effect of the stochastic terms on this energy transfers is implicitly accounted via the time-integration of the resolved variables $b$ and $w_a$.

\subsection{Girsanov transformation}\label{sec:Girsanov}

The previous formulations \eqref{eqs:QG-LU-mb} and \eqref{eqs:QG-LU} consist of only a zero-mean and temporally uncorrelated noise. However, this might not be enough in practice and including a mean or time-correlated component of the unresolved velocity field could be of crucial importance to obtain a relevant model. For instance, the eddy parameterization proposed by \citet{Zanna2017} is decomposed into a deterministic term based on the negative Laplacian of the PV transport and a zero-mean stochastic term. For the double-gyre circulation configuration, the considered deterministic parameterization allows to reproduce the eastwards jet for the coarse-resolution model, while the additional stochastic terms enhance the gyres circulation and improves the flow variability. Similarly, the random-forcing model proposed by \citet{Berloff2005} consists in a space-time correlated stochastic process to enhance the jet extension. 

The slow components of the sub-grid scales can be provided by adequate high-pass filtering of high-resolution data on the coarse grid. We aim in this work at investigating the incorporation of such slow components within the LU framework. However, the derivation of LU models \citep{Memin2014, Resseguier2017gafd1, Bauer2020} relies on the martingale properties of the centered noise and we need hence to properly handle non centred Brownian terms. The Girsanov transformation \citep{DaPrato2014} provides a theoretical tool that fully warrants such a superposition: by a change of the probability measure, the composed noise can be centered with respect to a new probability measure while the additional drift term appears, which pulls back time-correlated sub-grid-scale components into the dynamical system. The associated mathematical description is given as follows.

Let $\G_t$ be an $H$-valued $\filt_t$-predictable process satisfying $\Exp_{\sub\Pb} \big[ \exp (\alf \int_0^{\sub T} \|\G_t\|_{\sub H}^2\, \dt ) \big] < +\infty$, then the process $\{ \wt{\B}_t := \B_t + \int_0^t \G_s\, \ds\}_{\sub 0 \leq t \leq T}$ is an $H$-valued cylindrical Wiener process on $(\Omega, \filt, \{\filt_t\}_{\sub 0 \leq t \leq T}, \wt{\Pb})$ by a change of the probability measure:
\beq
\frac{\df \wt{\Pb}}{\df \Pb} = \exp \Big( - \int_0^T \bracket{\G_t, \df \B_t}_{\sub H} - \alf \int_0^T \|\G_t\|_{\sub H}^2\, \dt \Big) ,
\eeq
where $\bracket{\bs{f}, \bs{g}}_{\sub H} = \int_{\sub\dom} (\bs{f}\tp \bs{g}) (\x)\, \dx$ denotes the inner product on $H = (L^2 (\dom))^d$. Under this new probability measure $\wt{\Pb}$, the previous system \eqref{eqs:QG-LU} must be adapted to the modified noise $\wnoi$ with $\Exp_{\sub \wt{\Pb}} [\wnoi] = \bs{0}$ by a correction of the effective drift
\beq
\wt{\us} := \us - \bsig \G_t . 
\eeq
Hereafter, the resulting stochastic system will be adopted with $\bsig \G_t$ referred to as the \emph{Girsanov drift}.

\section{Numerical models}\label{sec:num-mod}

In this section, we first describe the discretization of the continuous stochastic QG system \eqref{eqs:QG-LU} (under a simplification) that will be used later for the numerical simulations. Then, we present numerical methods to estimate the spatial correlation functions of the unresolved flow component based on the spectral decomposition \eqref{eq:KL}.

\subsection{Discrete QG model}

\begin{figure}
\captionsetup{font=scriptsize}
\begin{center} 
\includegraphics[width=10cm]{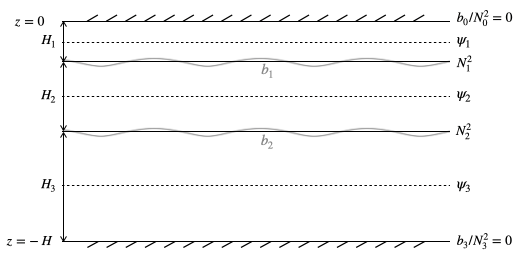}
\end{center}
\caption{Illustration for the vertical discretization of a three-layered QG system.} 
\label{fig:vert-disc}
\end{figure}

We consider here a vertically discretized QG model. This formulation is quite common in geophysical fluid dynamics and the derivation follows the standard methods \citep{Vallis2017, McWilliams2006}. As illustrated in Figure \ref{fig:vert-disc}, such model consists of $n$ ocean layers with constant thickness $H_k$ and density $\rho_k$ in each layer $k$. The prognostic variables $(q_k, \psi_k, \bu_k, \bsig_k, \ba_k)$ are assumed to be layer-averaged quantities. 

We first apply finite differences for the buoyancy frequency $N^2$ and anomaly $b$ on the interface between layers $k$ and $k+1$ (for $k=1,\ldots,n-1$):
\beq
b_k = \frac{f_0 (\psi_k - \psi_{k+1})}{(H_k + H_{k+1})/2},\ \quad N_k^2 = \frac{g_k'}{(H_k + H_{k+1})/2} ,
\eeq
where $g_k' := - g (\rho_k - \rho_{k+1})/\rho_0$ is introduced as the reduced gravity across the interface $k$. Applying subsequently finite differences for the stratification term $\pd_z (f_0 b / N^2)$ with the zero boundary conditions at ocean surface and bottom, $b_0 / N_0^2 = b_n / N_n^2 = 0$, the Helmholtz equation \eqref{eq:q-to-p} can be vertically discretized by 
\beqs\label{eqs:QG-layer}
\beq\label{eq:q-to-p-k}
\nabla^2 \psi_k + \frac{f_0^2}{H_k} \Big( \frac{\psi_{k-1} - \psi_{k}}{g_{k-1}'} - \frac{\psi_{k} - \psi_{k+1}}{g_{k}'} \Big) = q_k - \beta y .
\eeq
In this work, we adopt the dynamical core of the QG coupled model (Q-GCM) developed by \citet{Hogg2003}, which integrates numerically the state variables $q$ and $\psi$. In order to reduce the inaccuracy of numerical discretization, we do not include the PV sources and sinks (RHS terms of \eqref{eq:dPV}) in this work. 

These terms are naturally considered when working in $\bu$-$b$ formulations such as in the primitive equations dynamical core \citep{Bachman2019} or for multi-layers shallow water systems \citep{Adcroft2019}. In our case, for the type of noises considered (with moderated amplitude so as to respect the QG scaling performed in this work) these additional noise terms are small but leads to additional numerical difficulties. For stronger noise, with \emph{de facto} other scalings with potential modification of the hydrostatic balances these additional noise terms have to be considered in the PV equation.

Taking account the numerical dissipation and the vertical entrainment $e_k$ (across interfaces), the simplified stochastic evolution of PV reads
\beq\label{eq:dPV-k}
\wt{\sto}_t^k q_k = \left( (A_2 \nabla^2 - A_4 \nabla^4) (\nabla^2 \psi) + \frac{f_0}{H_k} (e_{k-1} - e_{k}) \right)\, \dt ,
\eeq
\eeqs
where $\wt{\sto}_t^k$ denotes the $k$-th layer stochastic transport operator under the probability measure $\wt{\Pb}$ (hence includes the Girsanov drift described in Section \ref{sec:Girsanov}, $ \nabla^4 =  \nabla^2 ( \nabla^2)$ is the bi-Laplacian operator associated with the dissipation coefficient $A_4$. 

As for the spatial discretization, we propose to use a a conservative flux form $\bdiv (\bu q)$ that corresponds exactly to the 9-points Arakawa Jacobian scheme \citep{Arakawa1981} of $\mr{J} (\psi, q)$, in which the zonal and meridional components of the advection flux are defined as $\sfrac{2}{3}\, \big( \ol{\ol{u}^x}^y \ol{q}^x \big) + \sfrac{1}{3}\, \big( \ol{\ol{u}^x \ol{q}^d}^y \big)$ and $\sfrac{2}{3}\, \big( \ol{\ol{v}^y}^x \ol{q}^y \big) + \sfrac{1}{3}\, \big( \ol{\ol{v}^x \ol{q}^d}^y \big)$. Here, $\ol{\bdot}^x, \ol{\bdot}^y$ and $\ol{\bdot}^d$ denote the central averaging operations in the zonal, meridional and diagonal directions respectively. We then apply it for the advection of PV by all the additional drift (such as the noise). On the other hand, we employ simply a central winding scheme to discretize the inhomogeneous diffusion term $\bdiv (\ba \grad  q)$. We adopt a stochastic Leapfrog scheme \citep{Ewald2005} for the time-stepping of \eqref{eq:dPV-k}. The inversion of the modified Helmholtz equation \eqref{eq:q-to-p} is carried out with a discrete Fourier transform method \citep{Hogg2003}. 

The vertical entrainment $e$ in the classical QG models \citep{Berloff2007, Marshall2012, Grooms2015, Zanna2017} accounts only for the Ekman pumping from the upper and bottom boundary layers, $e_0 = \pdiv \bs{\tau} / f_0$ and $e_n = \salf\, \delta_{ek} \nabla^2 \psi_n$, where $\bs{\tau}$ denotes a given surface wind stress and $\delta_{ek}$ stands for a prescribed thickness of the bottom Ekman layer. In the Q-GCM formulation \citep{Hogg2004}, an additional entrainment $e_1$ is considered due to the Ekman pumping of mixed layer temperature. This will be further detailed in Section \ref{sec:mix-layer}.

\subsection{Parameterizations of unresolved flow}\label{sec:EOF}

This section describes the numerical methods of the parameterization for the unresolved flow structure. We first review the empirical orthogonal functions (EOF) method which relies on an assumption of quasi-stationarity of the  noise covariance. To relax this stationarity constraint, we then propose a projection method  enabling to update on-line the noise spatial correlations.

Let $\{\bu_{\sub\text{HR}} (\x, t_i)\}_{i=1,\ldots,N}$ be the finite set of velocity snapshots provided by a high-resolution (HR) simulation. We first build the spatial local fluctuations $\bu_f (\x, t_i)$ of each snapshot at the coarse-grid points. In particular, for the QG system, one can first perform a high-pass filtering with a 2D Gaussian convolution kernel $G$ on each HR streamfunction $\psi_{\sub\text{HR}}$, to obtain the streamfunction fluctuations, $\psi_f (\x, t_i) = \big( (\mr{I} - \mr{G}) \star \psi_{\sub\text{HR}} \big) (\x, t_i)$ (only for the coarse-grid points $\x$). Then, the geostrophic velocity fluctuations can be derived by $\bu_f = \gradp_{\sub\text{LR}} \psi_f$. 

We next center the data set by $\bu_f' = \bu_f - \overline{\bu_f}^t$ (with $\overline{\bullet}^t$ the temporal mean) and perform the EOF procedure \citep{Sirovich1987} to get a set of orthogonal temporal modes $\{\alpha_n\}_{n=1,\ldots,N}$ and orthonormal spatial modes $\{\bp_n\}_{n=1,\ldots,N}$ satisfying 
\beq
\bu_f' (\x, t_i) = \sum_{n=1}^{N} \alpha_n (t_i) \bp_n (\x),
\eeq
where $\langle \bp_m, \bp_n \rangle_{\sub \mc{H}} = \delta_{m,n}$, $\overline{\alpha_m \alpha_n}^t = \lambda_n \delta_{m,n}$, $\mc{H} := L^2 (\dom)$ and $\dom = \area \times [-H, 0]$ denotes the 3D domain with the horizontal area $\area$ and the total depth $H$. 

Truncating subsequently the modes (with $M \ll N$) and rescaling them by a small-scale decorrelation time $\tau$, the $k$-th layer noise and its variance are built by
\beqs
\begin{align}
\bsig_k (\x)\, \df \wt{\B}_t &= \tau^{1/2} \sum_{n=1}^M \lambda_n^{1/2} \bp_{n,k} (\x)\, \df \beta_n (t), \label{eq:EOF-noi}\\ 
\ba_k (\x) &= \tau \sum_{n=1}^M \lambda_n \bp_{n,k} (\x) \bp_{n,k}\tp (\x) . \label{eq:EOF-a}
\end{align}
Note that this time scale $\tau$ is used to match the fact that the noise in \eqref{eq:dX} admits unit of meter. In practice, we often consider the coarse-grid simulation time-step $\Delta t_{\sub\text{LR}}$. 

In this work, we simply propose to represent the Girsanov drift by the temporal mean of the unresolved eddies, that is
\beq
\bsig_k (\x) \G_t = \sum_{n=1}^N \langle \overline{\bu_f}^t,  \bp_n \rangle_{\sub\mc{H}} \bp_{n,k} (\x) .
\eeq
\eeqs
We remark that in future works this stationary drift term could be generalized to slowly varying fluctuation components or to drift correction with respect to observation \citep{Dufee2022}.

The previous EOF procedure is an efficient off-line learning method, yet it relies on a strong stationary assumption, and hence leads to a sequence of random flow fields with no connection with the resolved dynamics. In the following, we propose an novel approach that project the EOF-based noise on a resolved state variable. In order to enforce the noise to act only on the resolved momentum \eqref{eq:QG-LU-moment} and provide efficient backscattering of KE, we consider constraining the noise along the iso-surfaces of the stratification, $\wwnoi \perp \grad \pd_z \wt{b}$ with $\wt{b} := f_0 b / N^2$. This projection procedure reads
\beq
\wt{\bsig}_k (\x, t)\, \df \wt{\B}_t = A^{1/2}\, \bs{\mr{P}}_k (\x, t)\ \bsig_k (\x)\, \df \wt{\B}_t,\ \quad \bs{\mr{P}}_k := \bs{\mr{I}}_2 - \frac{\grad (\pd_z \wt{b})_k \big( \grad (\pd_z \wt{b})_k \big)\tp}{|\grad (\pd_z \wt{b})_k|^2},
\eeq
where $\bsig_k \df \wt{\B}_t$ is generated as in \eqref{eq:EOF-noi} and $A$ is a scaling factor to ensure that the noise amplitude is projection invariant. From the definition of the noise quadratic variation \eqref{eq:bracket}, tensor $\ba$ defined through \eqref{eq:EOF-a} is updated as $\wt{\ba}_k = A\, \bs{\mr{P}}_k\, \ba_k\,  \bs{\mr{P}}_k\tp$. In the next section, we show that the resulting non-stationary noise leads to significant improvement of variability for coarse-resolution simulations.

\section{Numerical results}\label{sec:results}

In this section, we discuss and compare numerical results of several models run within different configurations. The objective consists in improving the variability of low-order models at low Reynolds numbers flow, and in the particular case of unresolved baroclinic instabilities.

\subsection{Model configurations}

We consider here a finite box ocean at mid-latitude driven by an idealized (stationary and symmetric) wind stress $\bs{\tau} = (-\tau_0 \cos (2\pi y)/L_y, 0)\tp$. A mixed horizontal boundary condition is used for the $k$-th layer streamfunction: $\psi_k |_{\pd\area} = f_k (t)$ and $\pd_n^2 \psi_k |_{\pd\area} = - (\alpha_{\text{bc}}/\Delta x) \pd_n \psi_k |_{\pd\area}$ (same for the 4-th order derivative). Here, $f_k$ is a time-dependent function constrained by mass conservation \citep{McWilliams1977}, $\Delta x$ stands for the horizontal resolution and $\alpha_{\text{bc}}$ is a nondimensional coefficient associated to the slip boundary conditions \citep{Haidvogel1992}. 

A deterministic eddy-resolving ($\Delta x = 5$ km) model is first simulated and referred to as the references (\emph{REF}). It is then  compared to several coarse-resolution ($\Delta x = 40$ km, 80 km, 120 km) models: the benchmark deterministic model (\emph{DET}), two stochastic models with a stationary EOF-based noise (\emph{STO-EOF}) and a non-stationary projection noise (\emph{STO-EOF-P}). The eddy-resolving model starts from a quiescent initial condition, whereas spin-up conditions downsampled from \emph{REF} (after 90-years integration) are adopted for all the coarse-resolution models. The common parameters for all the simulations are listed in Table~\ref{tab:params-pub}, whereas resolution dependent parameters are presented separately in Table~\ref{tab:params-pri}. 

\begin{table}
\captionsetup{font=footnotesize}
\begin{center}
\resizebox{0.65\textwidth}{!}{%
\begin{tabular}{ccc}
\hline
Parameters                     & Value                                                  & Description                      \\ \hline
$X \times Y$                   & $(3840\times 4800)$ km                    & Domain size               \\
$H_k$                             & $(350, 750, 2900)$ m                       & Mean layer thickness   \\
$\rho$                             & $1000$ kg m$^3$                                & Density   \\
$g_k'$                             & $(0.025, 0.0125)$ m s$^{-2}$              & Reduced gravity       \\
$\delta_{\text{ek}}$         & $2$ m                                                & Bottom Ekman layer thickness   \\
$\tau_0$                         & $2\times 10^{-5}$ m$^2$ s$^{-2}$           & Wind stress magnitude \\ 
$\alpha_{\text{bc}}$      & $0.2$                                                  & Mixed boundary condition coefficient               \\
$f_0$                              & $9.375\times 10^{-5}$ s$^{-1}$            & Mean Coriolis parameter   \\
$\beta$                           & $1.754\times 10^{-11}$ (m s)$^{-1}$    & Coriolis parameter gradient   \\
$L_d$                             & $(39, 22)$ km                                   & Baroclinic Rossby radii   \\
\hline
\end{tabular}
}
\caption{Common parameters for all the models.}
\label{tab:params-pub}
\end{center}
\end{table}

\begin{table}
\captionsetup{font=footnotesize}
\begin{center}
\resizebox{0.3\textwidth}{!}{%
\begin{tabular}{ccc}
\hline
$\Delta x$ (km)      & $\Delta t$ (s)    & $A_4$ (m$^4$ s$^{-1}$)  \\ \hline
5                             & 600                  & $2\times 10^{9}$  \\
40                           & 1200                & $5\times 10^{11}$   \\
80                           & 1440                & $5\times 10^{12}$   \\
120                         & 1800                & $1\times 10^{13}$    \\
\hline
\end{tabular}
}
\caption{Values of grid varying parameters.} 
\label{tab:params-pri}
\end{center}
\end{table}

\begin{figure}
\captionsetup{font=footnotesize}
\begin{center}
\includegraphics[width=3.8cm]{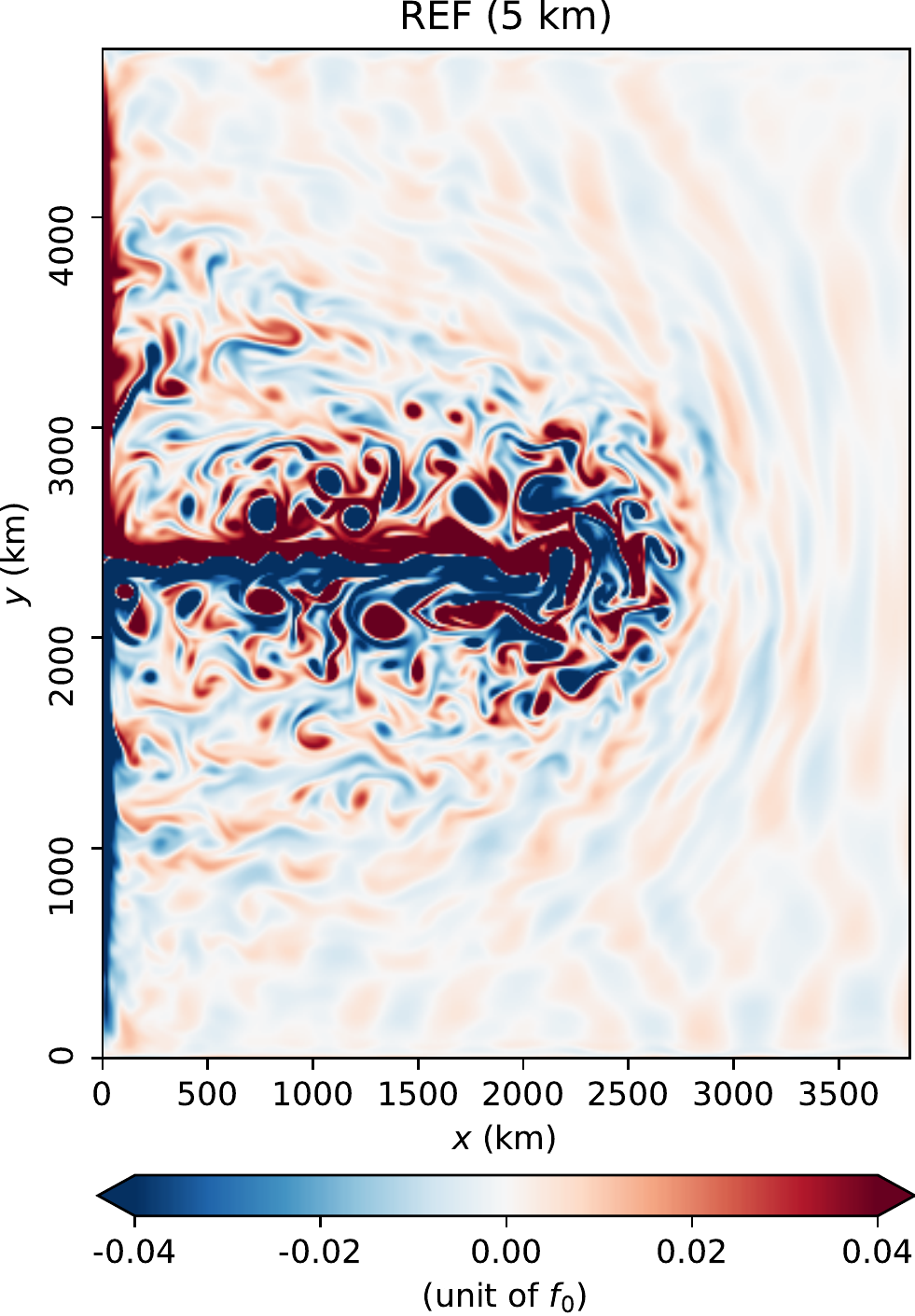}
\includegraphics[width=3.8cm]{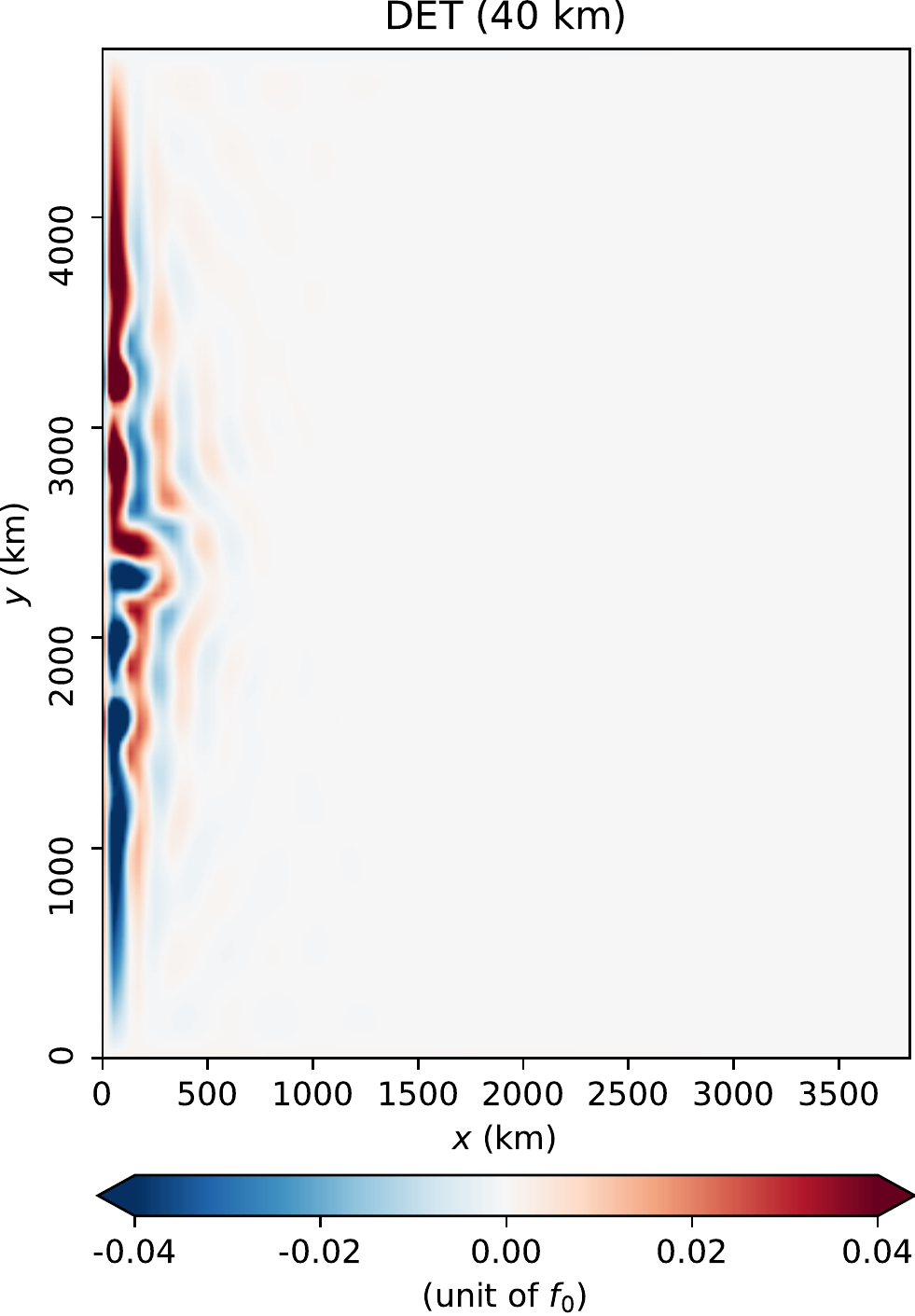}
\includegraphics[width=3.8cm]{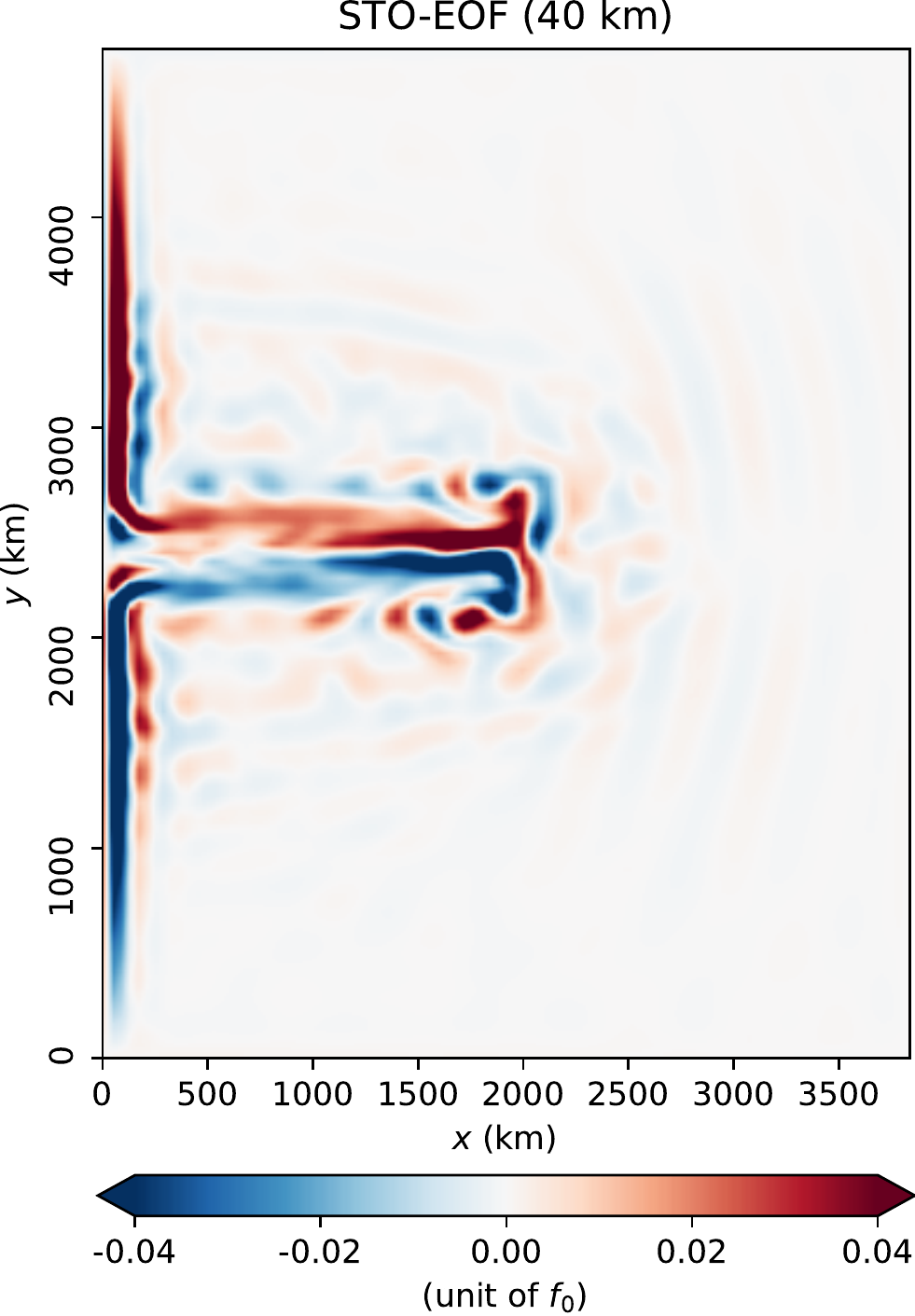}
\includegraphics[width=3.8cm]{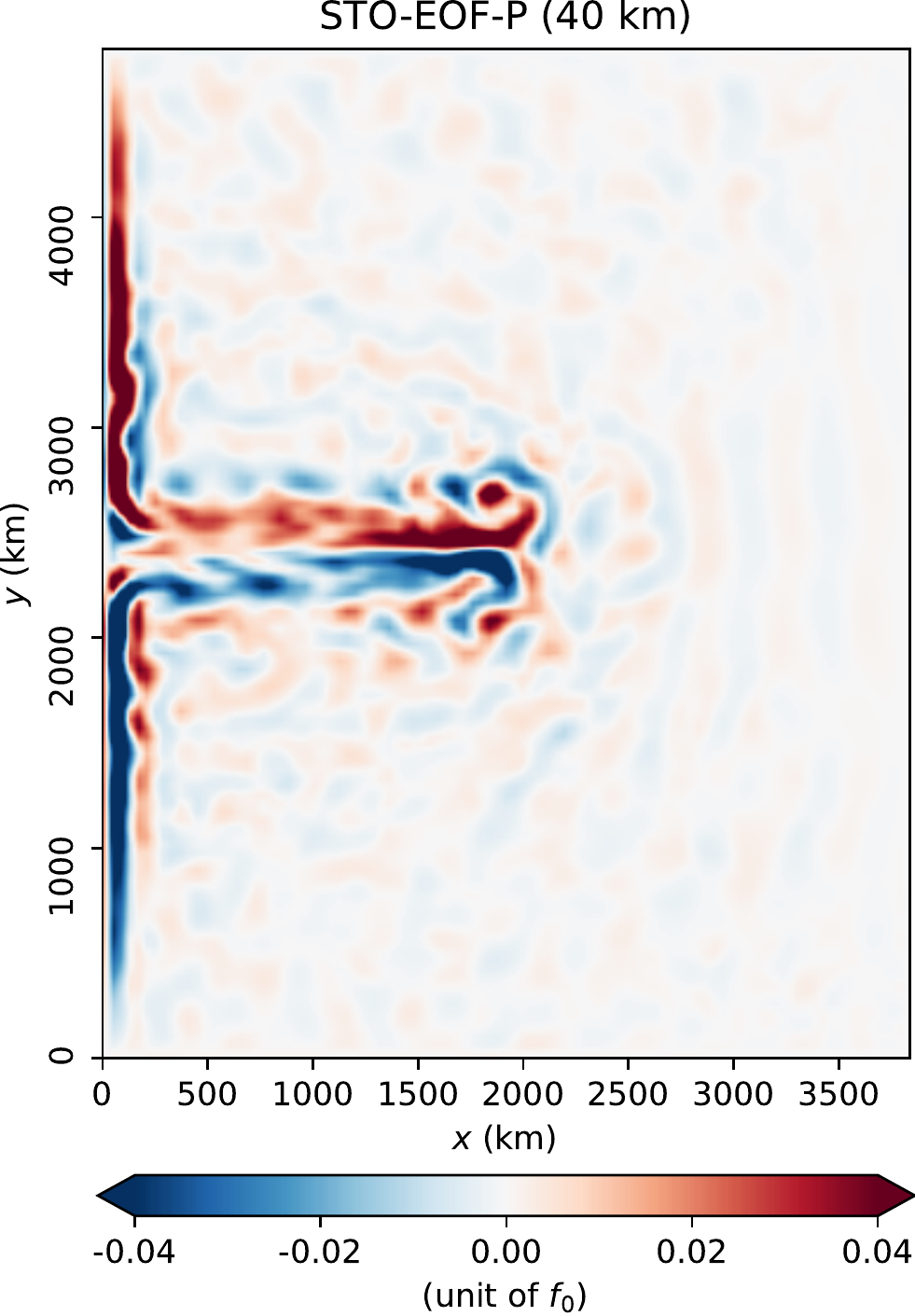} \\
\par\medskip
\includegraphics[width=3.8cm]{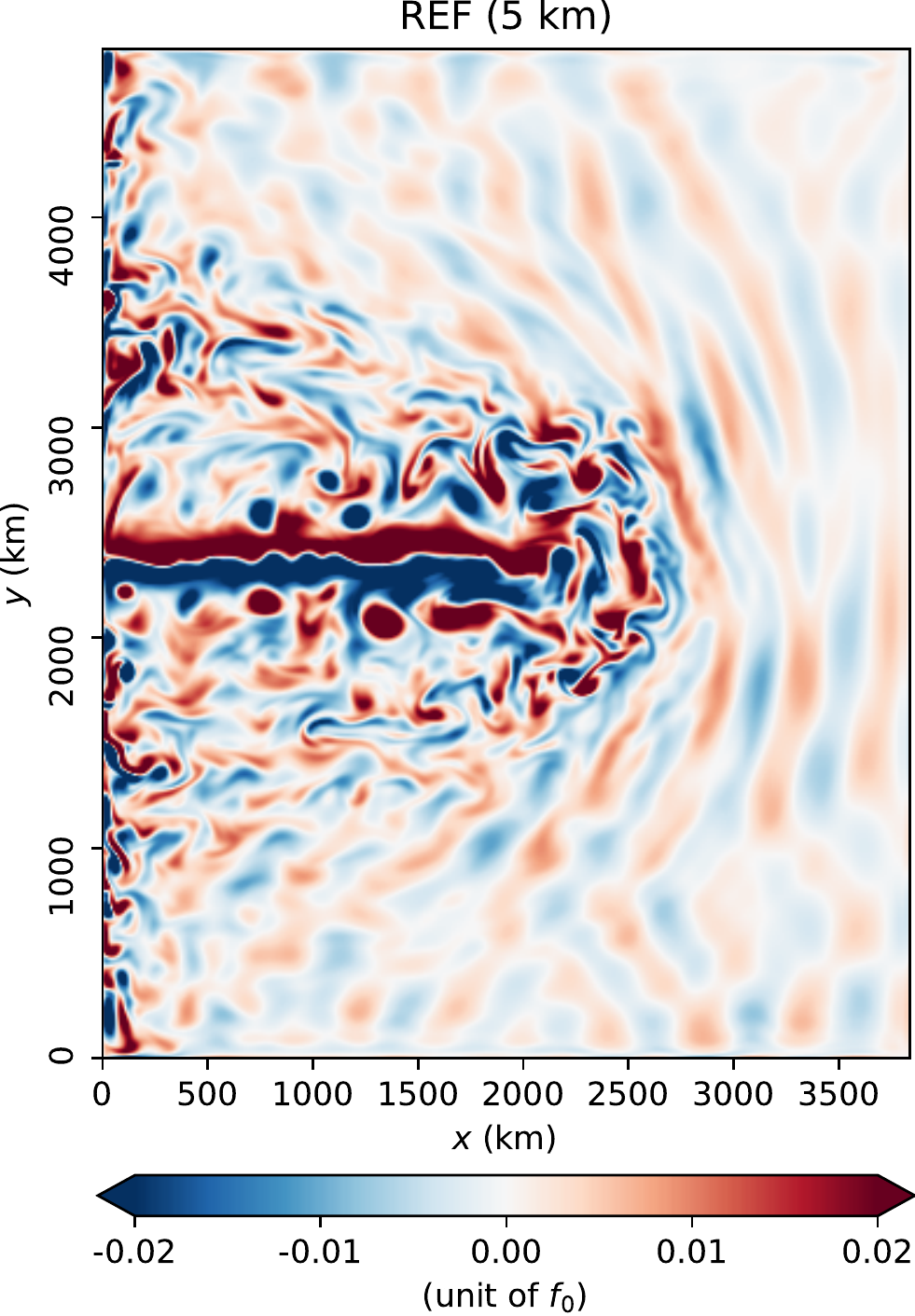}
\includegraphics[width=3.8cm]{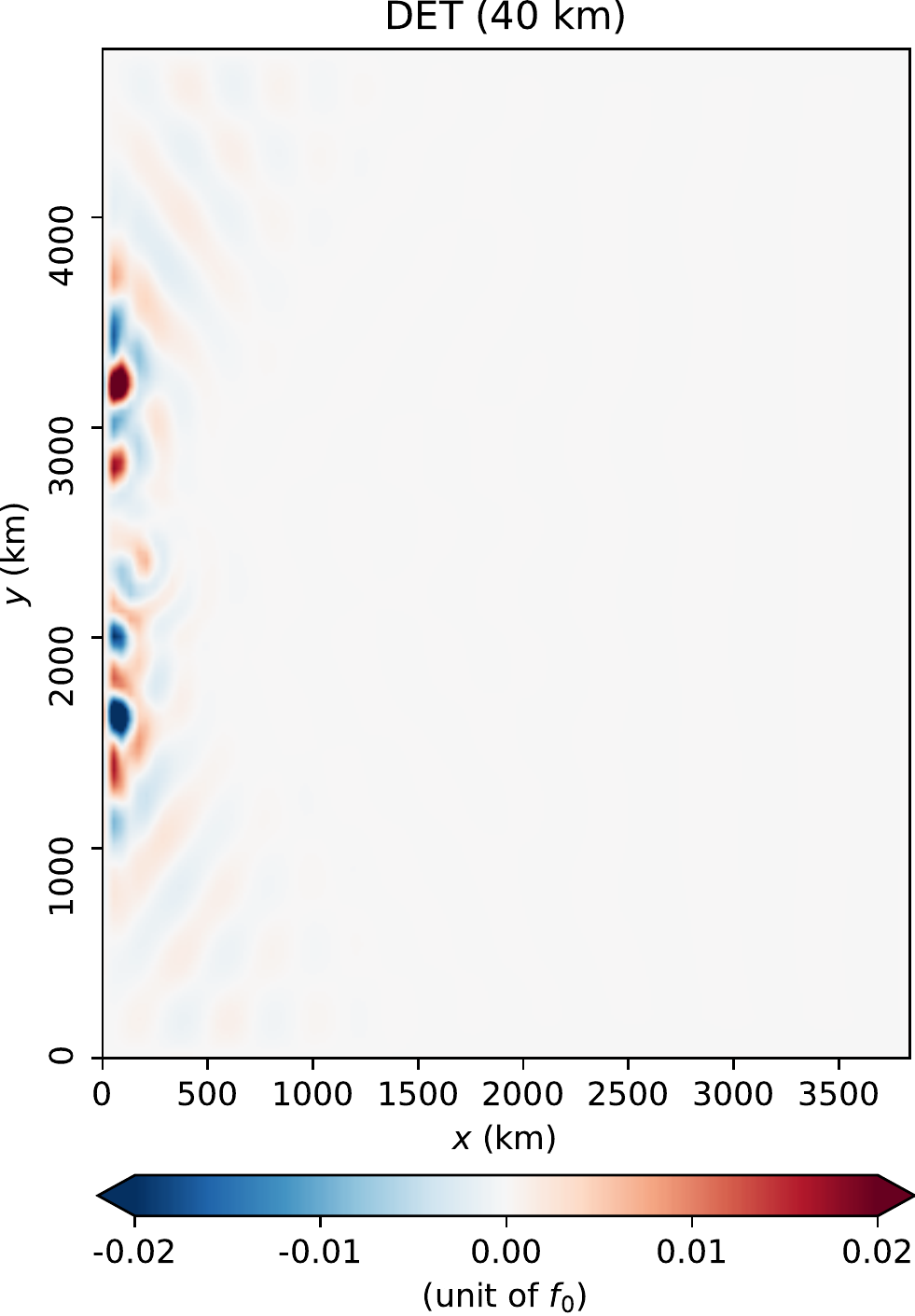}
\includegraphics[width=3.8cm]{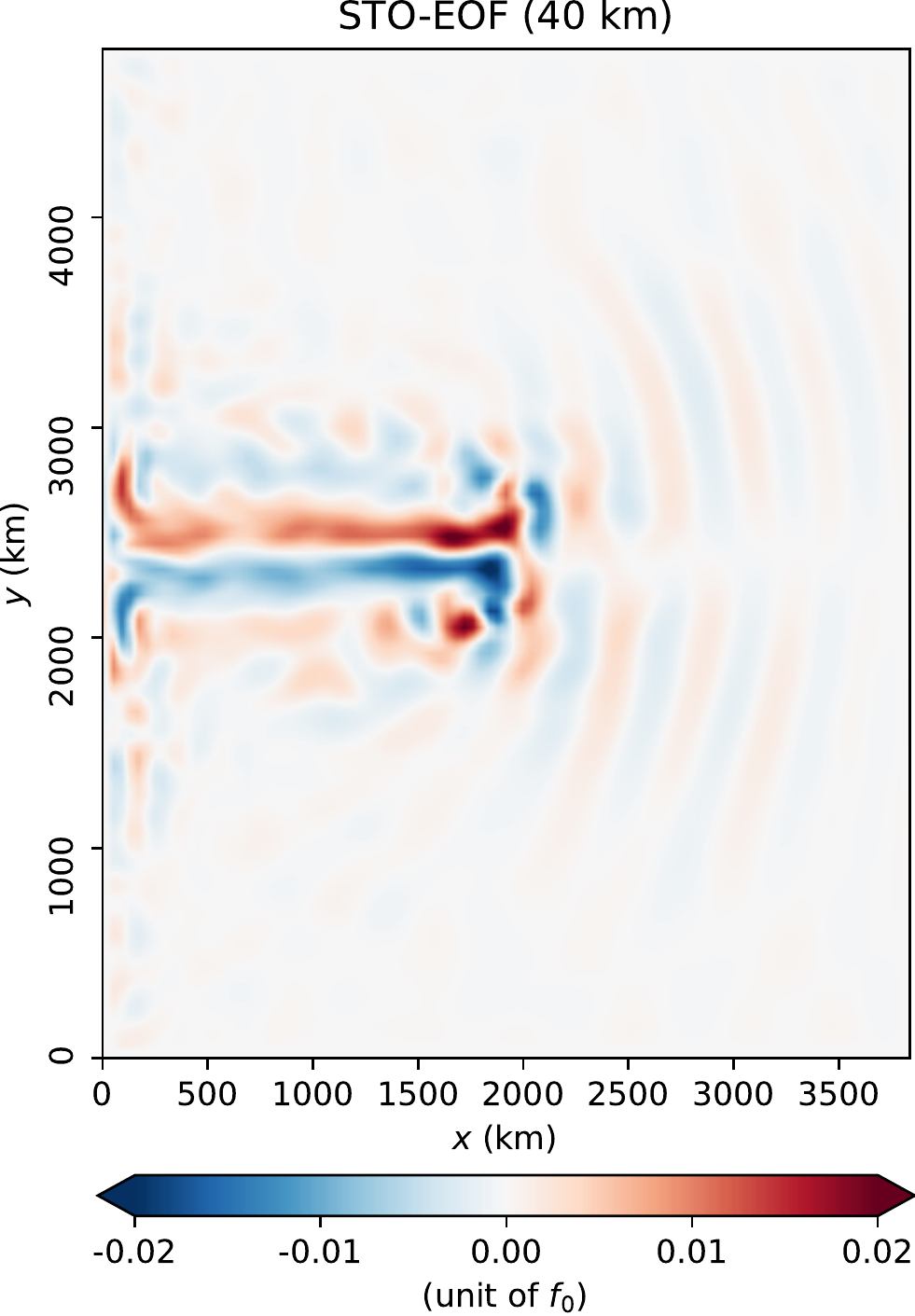}
\includegraphics[width=3.8cm]{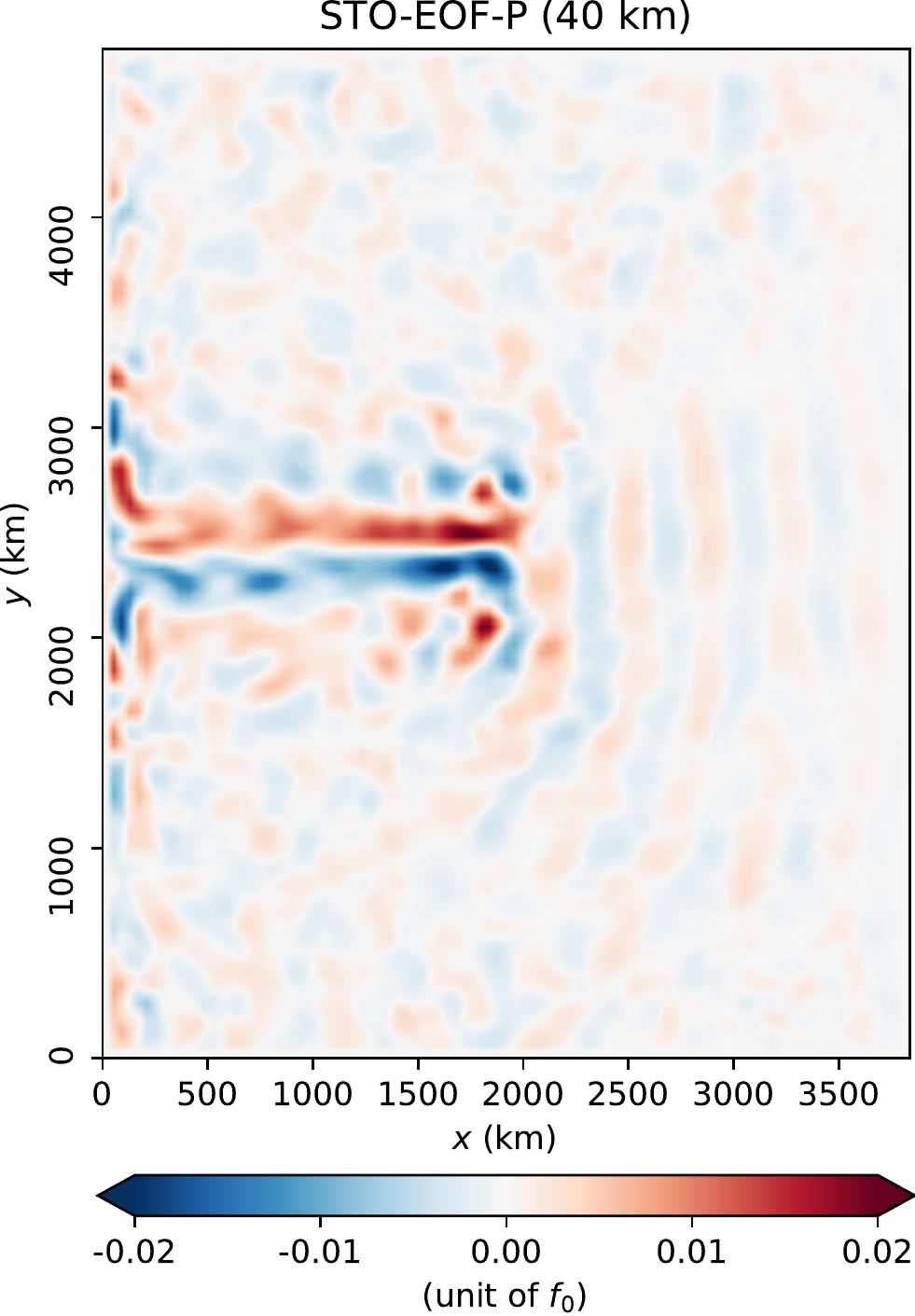}
\end{center}
\caption{Snapshots of relative vorticity (divided by $f_0$) for the two upper layers (by rows) provided by different simulations (by columns) after 60-years integration.}
\label{fig:spot-vort}
\end{figure}

Snapshots of the relative vorticity on the two top layers are shown in Figure~\ref{fig:spot-vort}. The dynamics of \emph{REF} model is mainly characterized by the energetic eastward jet with adjacent recirculations and by the fast westward Rossby waves. The former results from the most active mesoscale eddies effect through baroclinic instability. However, this instability cannot be resolved once the horizontal resolution is similar or greater than the baroclinic deformation radius maximum (39 km here). For instance, without any eddy parameterization, the \emph{DET} (40 km) simulation only generates a smooth field due to an over-dissipation effect. 

\begin{figure}
\captionsetup{font=footnotesize}
\begin{center}
\includegraphics[width=3.8cm]{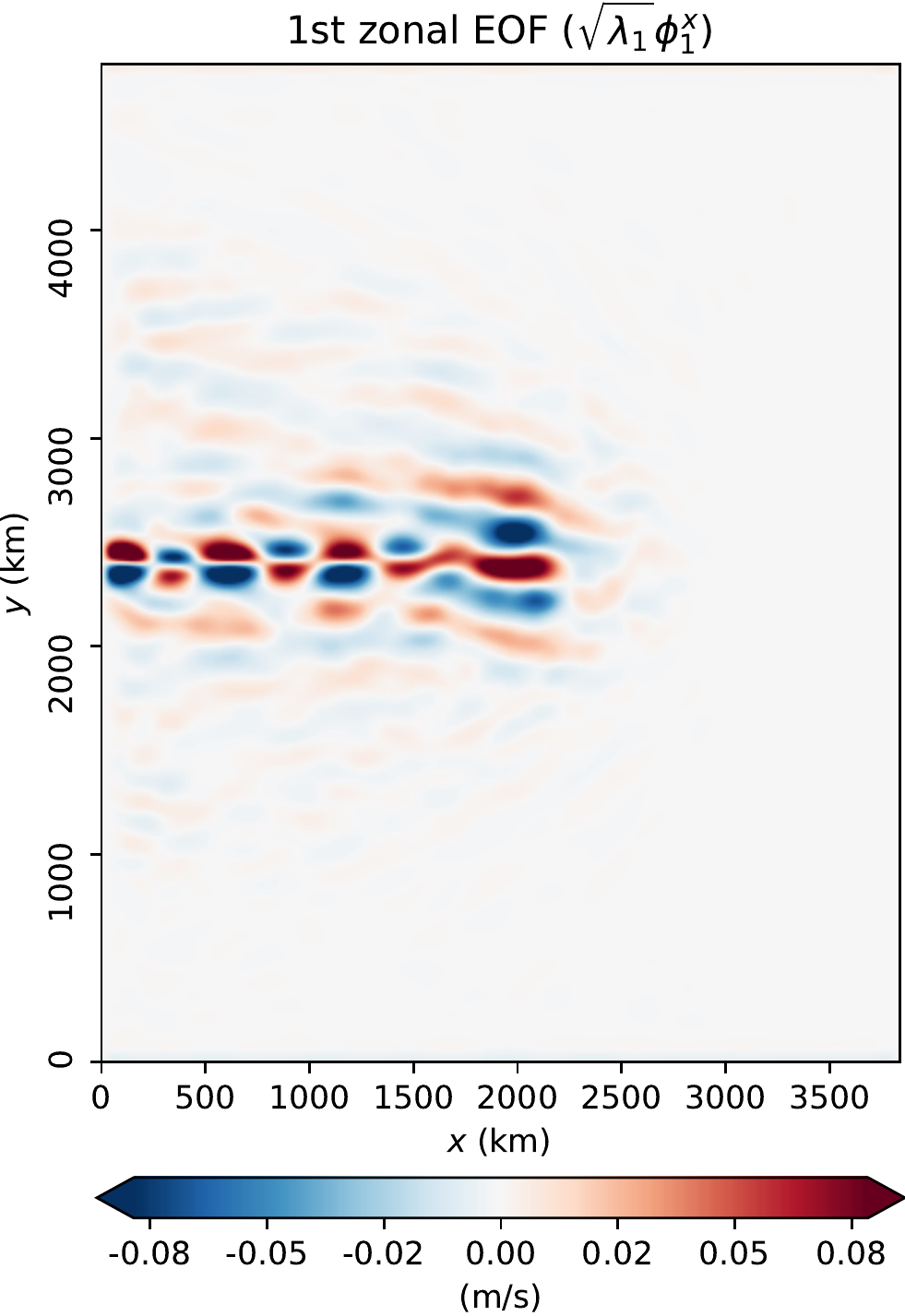}
\includegraphics[width=3.8cm]{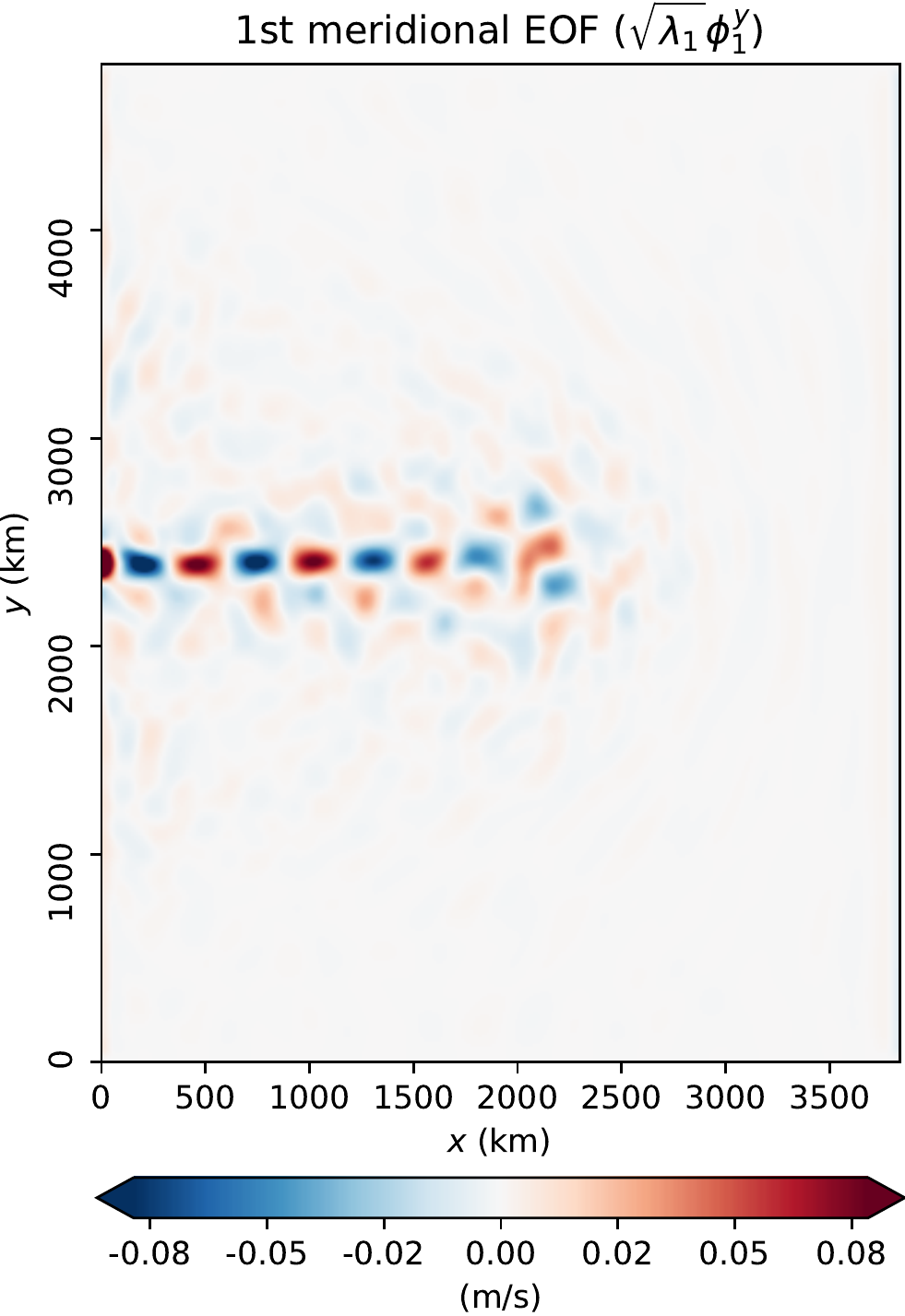}
\includegraphics[width=3.8cm]{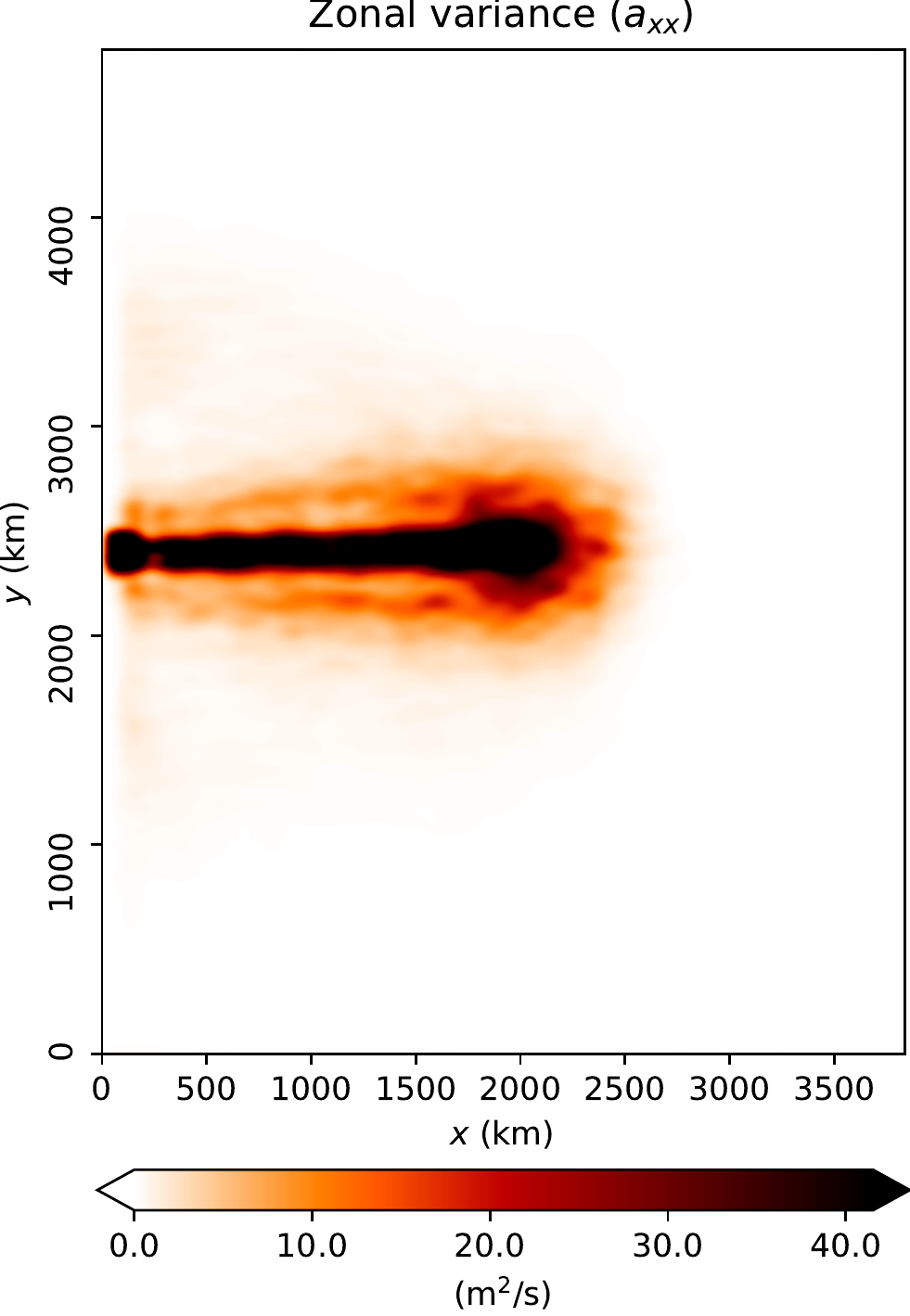}
\includegraphics[width=3.8cm]{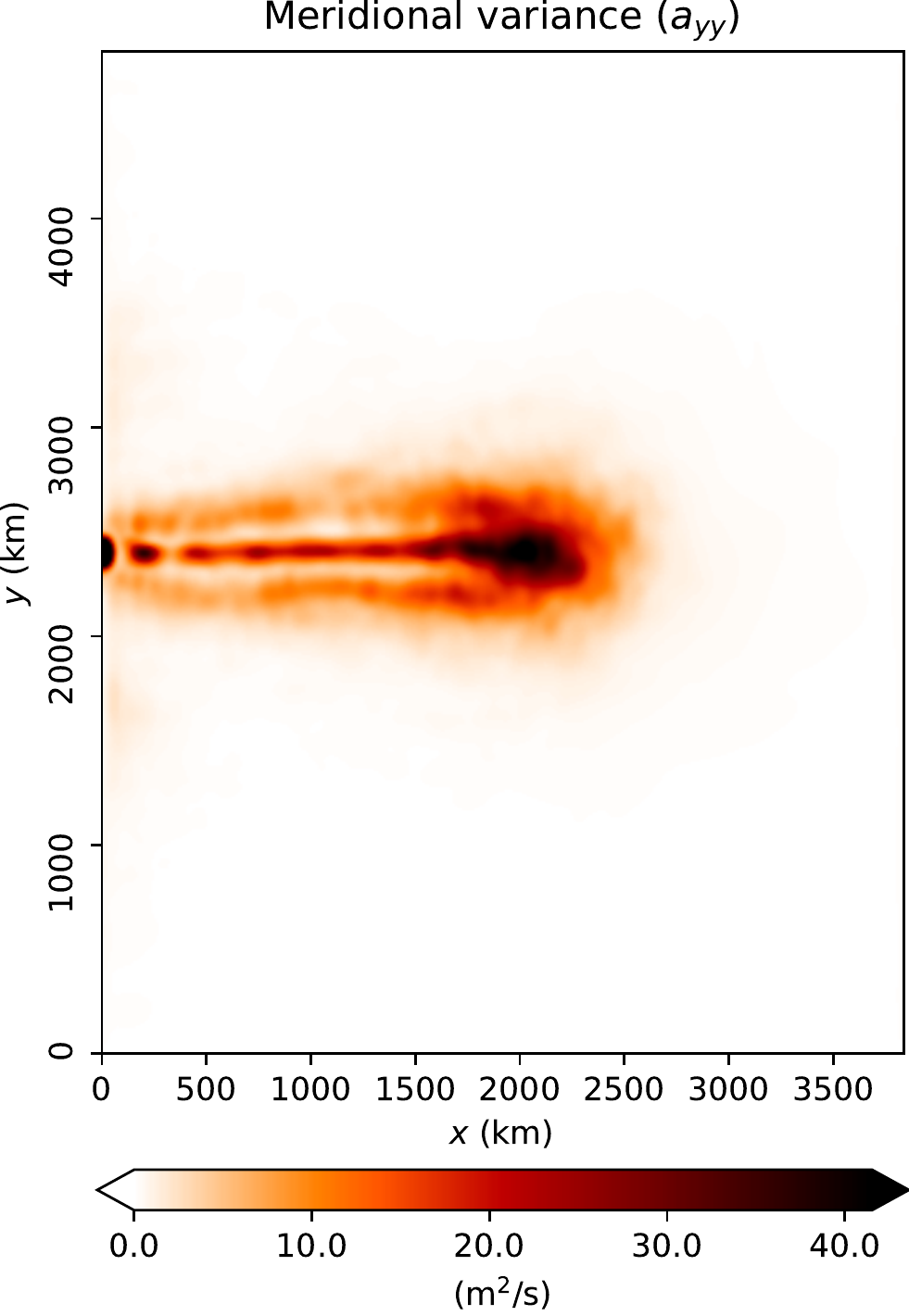}
\end{center}
\caption{Illustration for noise's first EOF (zonal and meridional components in first two sub-figures) and for noise's variance (diagonal components in last two sub-figures) at resolution 40 km.}
\label{fig:spot-noi}
\end{figure}

To calibrate the structure of the unresolved flow (noise and Girsanov drift), the \emph{REF} data are sampled at each day during the first ten years after the spin-up, this is followed by the procedure presented in Section \ref{sec:EOF} with a fixed 200 km wide Gaussian filter. Considering the computational cost, we employ only 50 EOFs to build the noises \eqref{eq:EOF-noi} for the stochastic simulations at different resolutions. However, the eigenvalues of these truncated EOFs are amplified in order to capture 95 \% energy of the full set of modes. An example of the first EOF ($\sqrt{\lambda_1} \bp_1$) and the noise variance ($\ba$ defined in \eqref{eq:EOF-a}) is illustrated by Figure~\ref{fig:spot-noi}. The noise structure is mainly characterized by the most energetic fluctuations in the jet region. 

Including the additional advection of PV by the calibrated unresolved flow component, both \emph{STO-EOF} and \emph{STO-EOF-P} models are able to reproduce the eastward jet on the coarse mesh, and the latter seems to enhance fluctuations everywhere. For instance, such result is illustrated at a resolution of 40 km by Figure~\ref{fig:spot-vort}. The ability of these coarse models to reproduce the statistical properties of the \emph{REF} model will be diagnosed and analyzed more precisely in the next section. 

\subsection{Statistical diagnostics}

To investigate the prediction of the long-term statistics, we sample the data (streamfunction) of each run at 15-day intervals over the 120 years (after the spin-up) and compute the statistics over the last 100 years.

\begin{figure}
\captionsetup{font=footnotesize}
\begin{center}
\includegraphics[width=3.8cm]{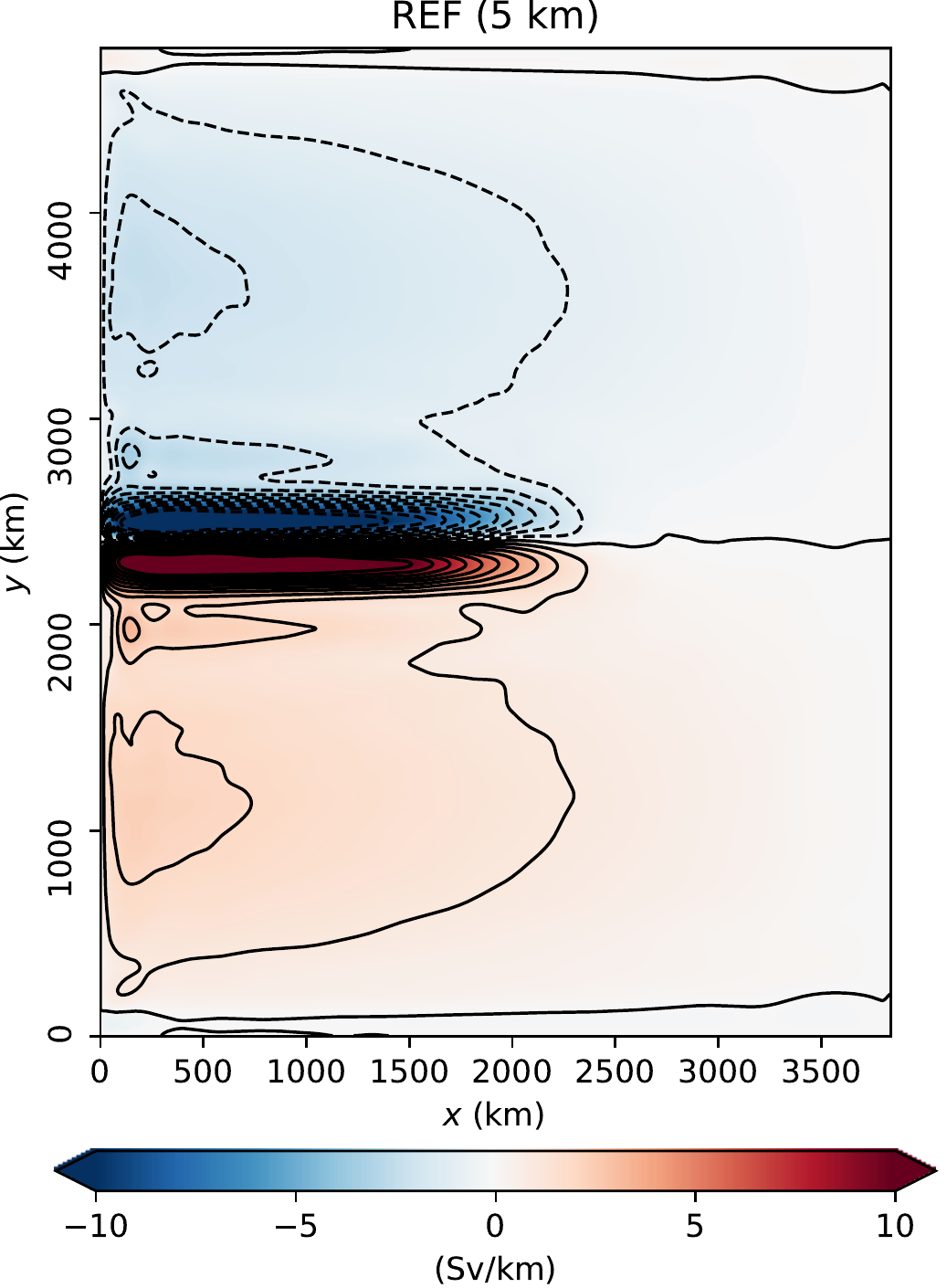}
\includegraphics[width=3.8cm]{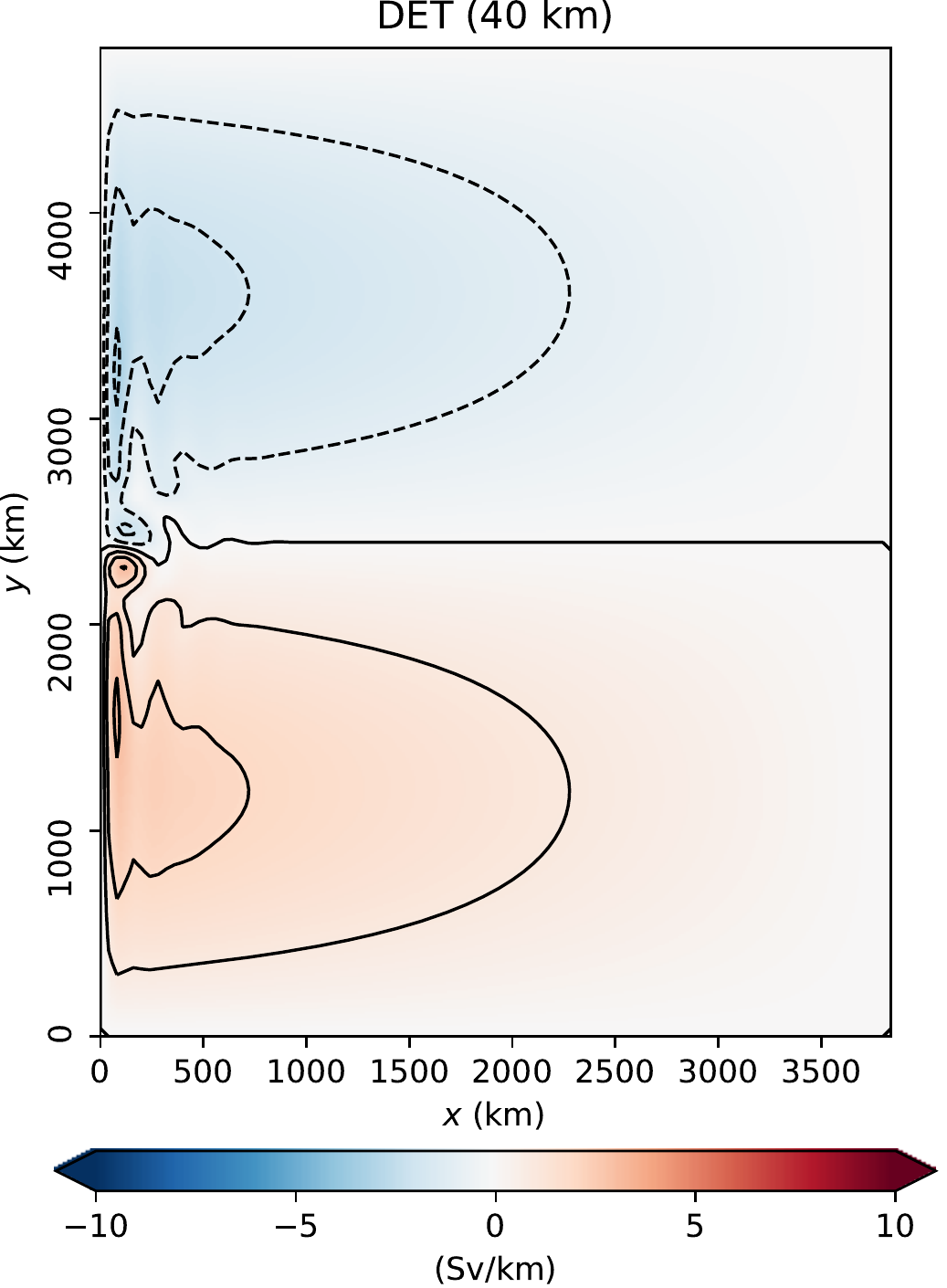} 
\includegraphics[width=3.8cm]{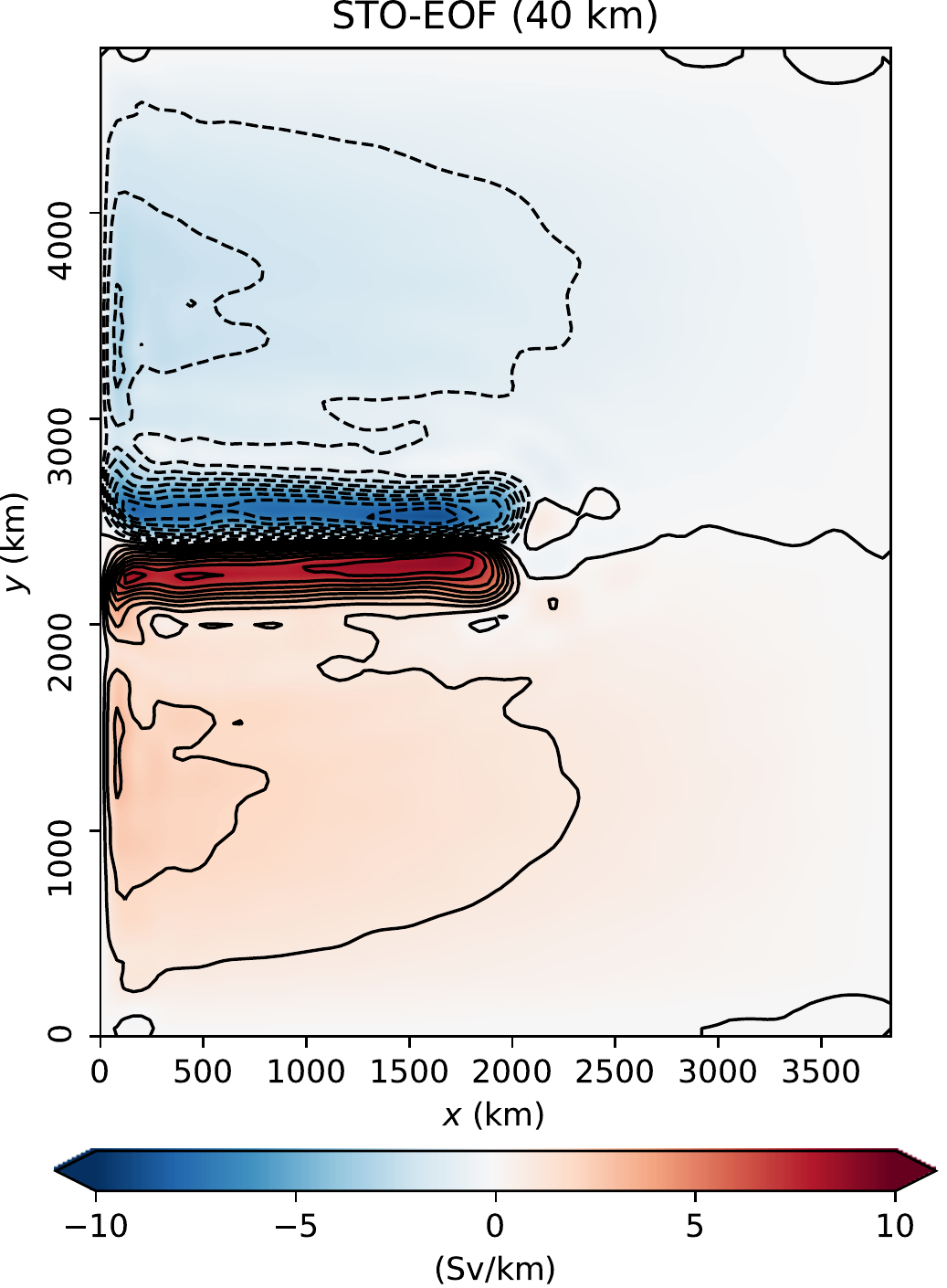}
\includegraphics[width=3.8cm]{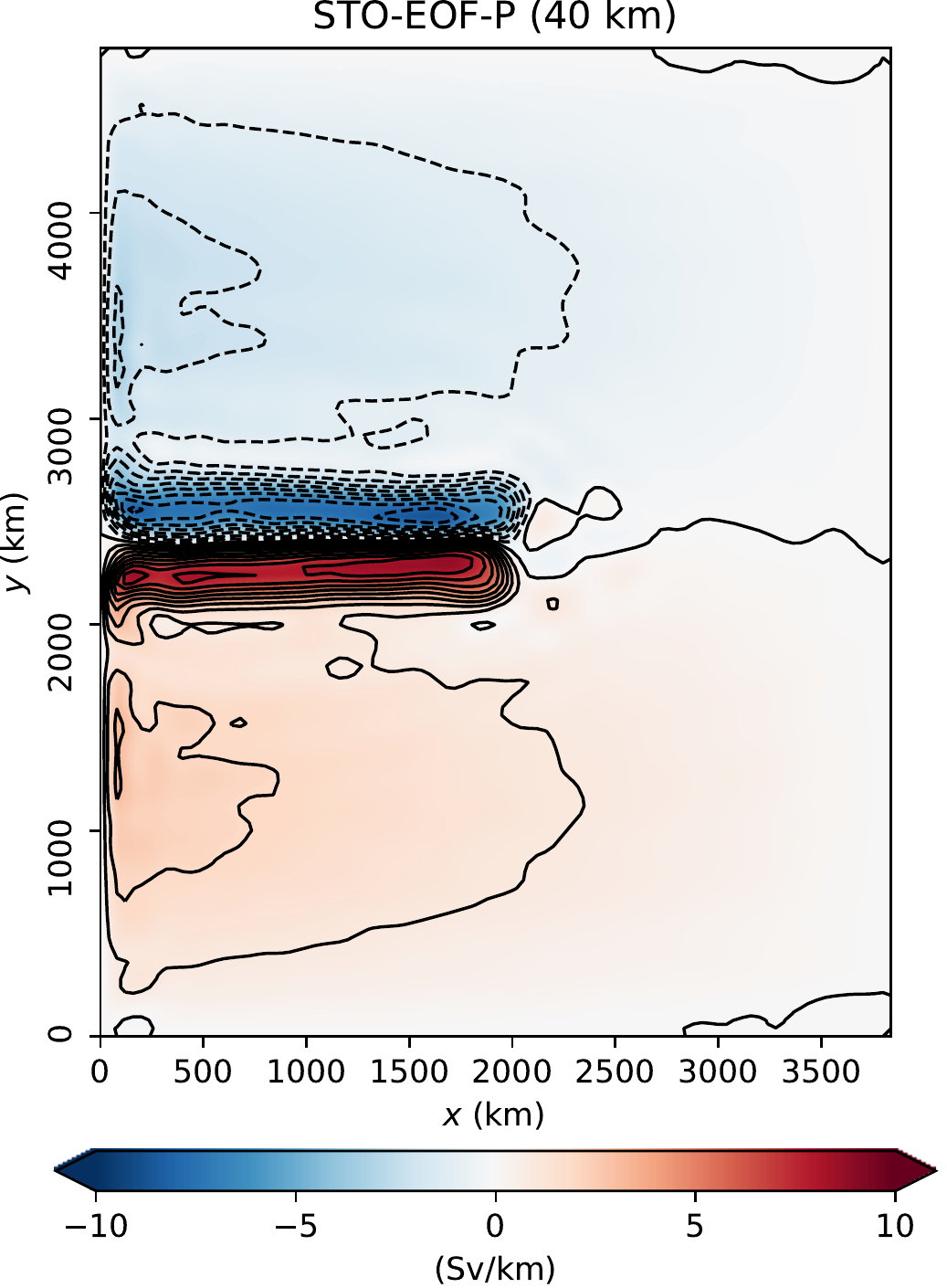} \\
\par\medskip
\includegraphics[width=3.8cm]{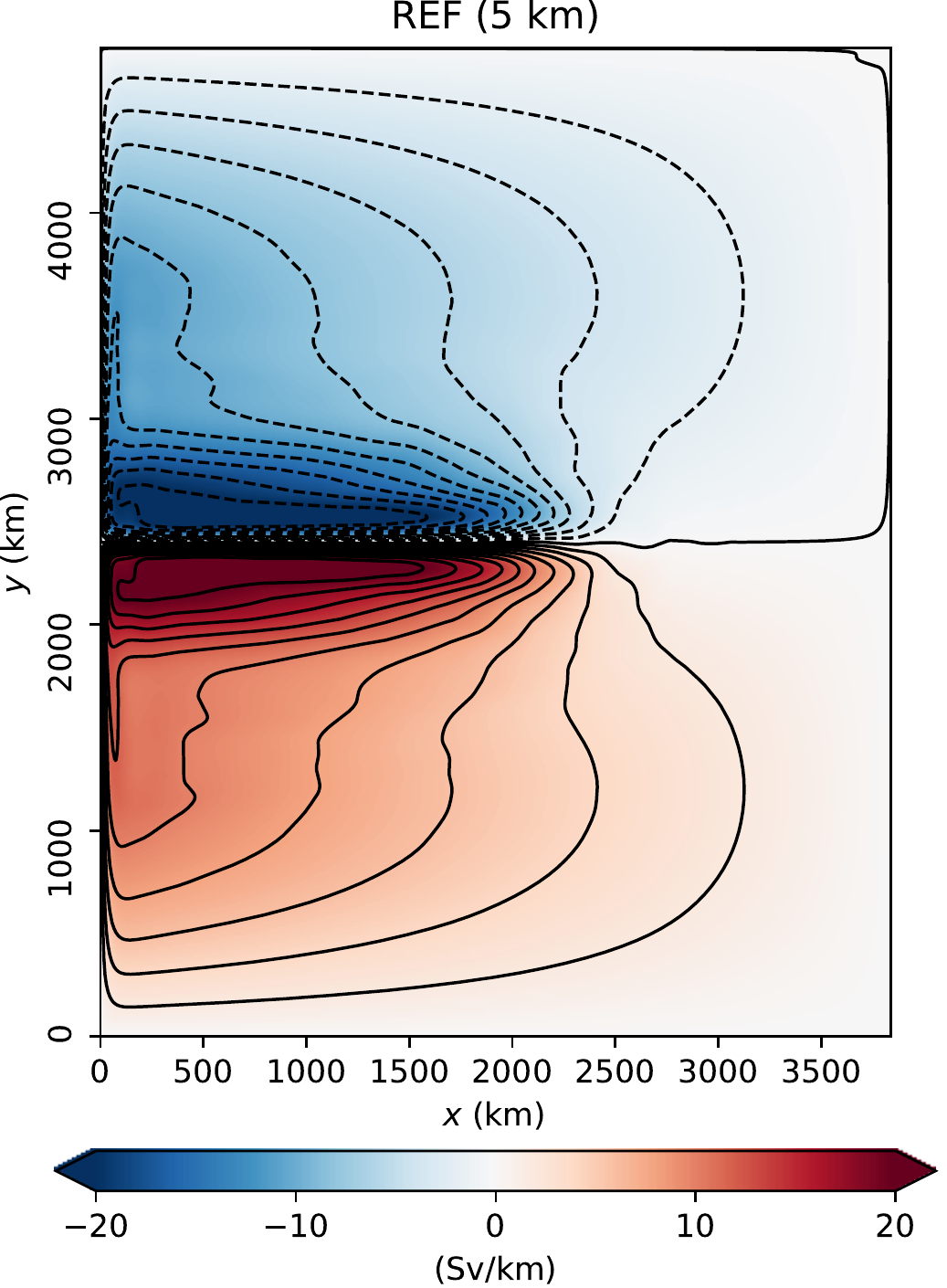}
\includegraphics[width=3.8cm]{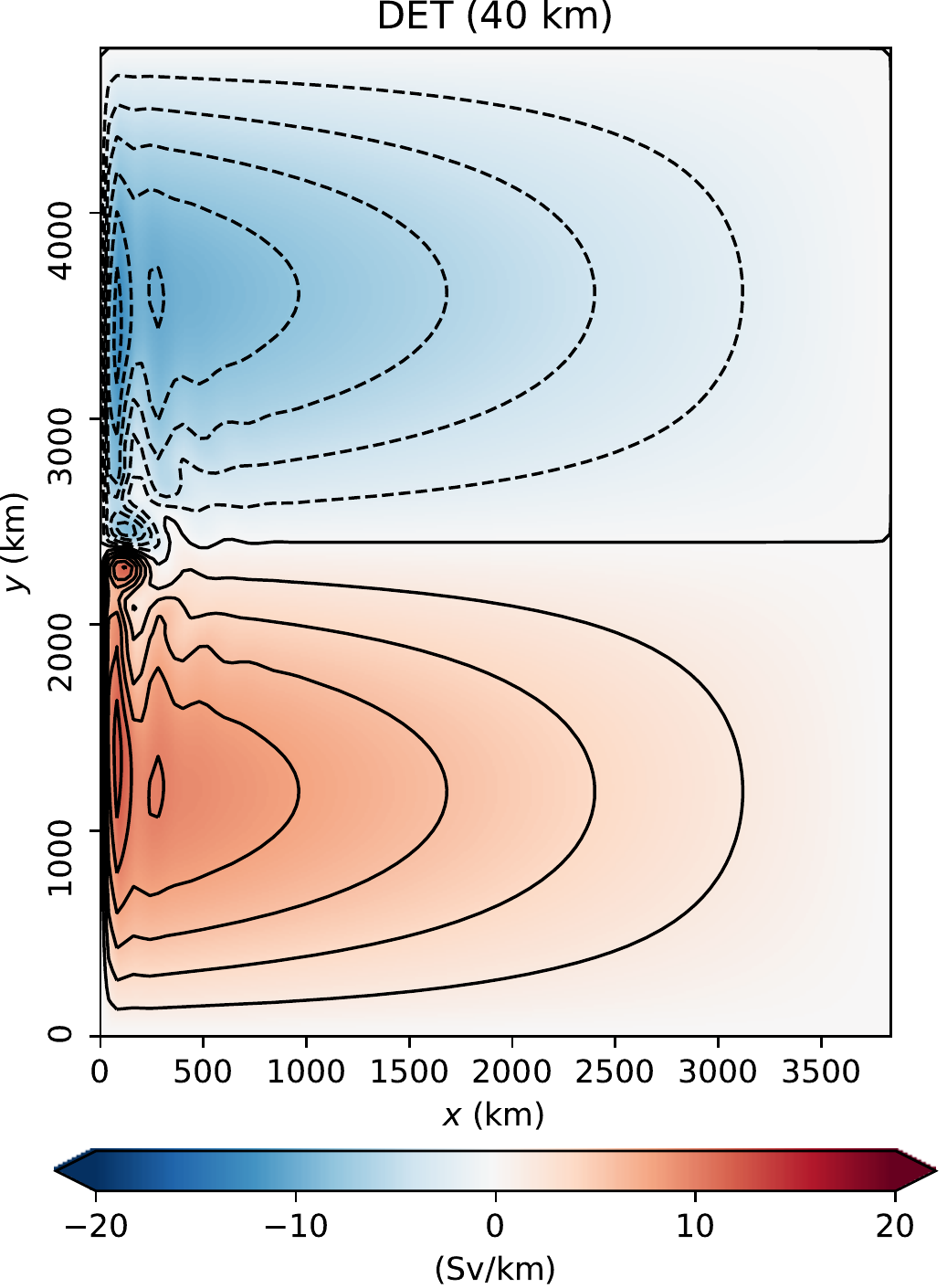} 
\includegraphics[width=3.8cm]{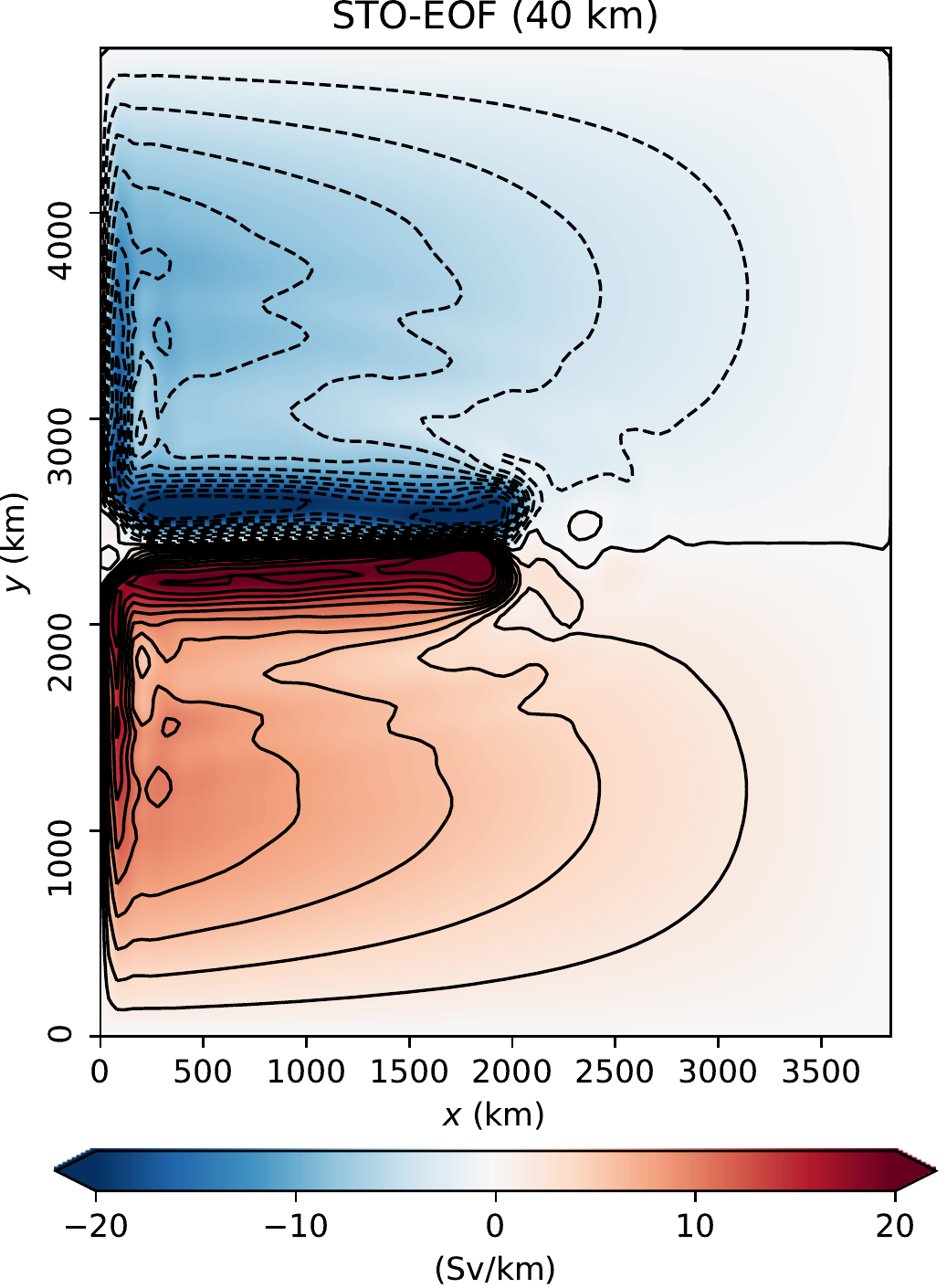}
\includegraphics[width=3.8cm]{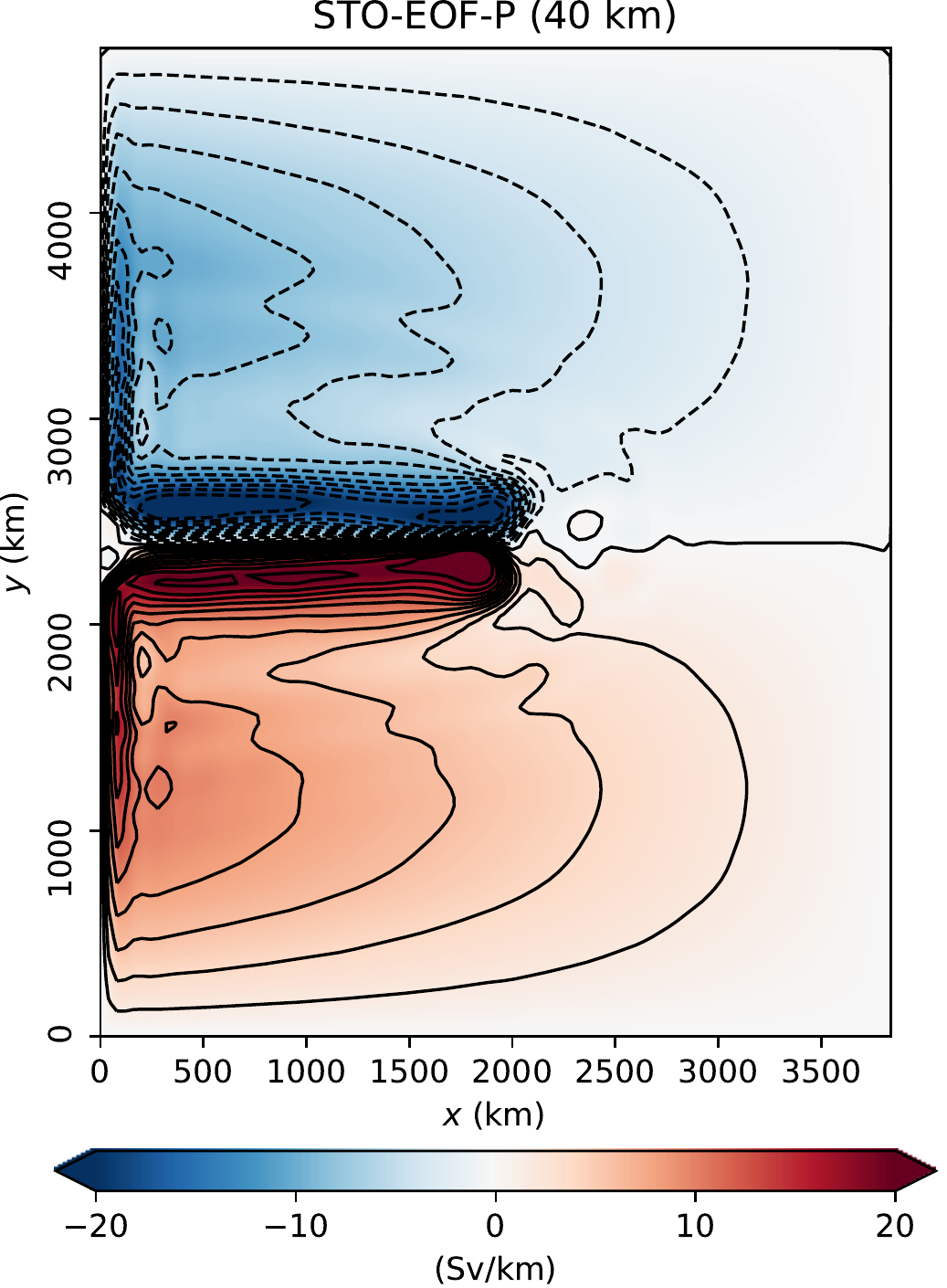}
\end{center}
\caption{Comparison of time-mean contour of barotropic (top row) and first baroclinic (bottom row) streamfunctions for different models (by columns). Note that 1 Sv = $10^6$ m$^3$/s.}
\label{fig:mean-pm40}
\end{figure}

We first compare the temporal mean of the modal streamfunctions for both coarse models to that subsampled (in space) of the eddy-resolving model. From Figure \ref{fig:mean-pm40}, we observe that both barotropic and baroclinic modal mean of the \emph{REF} characterize the eastwards jet, and the two stochastic models (at 40 km) enable to reproduce qualitatively the local structures of both vertical modes predicted by the \emph{REF}. Conversely, the \emph{DET} model can only capture the symmetric double-gyre structure. 

To quantify the models' statistics at different resolutions, we then employ the spatial root-mean-square error (RMSE) of the temporal mean and the standard deviation (std) between each coarse model and the subsampled \emph{REF} one. To assess in a single measure the mean and the variance reconstruction, we adopt the Gaussian relative entropy (GRE) proposed by \citet{Grooms2015} that is defined as
\beq\label{eq:GRE}
\text{GRE} = \frac{1}{|\dom|} \int_{\dom} \alf \left( \frac{\big( \overline{\psi^{\sub\text{M}}}^t - \overline{\psi^{\sub\text{R}}}^t \big)^2}{\sigma^2 [\psi^{\sub\text{M}}]} + \frac{\sigma^2 [\psi^{\sub\text{R}}]}{\sigma^2 [\psi^{\sub\text{M}}]} - 1 - \log \Big( \frac{\sigma^2 [\psi^{\sub\text{R}}]}{\sigma^2 [\psi^{\sub\text{M}}]} \Big) \right)\, \dx,
\eeq
where $\sigma$ denotes the std, $\psi^{\sub\text{M}}$ (resp. $\psi^{\sub\text{R}}$) stands for the streamfunction of the coarse model (resp. for that subsampled from the \emph{REF}). 

\begin{figure}
\captionsetup{font=footnotesize}
\begin{center}
\includegraphics[width=4.8cm]{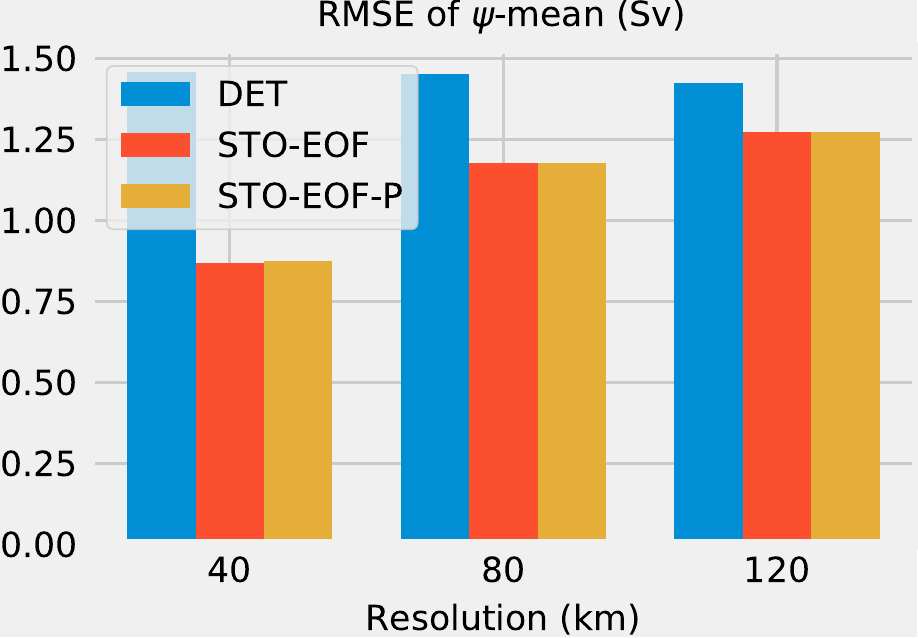} 
\includegraphics[width=4.8cm]{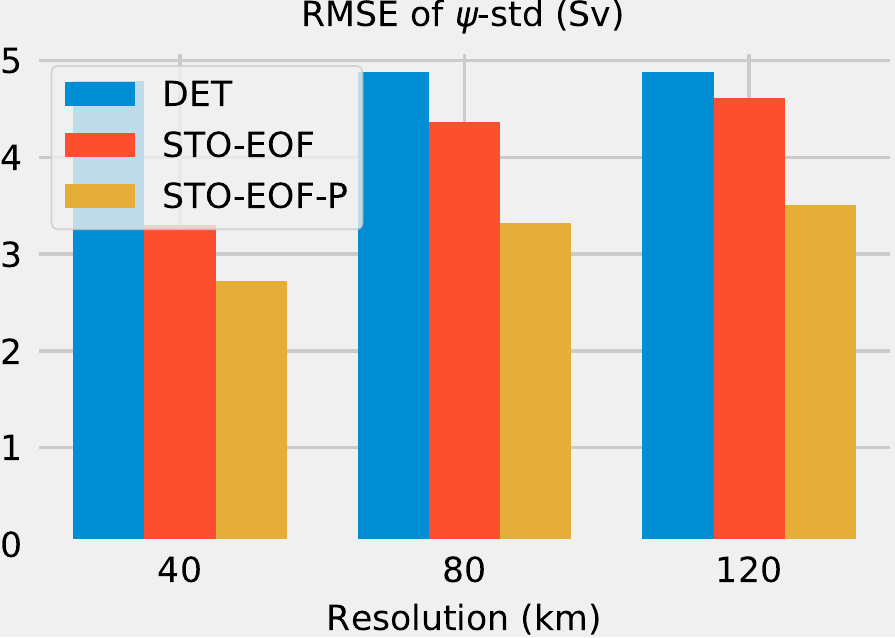} 
\includegraphics[width=4.8cm]{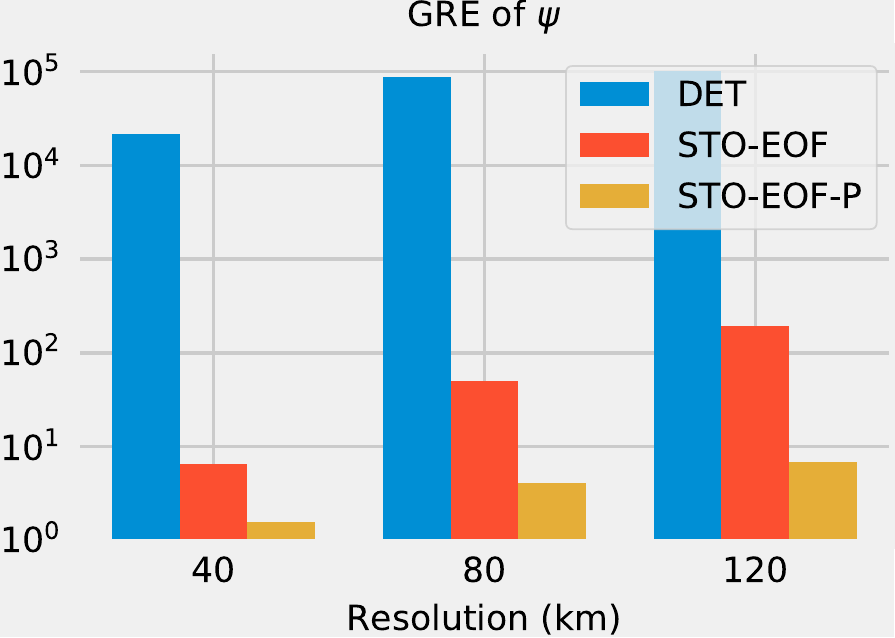}
\end{center}
\caption{Comparison of variability measures (by sub-figures) for different coarse models (by colors and groups in each sub-figure). Note that the $y$-axis of the last sub-figure is in $\log$-scale due to poor variability of the \emph{DET} models.}
\label{fig:stat}
\end{figure}

These three global criterion are computed for all the coarse models at different resolutions and the results are summarized in Figure \ref{fig:stat}. For all the resolutions tested the two random models have lower errors than the \emph{DET} model both in terms of mean and std. In addition, the \emph{STO-EOF-P} models enable to reduce the RMSE of the std and the GRE, which means an improvement on the prediction of variability.

\begin{figure}
\captionsetup{font=footnotesize}
\begin{center}
\includegraphics[width=5cm]{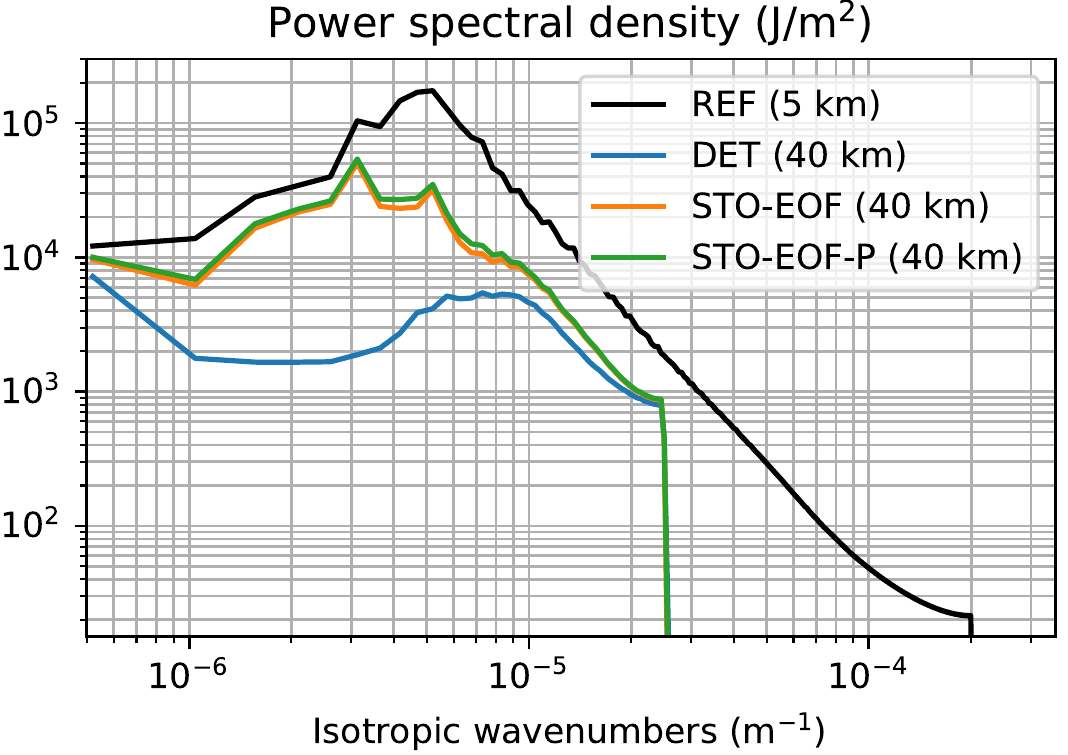}
\includegraphics[width=5cm]{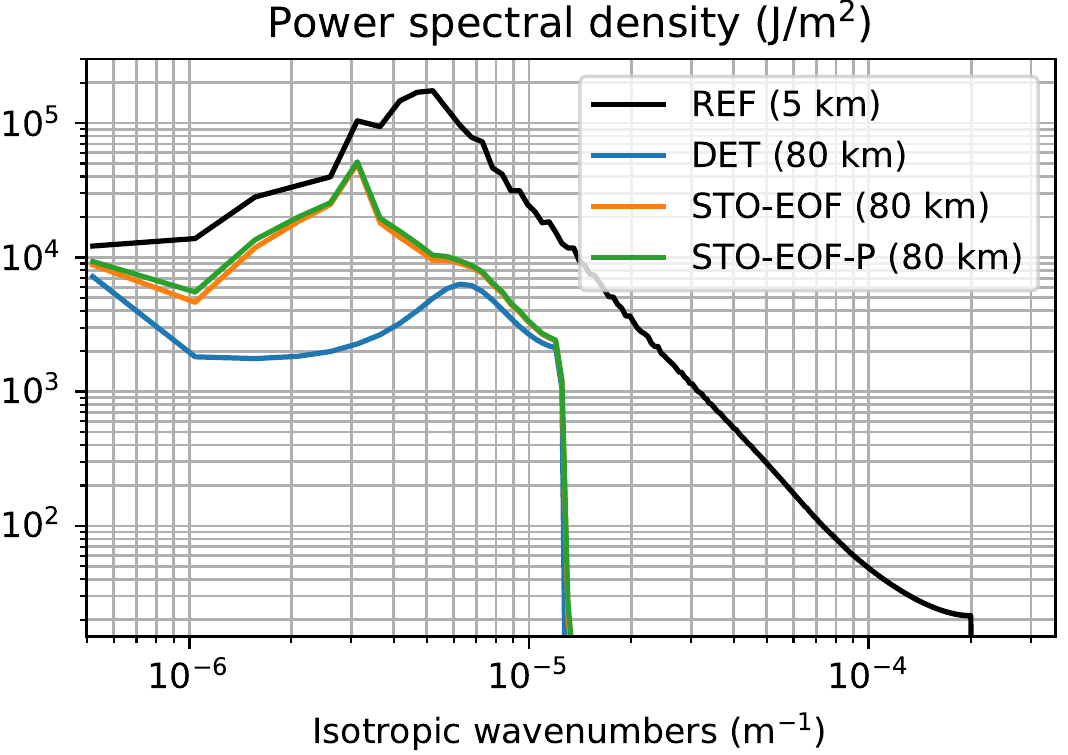} 
\includegraphics[width=5cm]{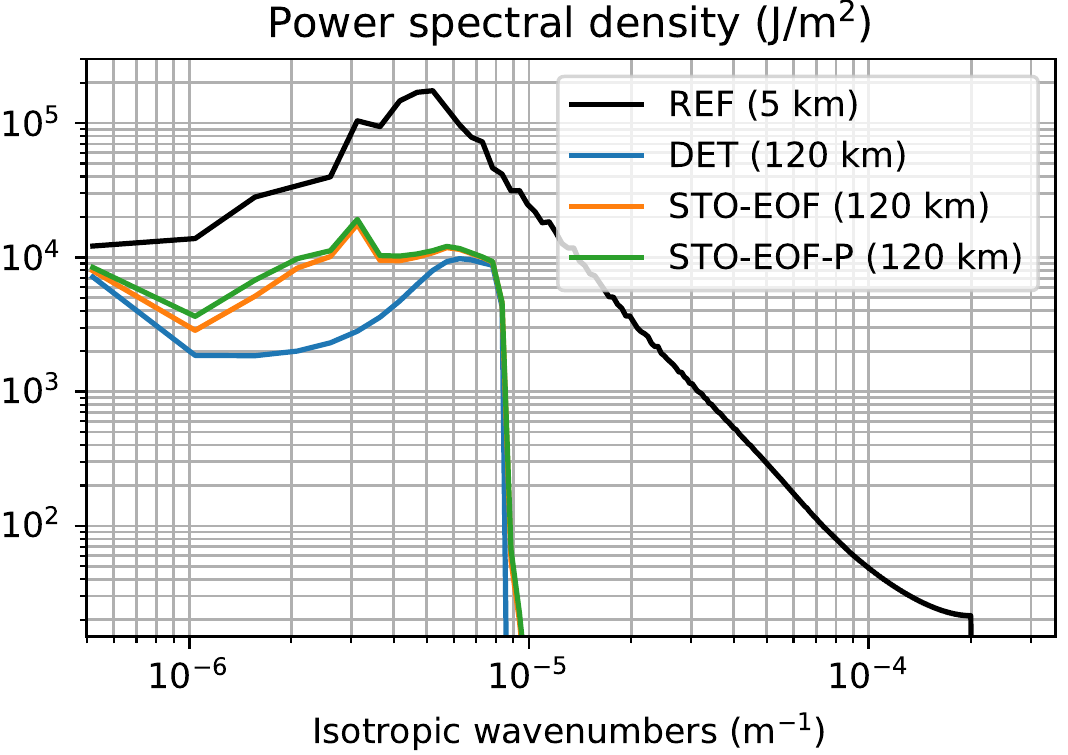} 
\end{center}
\caption{Temporal mean of vertically integrated KE spectra for different models (by colors) at different resolutions (by sub-figures).}
\label{fig:spec-KE}
\end{figure}

We investigate subsequently the models' variability based on the energy analysis. The time-averaged KE spectra provided by different models are compared in Figure~\ref{fig:spec-KE}. Unsurprisingly, compared to the \emph{REF}, the \emph{DET} coarse models produce a severe lack of resolved KE \citep{Arbic2013, Kjellsson2017} due to the excessive dissipation without any eddy parameterization. Both stochastic models bring more energy over all the wavenumbers, particularly at large scales. Moreover, when the resolution increases (from 120 km to 40 km), their spectra slopes in the inertial range become more and more closer to that of the \emph{REF} model.

\begin{figure}
\captionsetup{font=footnotesize}
\begin{center}
\includegraphics[width=3.8cm]{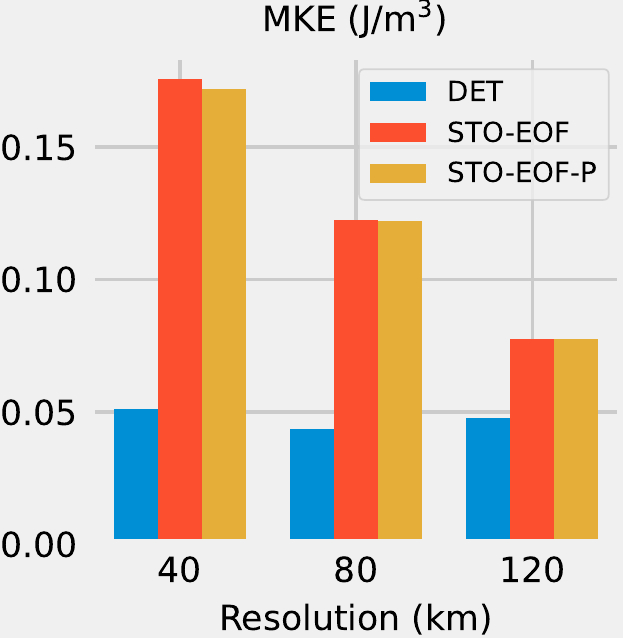}
\includegraphics[width=3.8cm]{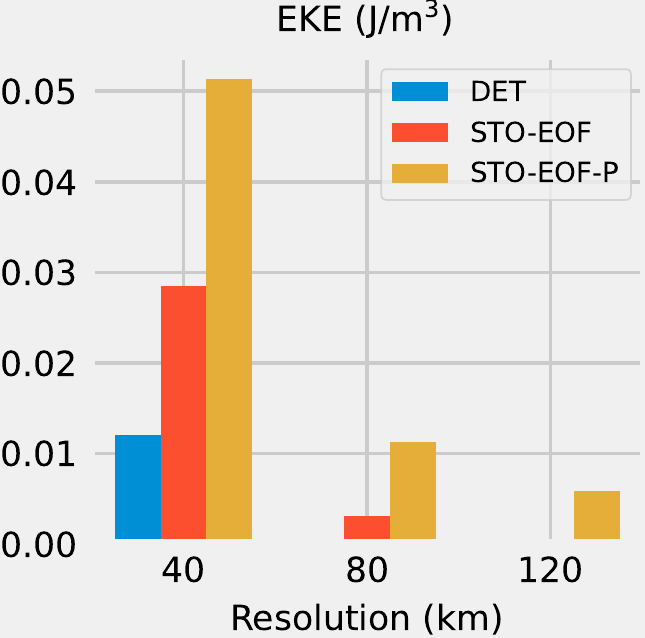} 
\includegraphics[width=3.8cm]{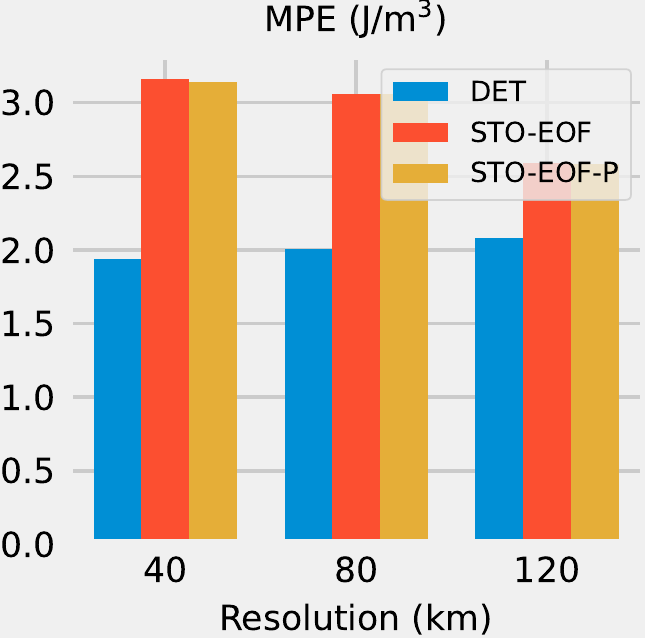}
\includegraphics[width=3.8cm]{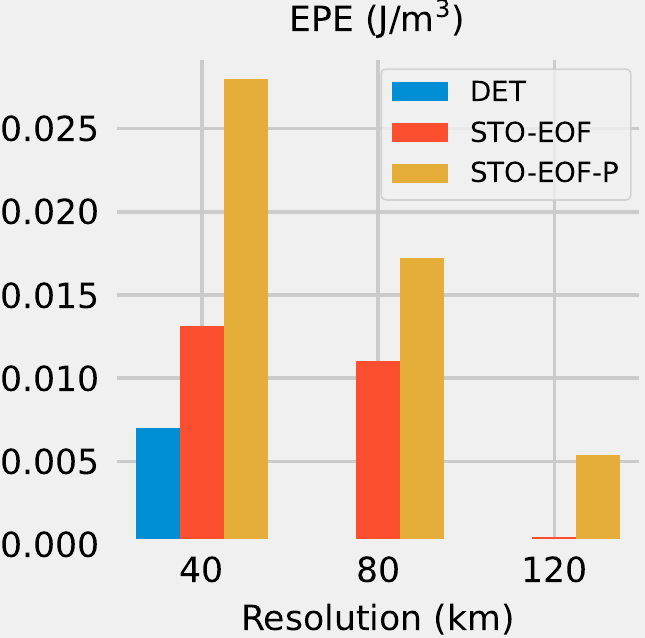}
\end{center}
\caption{Comparison of the global MKE, EKE, MPE and EPE (by sub-figures) for different coarse models (by colors and groups in each sub-figure). The corresponding values of the \emph{REF} model: MKE = 0.23 J/m$^3$, EKE = 0.62 J/m$^3$, MPE = 4.01 J/m$^3$ and EPE = 0.34 J/m$^3$.}
\label{fig:KEdecomp}
\end{figure}

We compare finally the energy decomposition as described in Section \ref{sec:cont-mod} for all the models. However, we focus only on a single realization here and the eddy-mean decomposition of $\bu$ and $b$ is simply build from a global time-average. From Figure~\ref{fig:KEdecomp}, we observe that both random models have higher MKE and MPE than the \emph{DET} model. In particular, at resolution 40 km, they are almost at the same order as that of the \emph{REF} model. Compared to the \emph{STO-EOF} models, \emph{STO-EOF-P} models produce globally higher EKE and EPE at all the different resolutions. However, they remain much lower than that of the \emph{REF} ones. 

\begin{figure}
\captionsetup{font=footnotesize}
\begin{center}
\includegraphics[width=3.8cm]{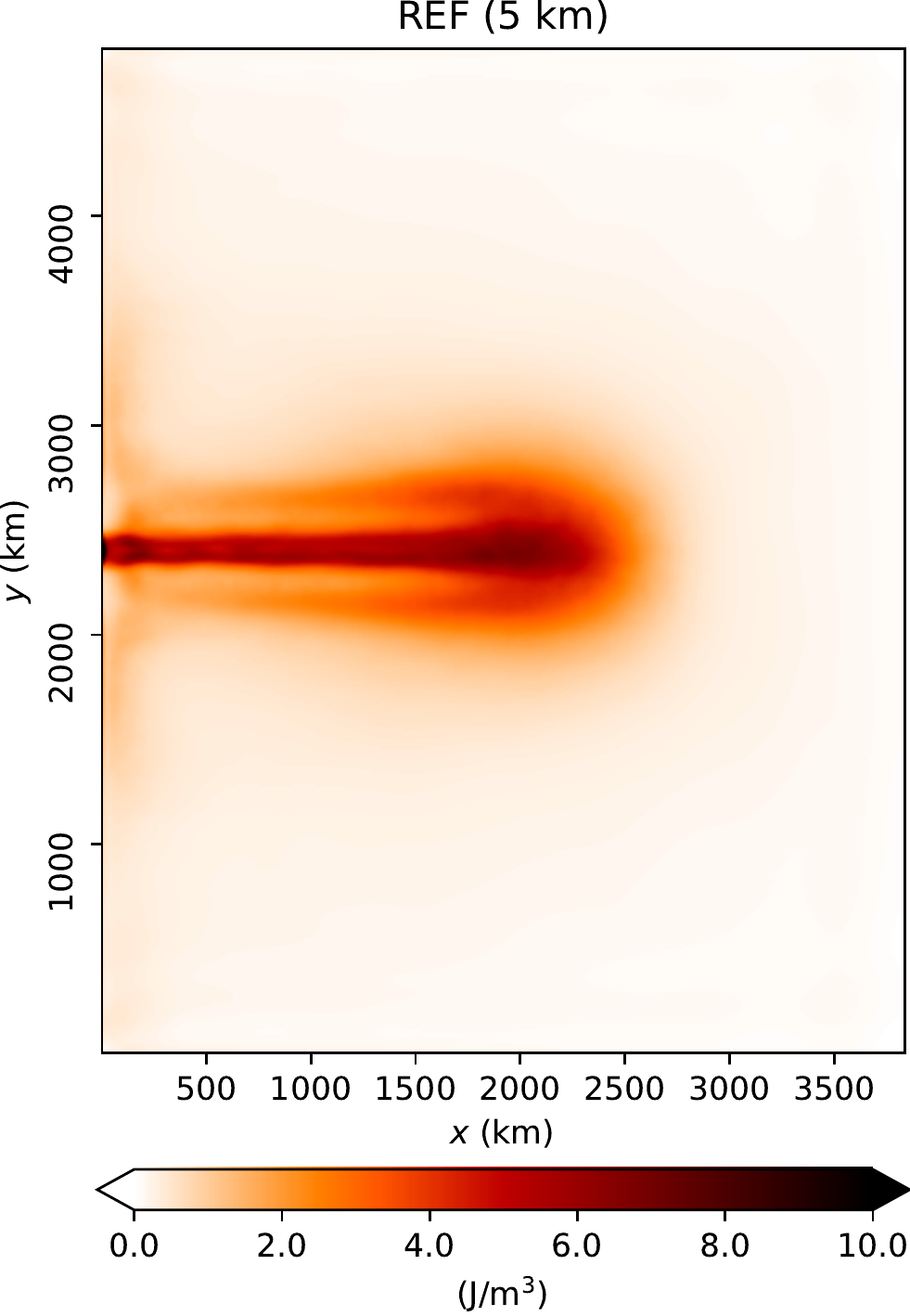}
\includegraphics[width=3.8cm]{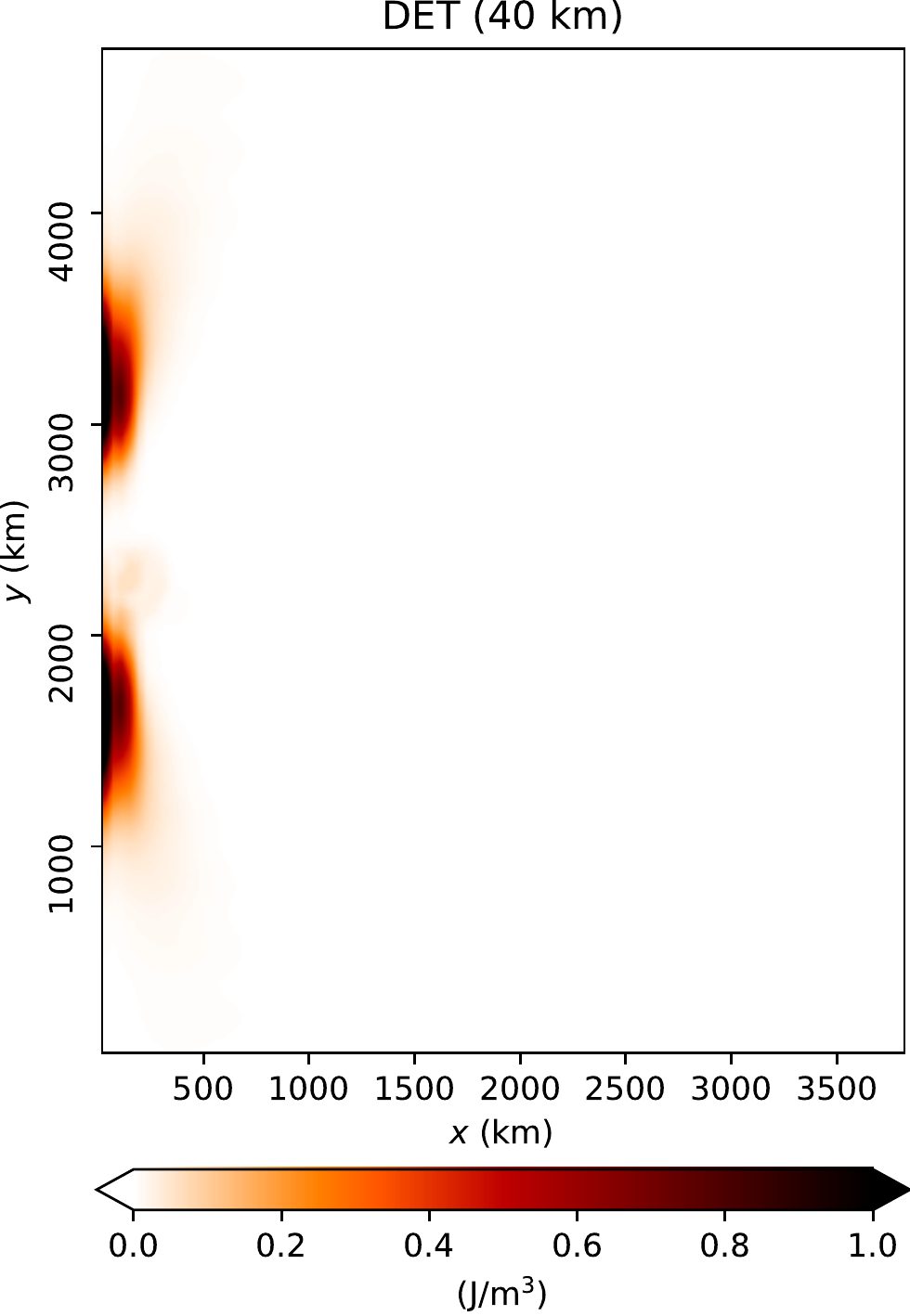}
\includegraphics[width=3.8cm]{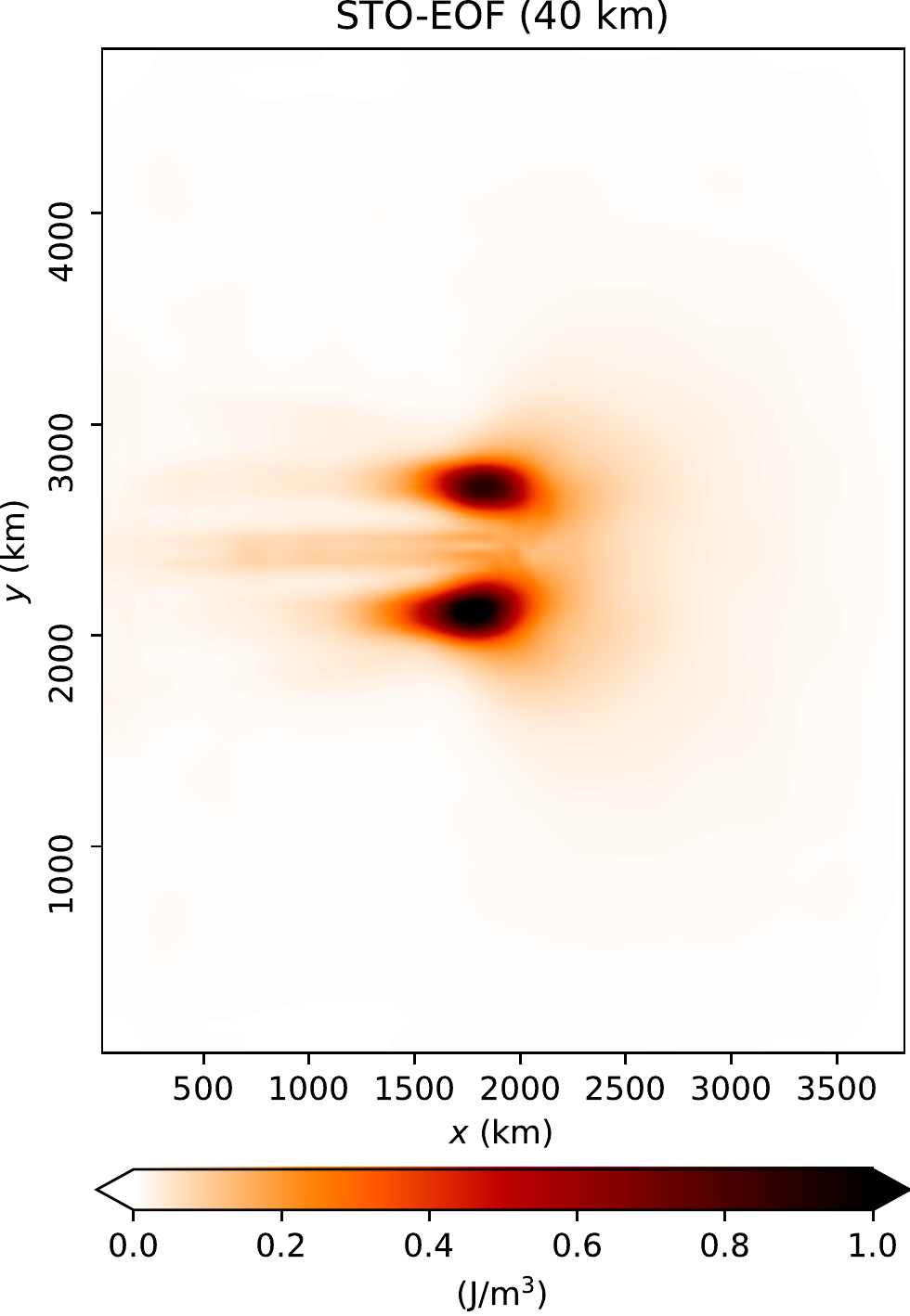}
\includegraphics[width=3.8cm]{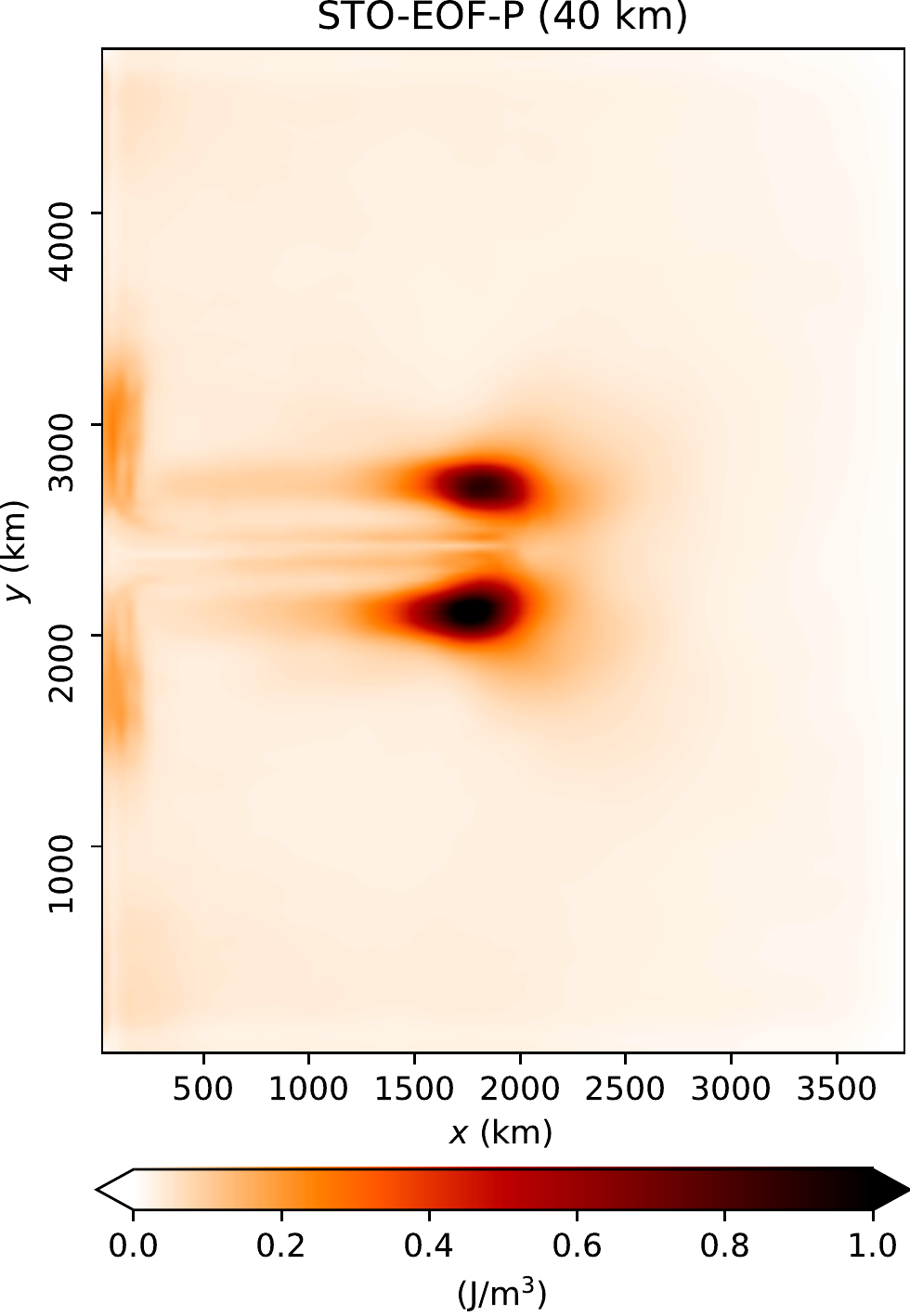} \\
\par\medskip
\includegraphics[width=3.8cm]{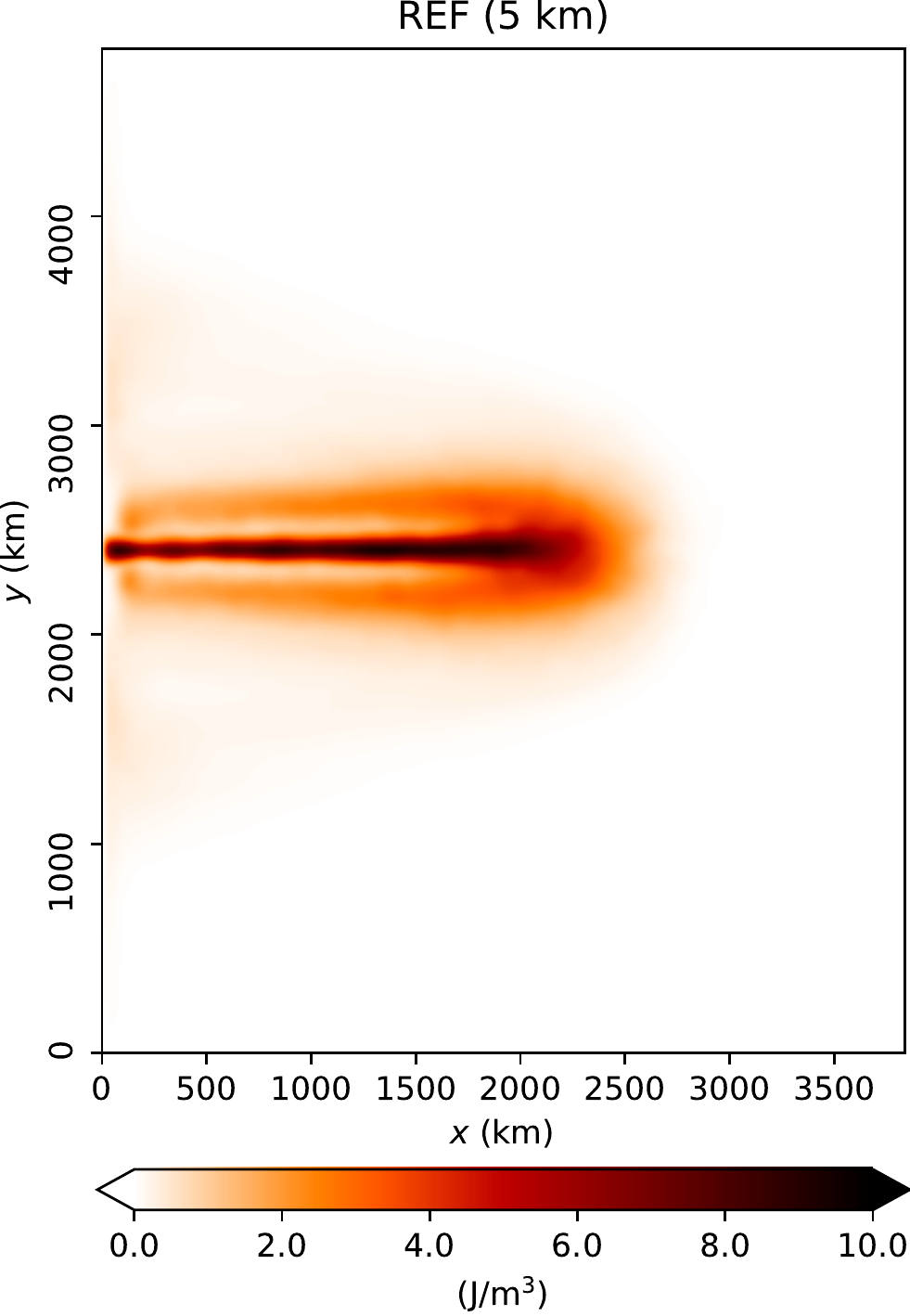}
\includegraphics[width=3.8cm]{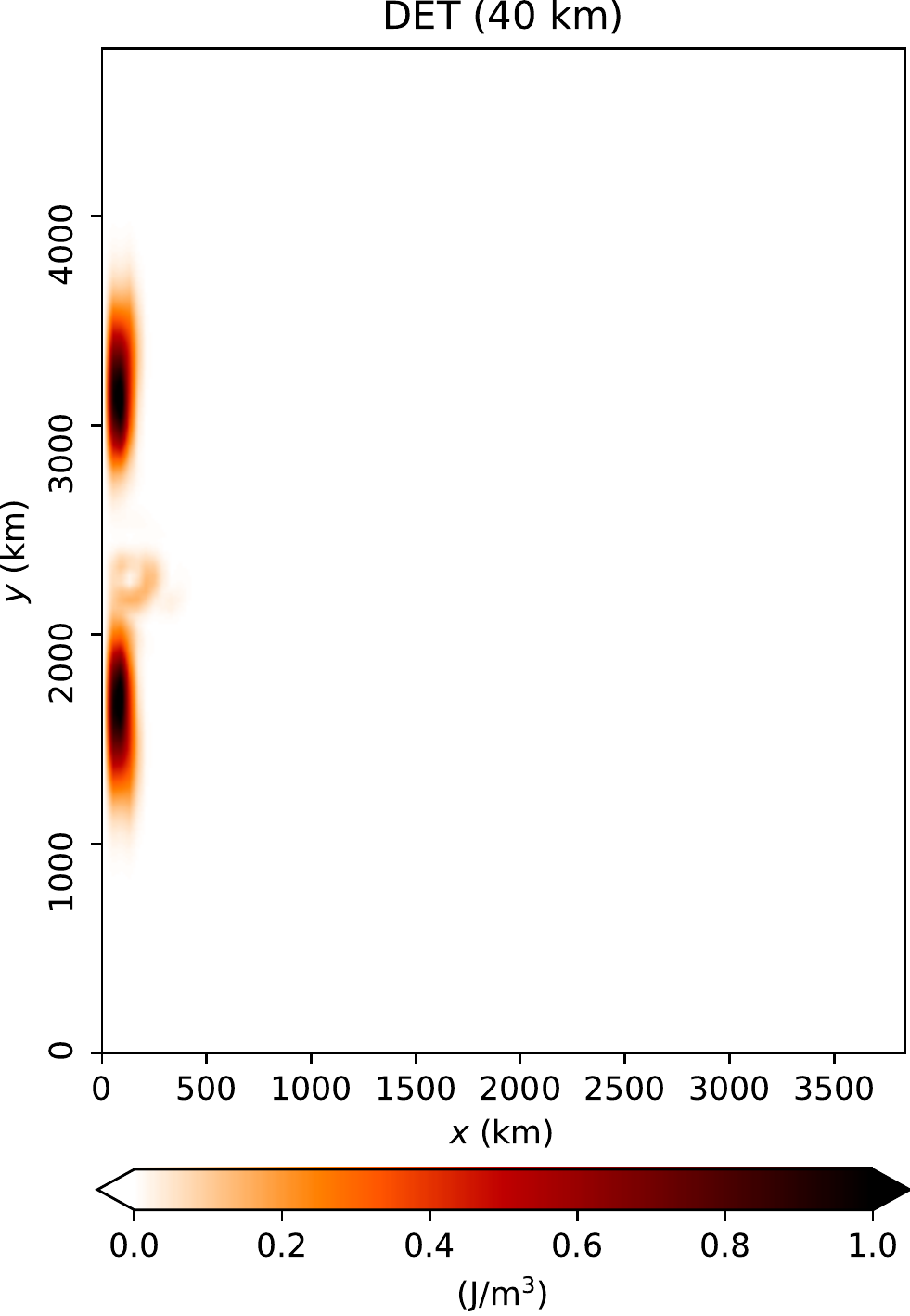}
\includegraphics[width=3.8cm]{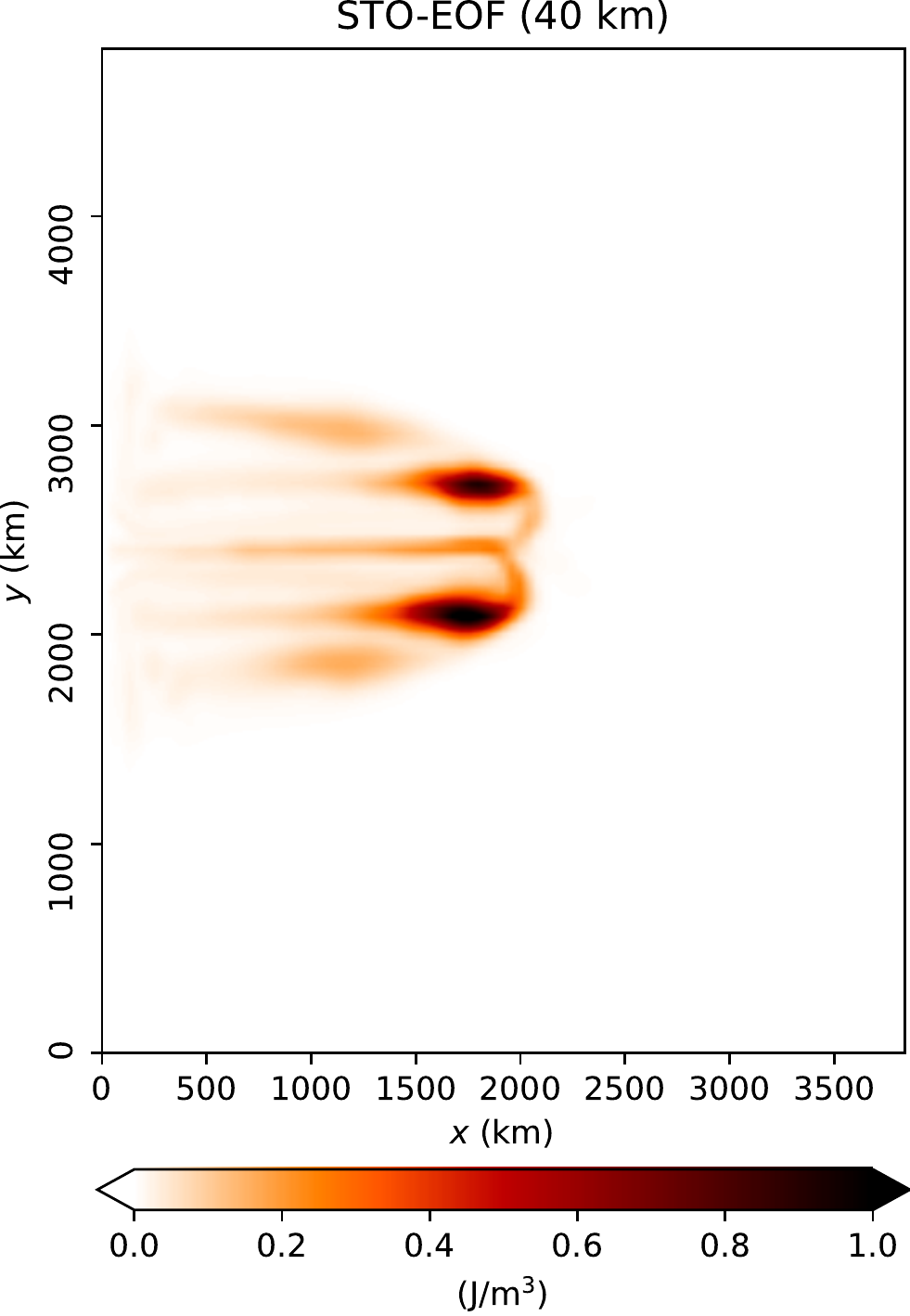}
\includegraphics[width=3.8cm]{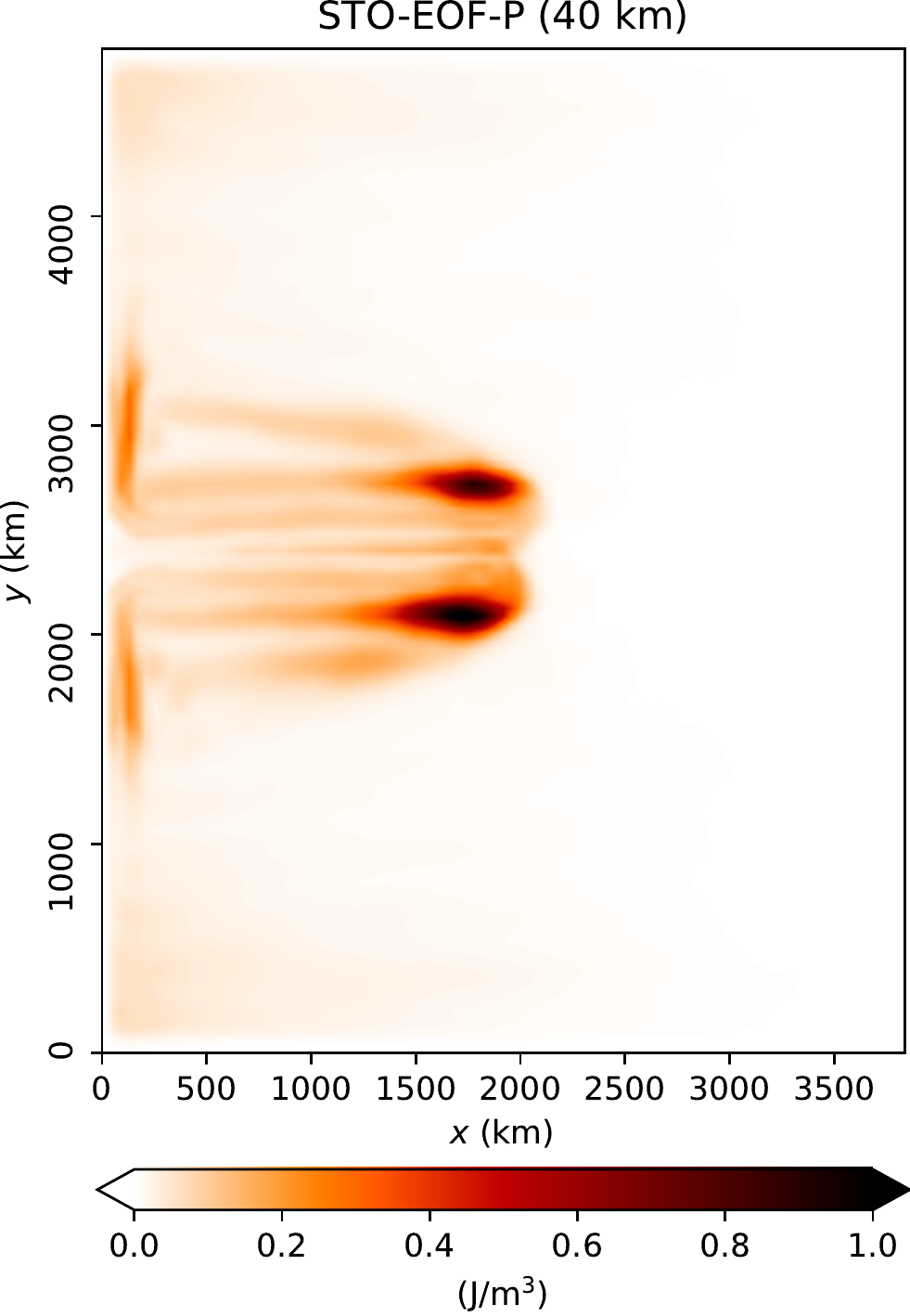}
\end{center}
\caption{Comparison of the vertically-integrated EKE density (top row) and EPE density (bottom row) for different models (by columns). Note that the colorbar of coarse-resolution (40 km) models is different than that of the \emph{REF} model (5 km).}
\label{fig:denergy}
\end{figure}

Furthermore, the horizontal density of the EKE and EPE provided by different models are compared in Figure~\ref{fig:denergy}. It shows that the \emph{STO-EOF-P} model improves locally the eddy energy almost everywhere except in the jet region. A more precise parameterization of the unresolved flow is required to explore in future works to improve locally the energy transfers.

In summary, the diagnostic results of this section demonstrate that including the unresolved flow structure inferred from the data into the proposed random models enable to reproduce on the coarse mesh the mean flow of the eddy-resolving simulation. In addition, as performed here through a projection onto the iso-surfaces of vertical stratification, providing an adequate dynamics for the unresolved flow structure is of crucial importance to increase the variability of the coarse models.

\subsection{Including mixed layer dynamics}\label{sec:mix-layer}

In this section, we follow further the Q-GCM formulation \citep{Hogg2004} to embed a mixed layer (of thickness $H_m = 100$ m) in the upper QG layer. This brings an additional diabatic forcing into the PV equation \eqref{eq:dPV-k} due to the Ekman pumping of sea surface temperature (SST). The vertical velocity across the first ocean interface is $e_1 = - \salf\, (\Delta_m T / \Delta_1 T)\, e_0$, where $\Delta_m T = T_m - T_1$ (resp. $\Delta_1 T = T_1 - T_2$) denotes the temperature difference across the bottom boundary of the mixed layer (resp. of the upper QG layer). Here, the QG potential temperature are fixed to be $(287, 282, 276)$ K. 

The additional forcing $e_1$ evolves in time as the SST does. In the present work, we only considered the geostrophic component of the unresolved noise in the mixed layer. In this case, the stochastic evolution of SST reads
\beq\label{eq:sst}
\wt{\sto}_t^m T_m + T_m \bdiv \bu_m\, \dt = \left( (K_2 \nabla^2 - K_4 \nabla^4) T_m + \frac{T_m + T_1}{2 H_m} e_{0} \right)\, \dt ,
\eeq
where $\bu_m = \gradp \psi_1 - \bs{\tau}^{\perp} / (f_0 H_m)$ is the divergent mixed layer velocity, $\wt{\sto}_t^m$ is the mixed layer stochastic transport operator under probability measure $\wt{\Pb}$ (associated to $\bu_m$), $K_2 = 100$ m$^2$/s and $K_4 = A_4$ (see values in Table~\ref{tab:params-pri}). Note that this shallow-water type equation is derived by vertical integration of a 3D heat equation. Equation \eqref{eq:sst} is numerically discretized by the central winding scheme (in space) and the stochastic Leapfroq scheme (in time).

\begin{figure}
\captionsetup{font=footnotesize}
\begin{center}
\includegraphics[width=3.8cm]{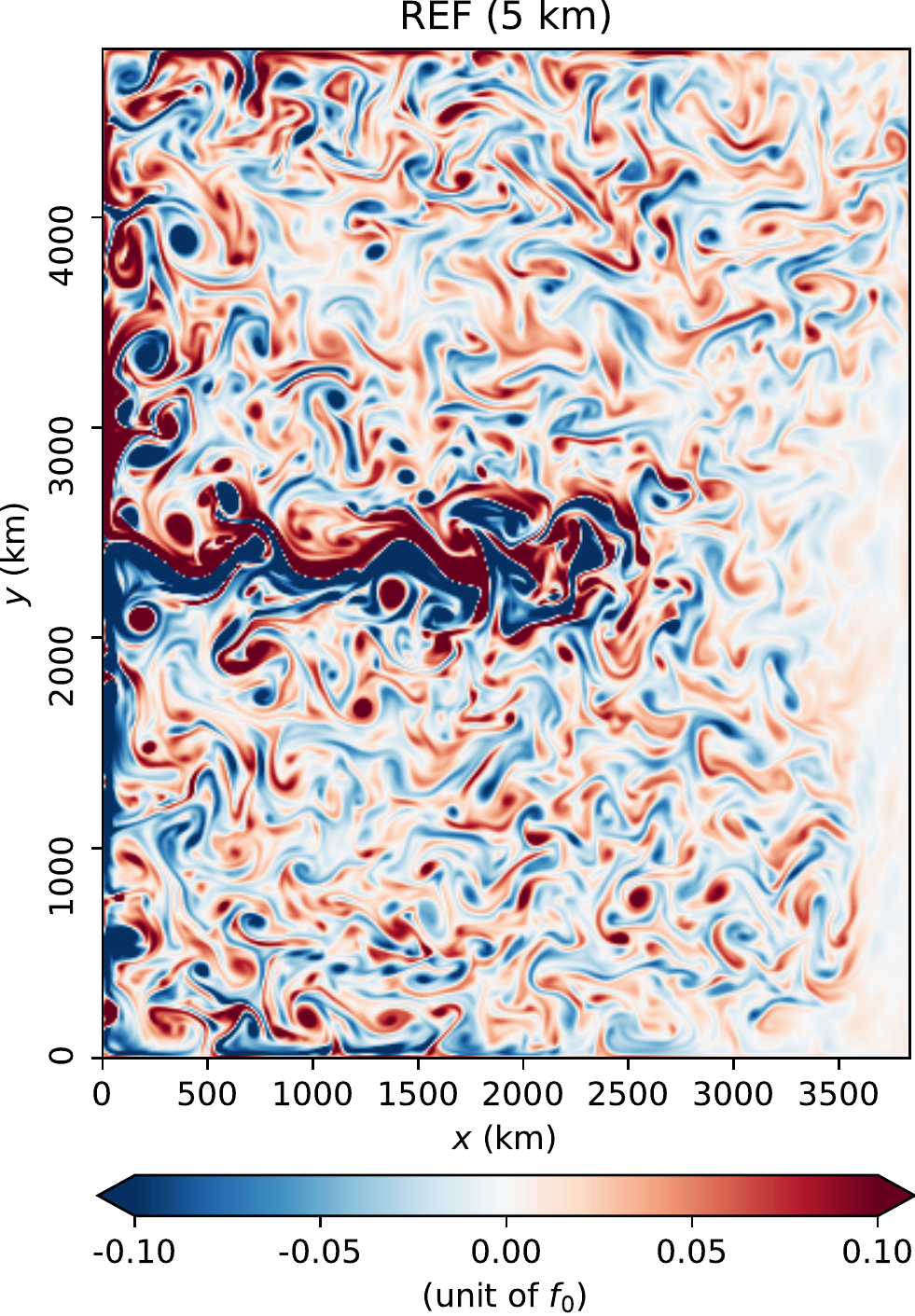}
\includegraphics[width=3.8cm]{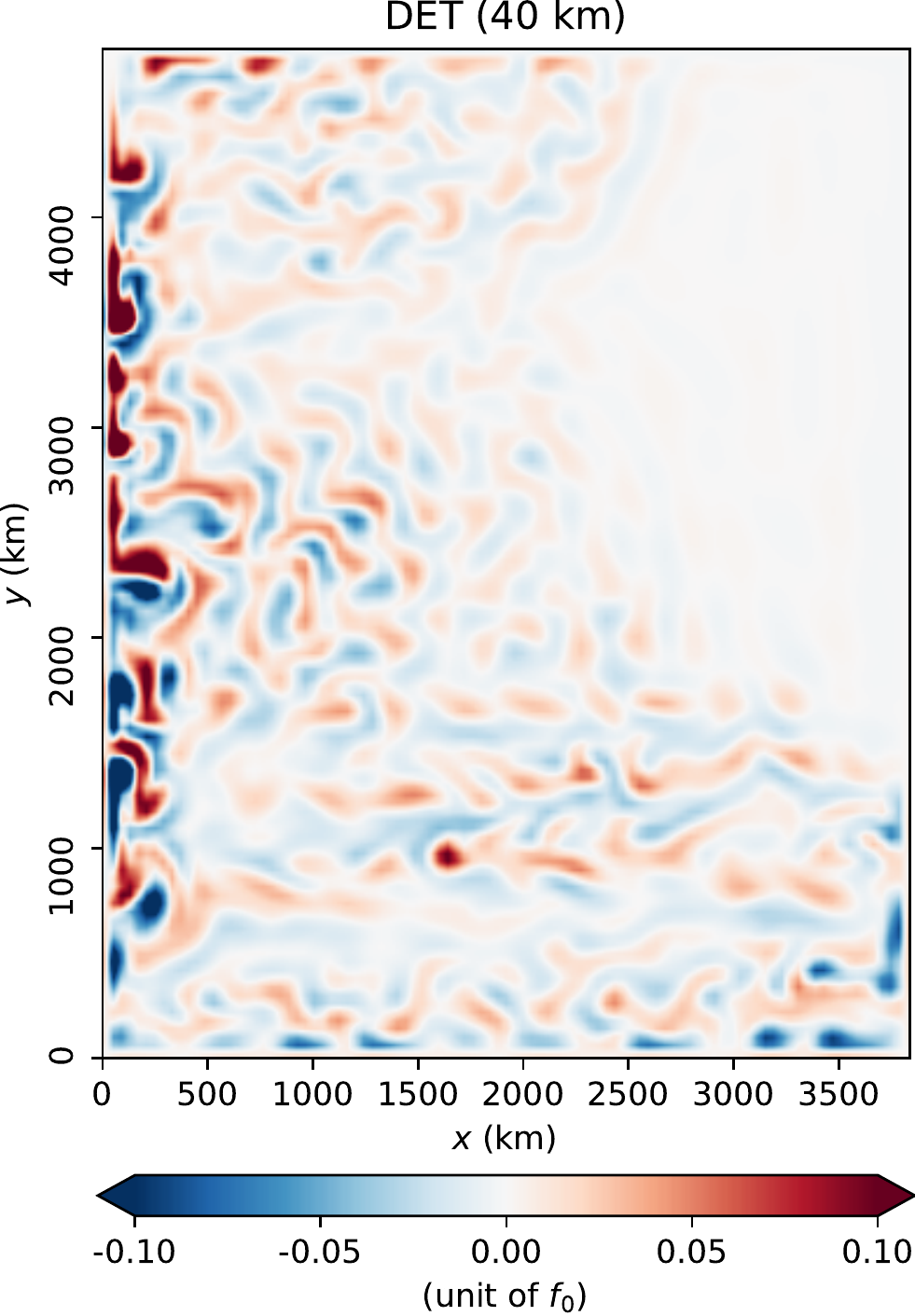}
\includegraphics[width=3.8cm]{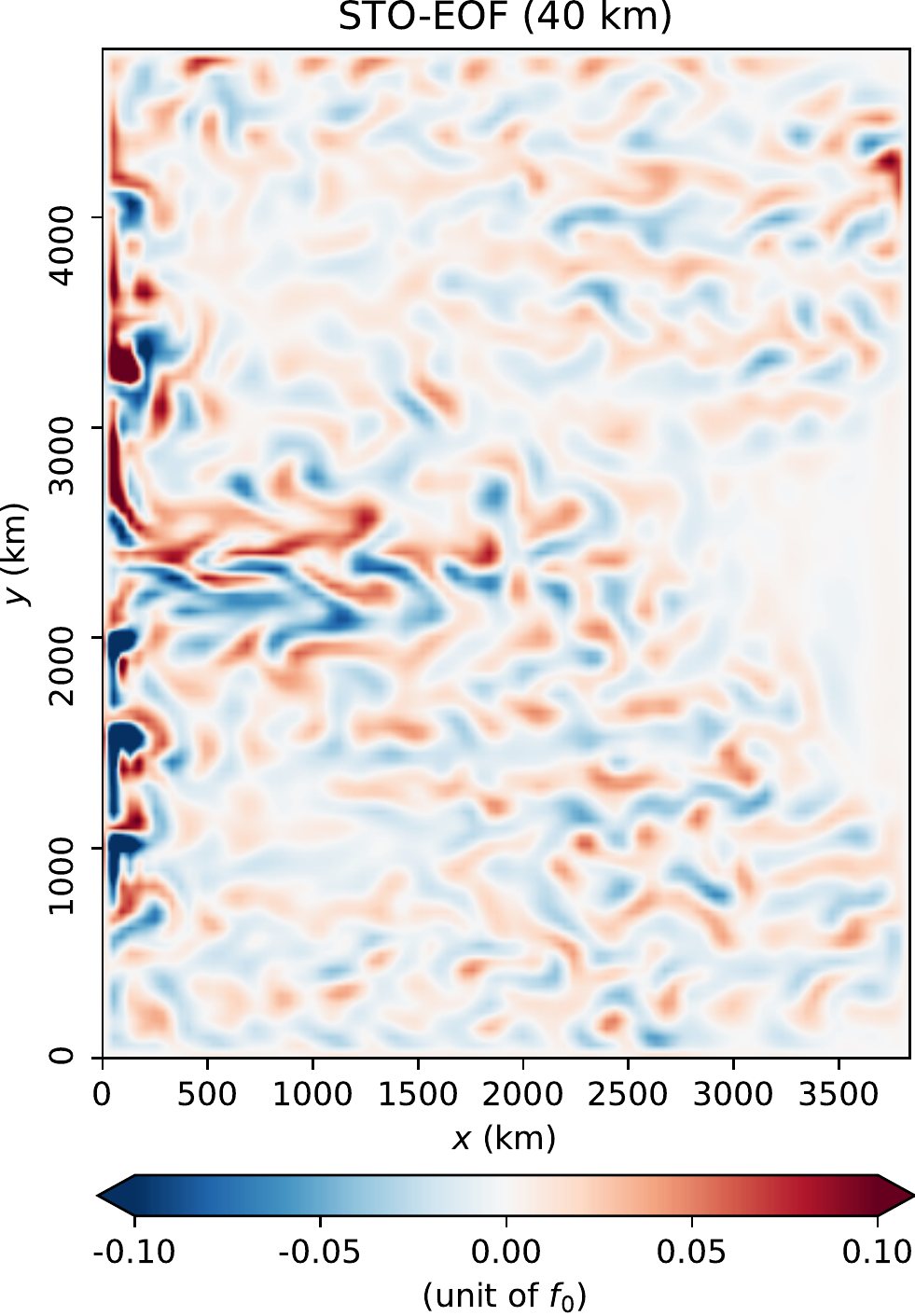}
\includegraphics[width=3.8cm]{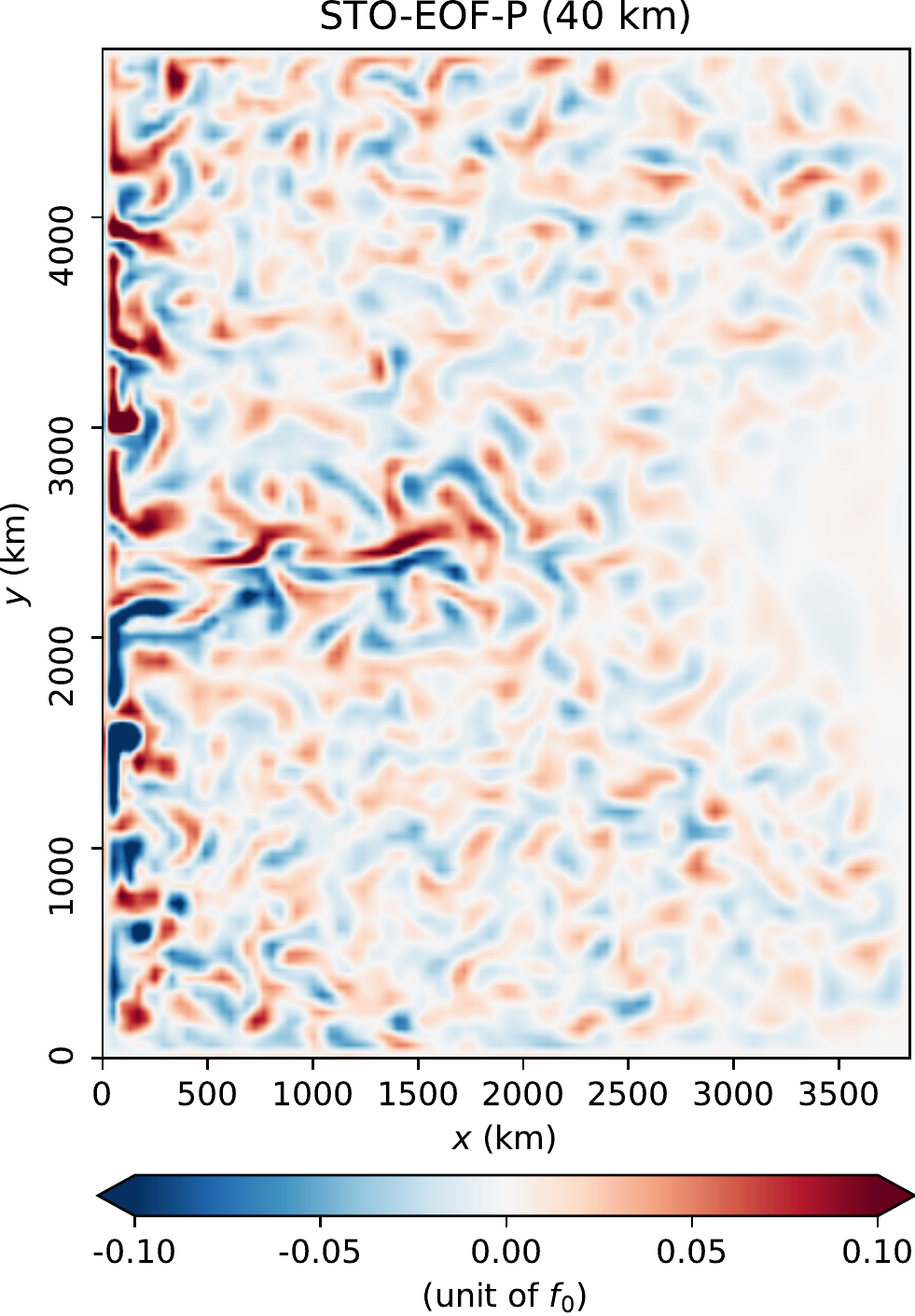} \\
\par\medskip
\includegraphics[width=3.8cm]{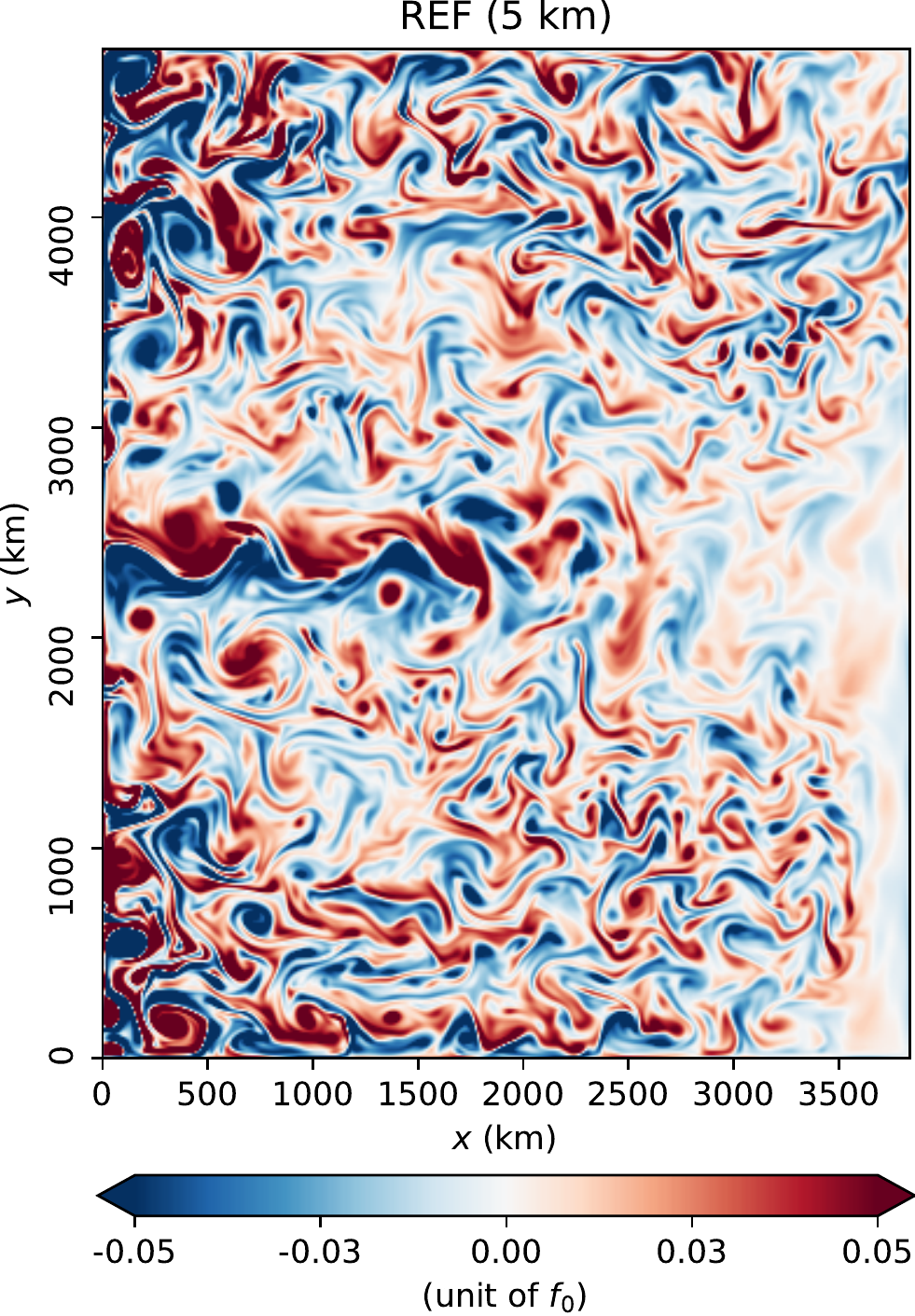}
\includegraphics[width=3.8cm]{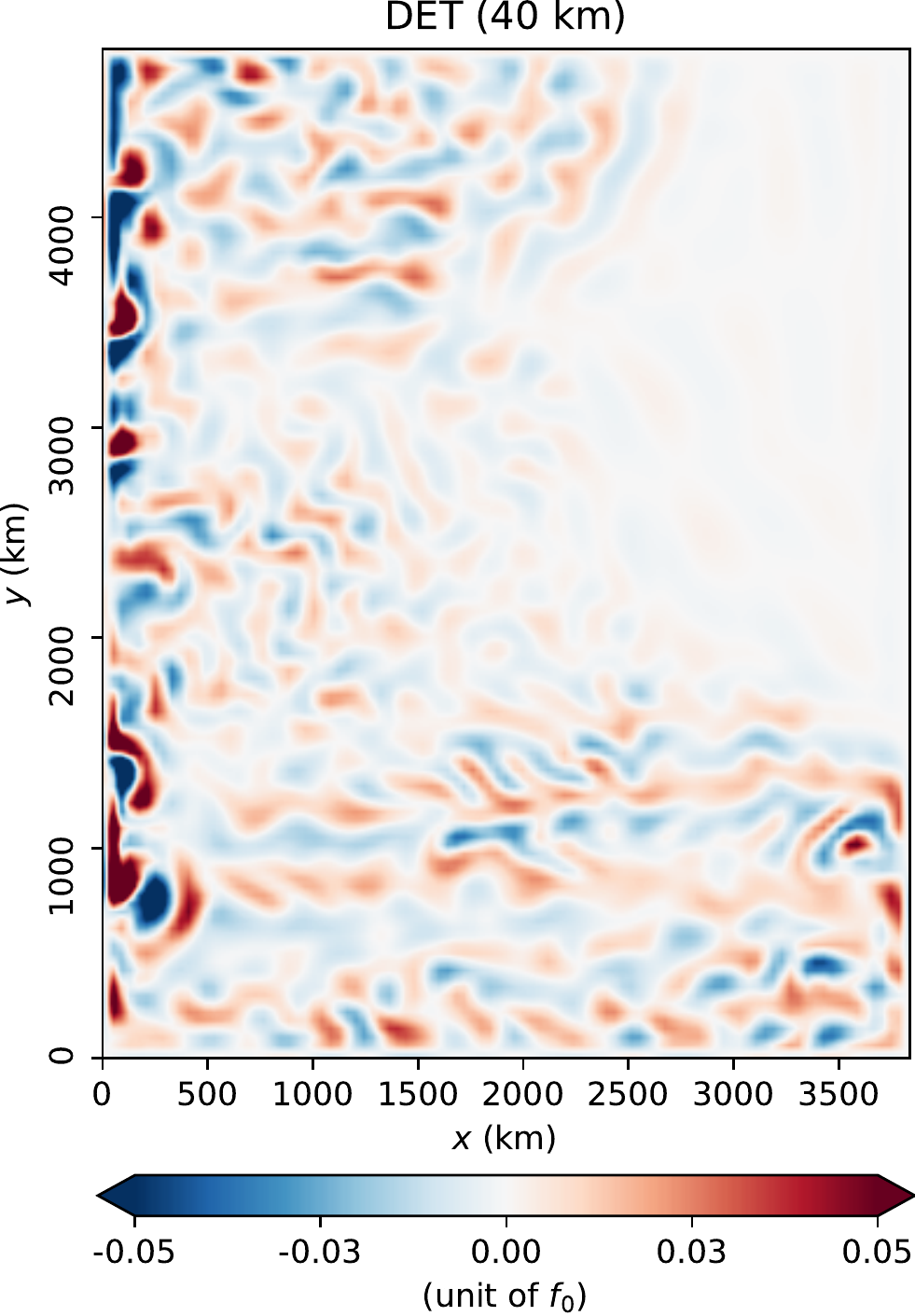}
\includegraphics[width=3.8cm]{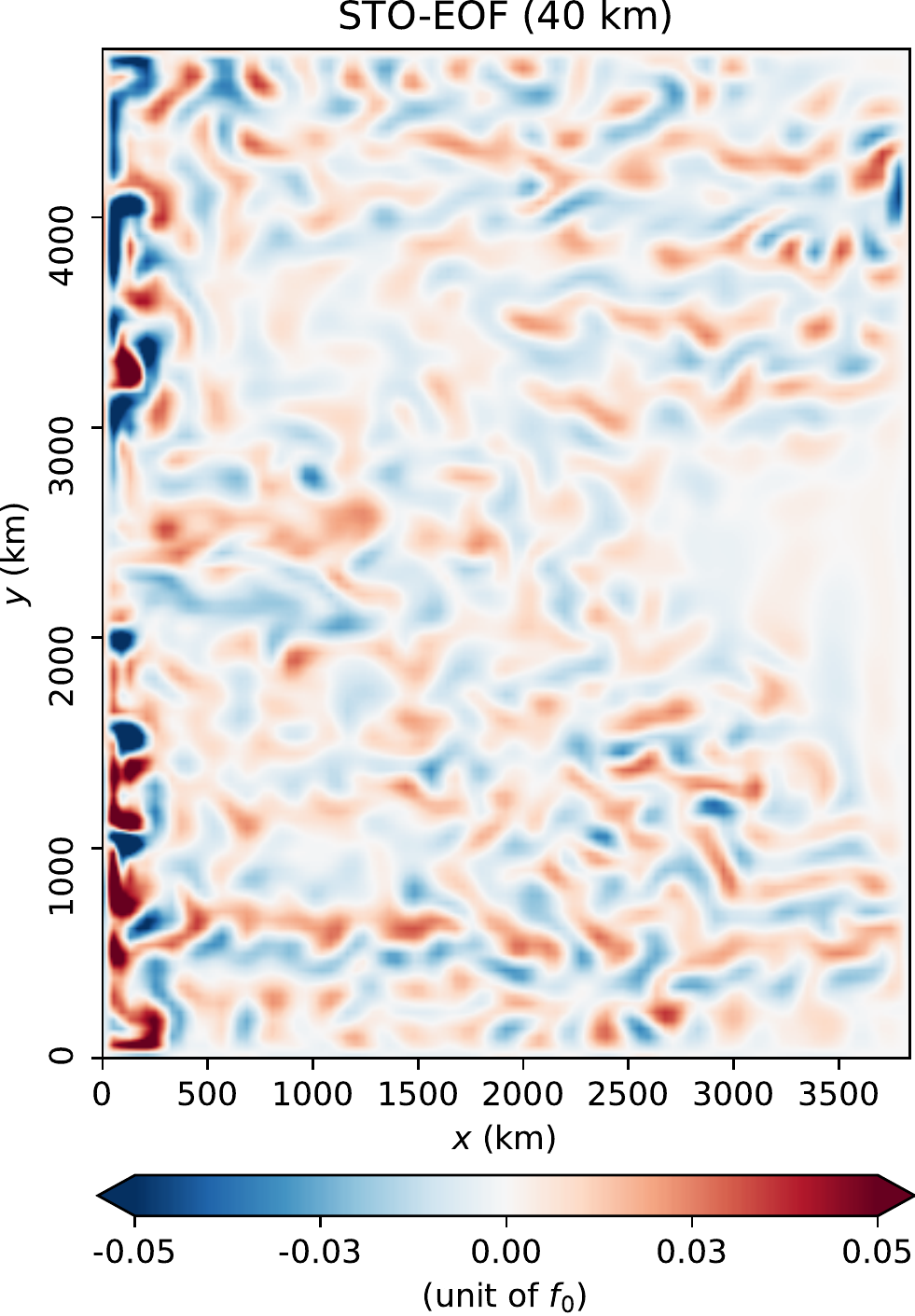}
\includegraphics[width=3.8cm]{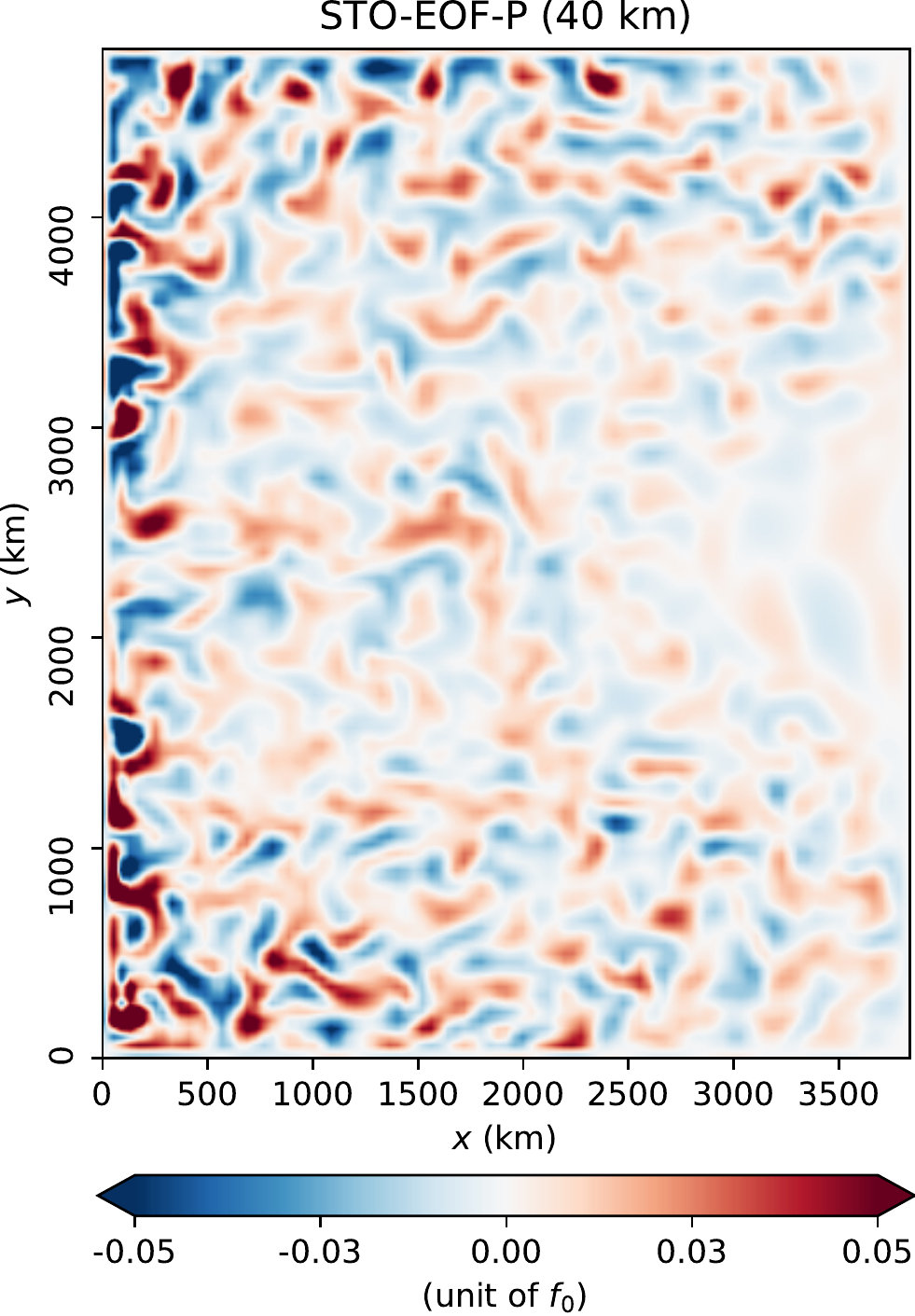} 
\end{center}
\caption{Snapshots of relative vorticity (divided by $f_0$) for the two upper layers (by rows) provided by different simulations (by columns) after 60-years integration.}
\label{fig:spot-PV-sst}
\end{figure}

As shown in Figure \ref{fig:spot-PV-sst}, the inclusion of the additional forcing coming from the SST evolution generates more small-scale eddies almost everywhere in \emph{REF}. The improvements of the two stochastic models observed in the previous test case still hold here, namely the zonal jet is reproduced and perturbed. In addition, the variability of the gyres seems also to be enhanced in this case. 

\begin{figure}
\captionsetup{font=footnotesize}
\begin{center}
\includegraphics[width=3.8cm]{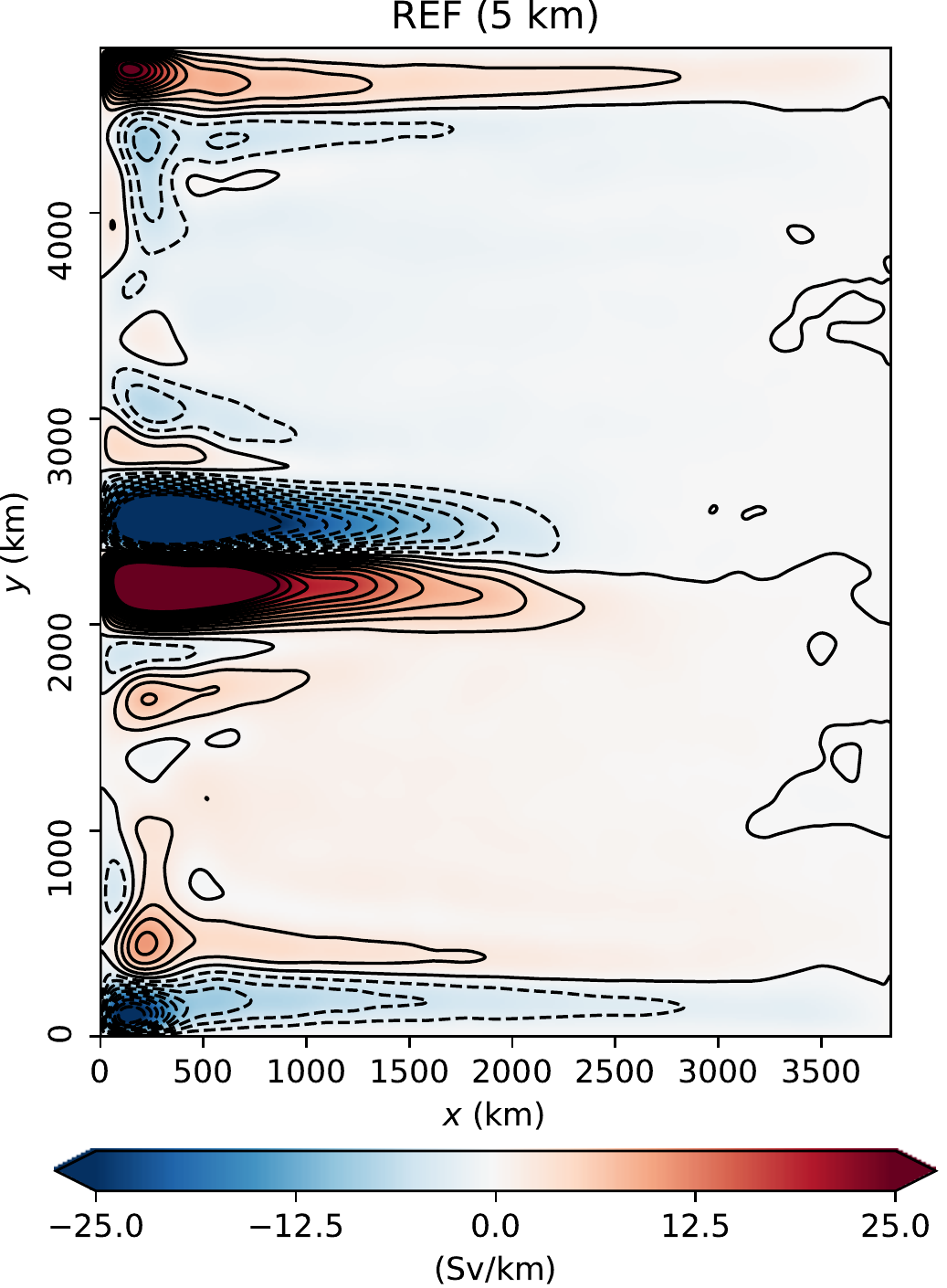}
\includegraphics[width=3.8cm]{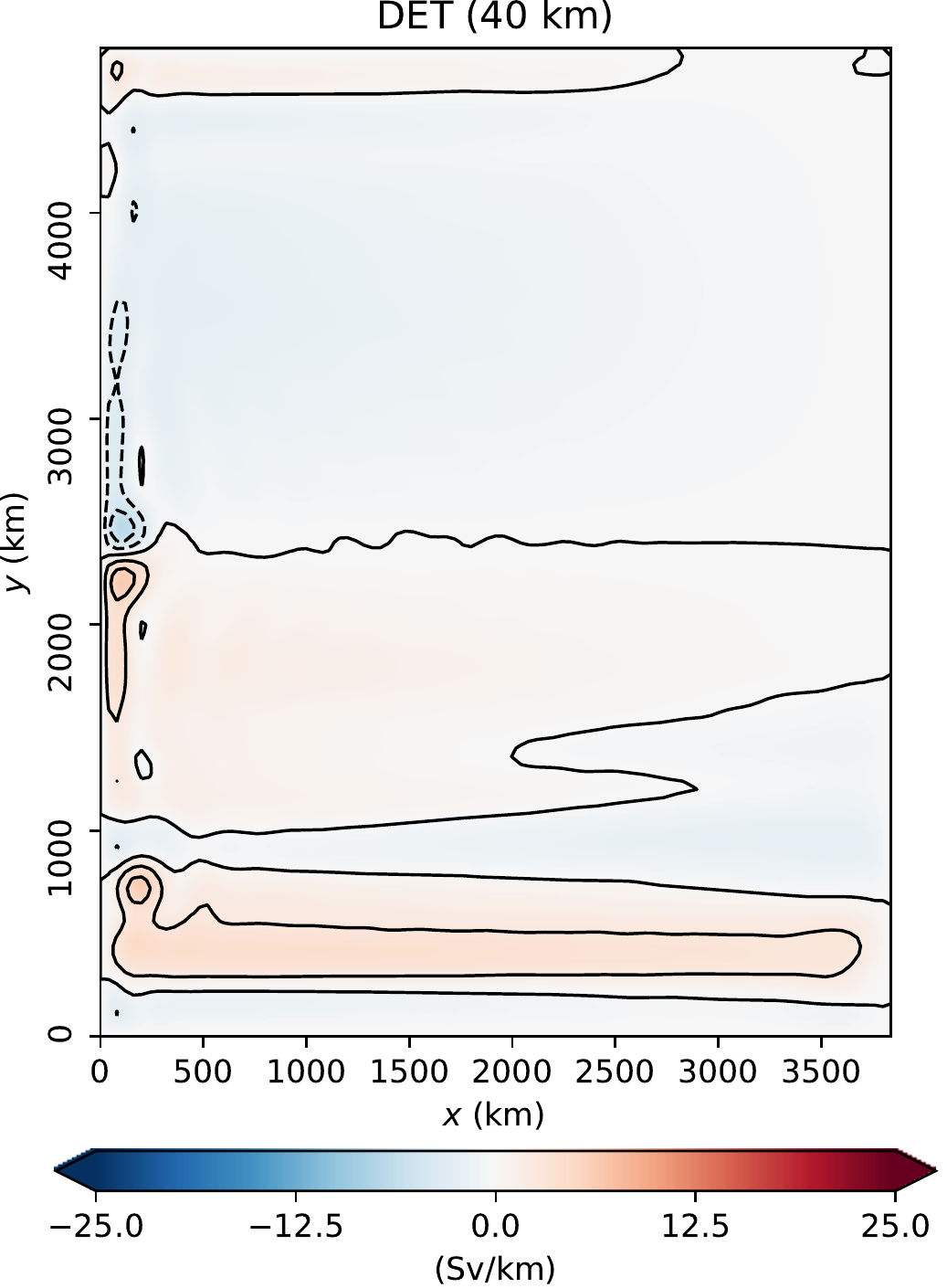} 
\includegraphics[width=3.8cm]{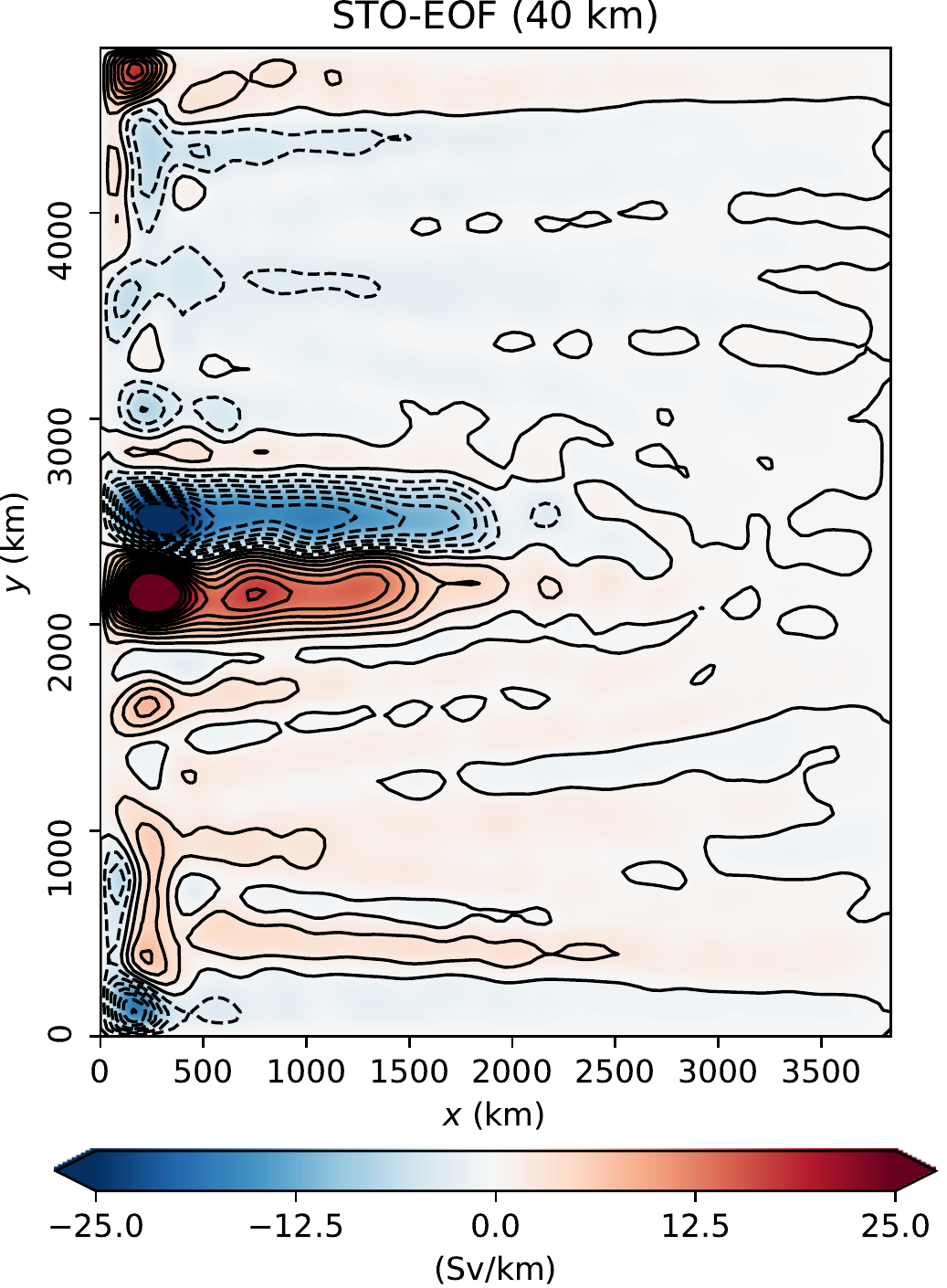}
\includegraphics[width=3.8cm]{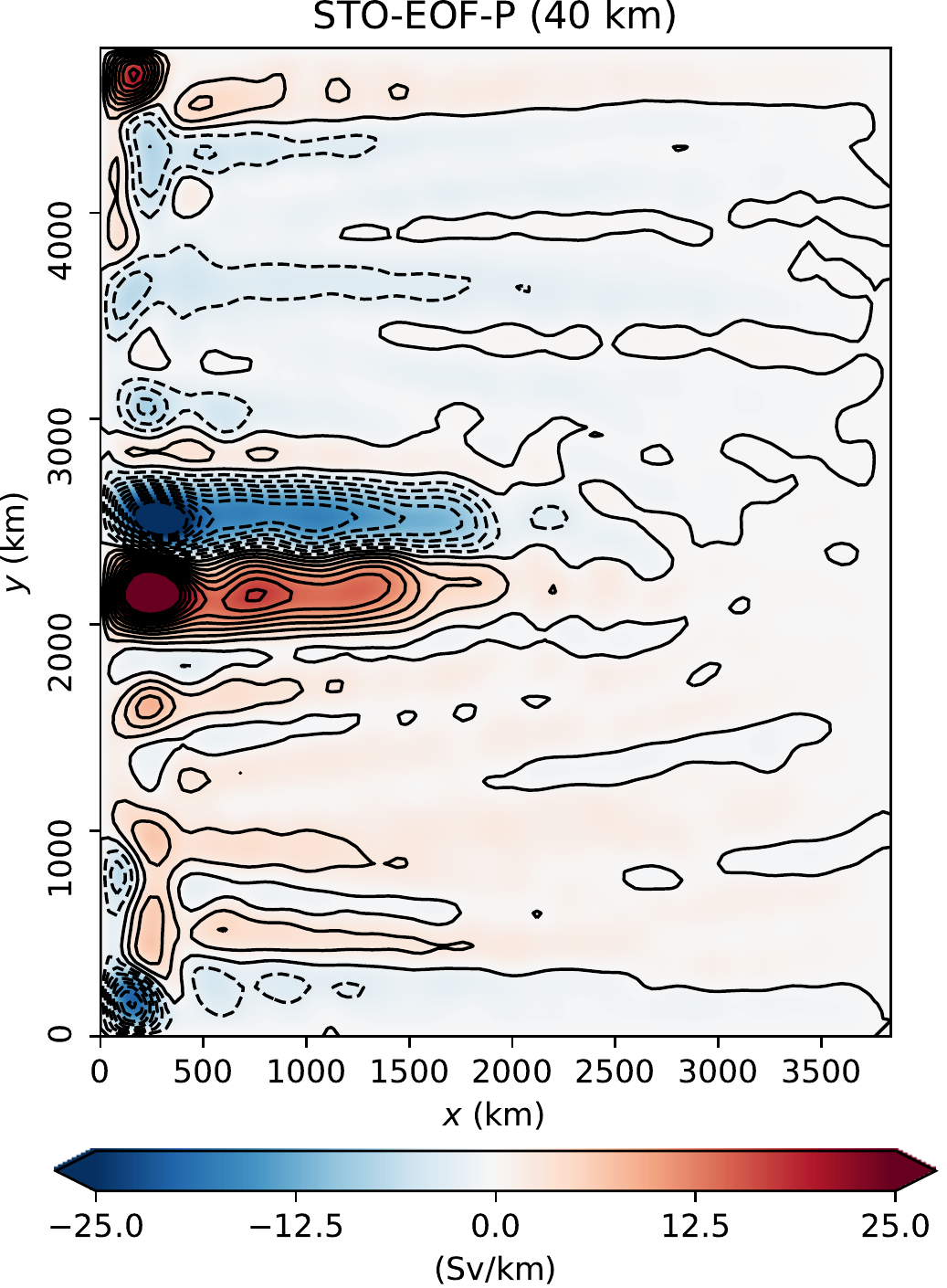} \\
\par\medskip
\includegraphics[width=3.8cm]{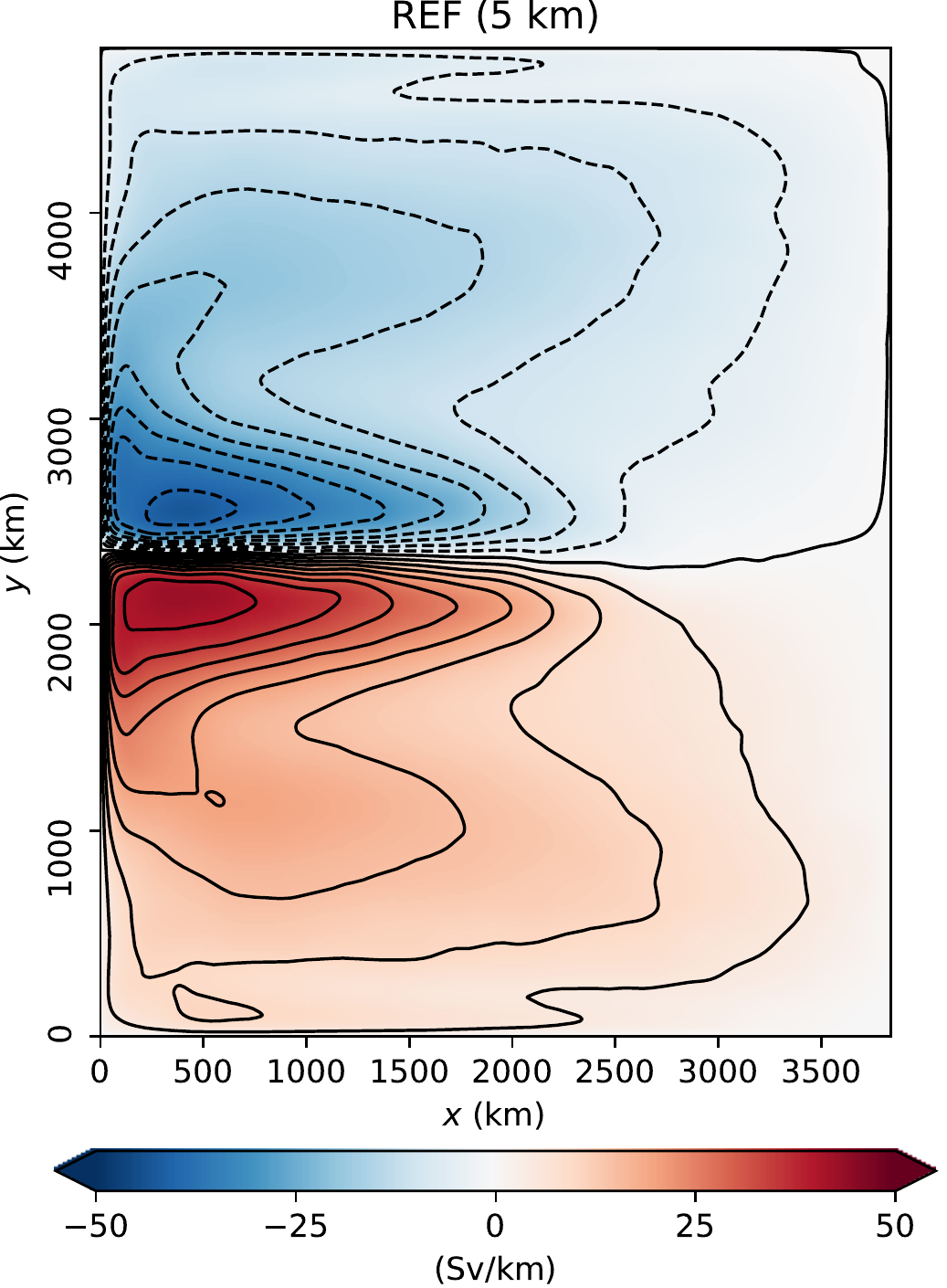}
\includegraphics[width=3.8cm]{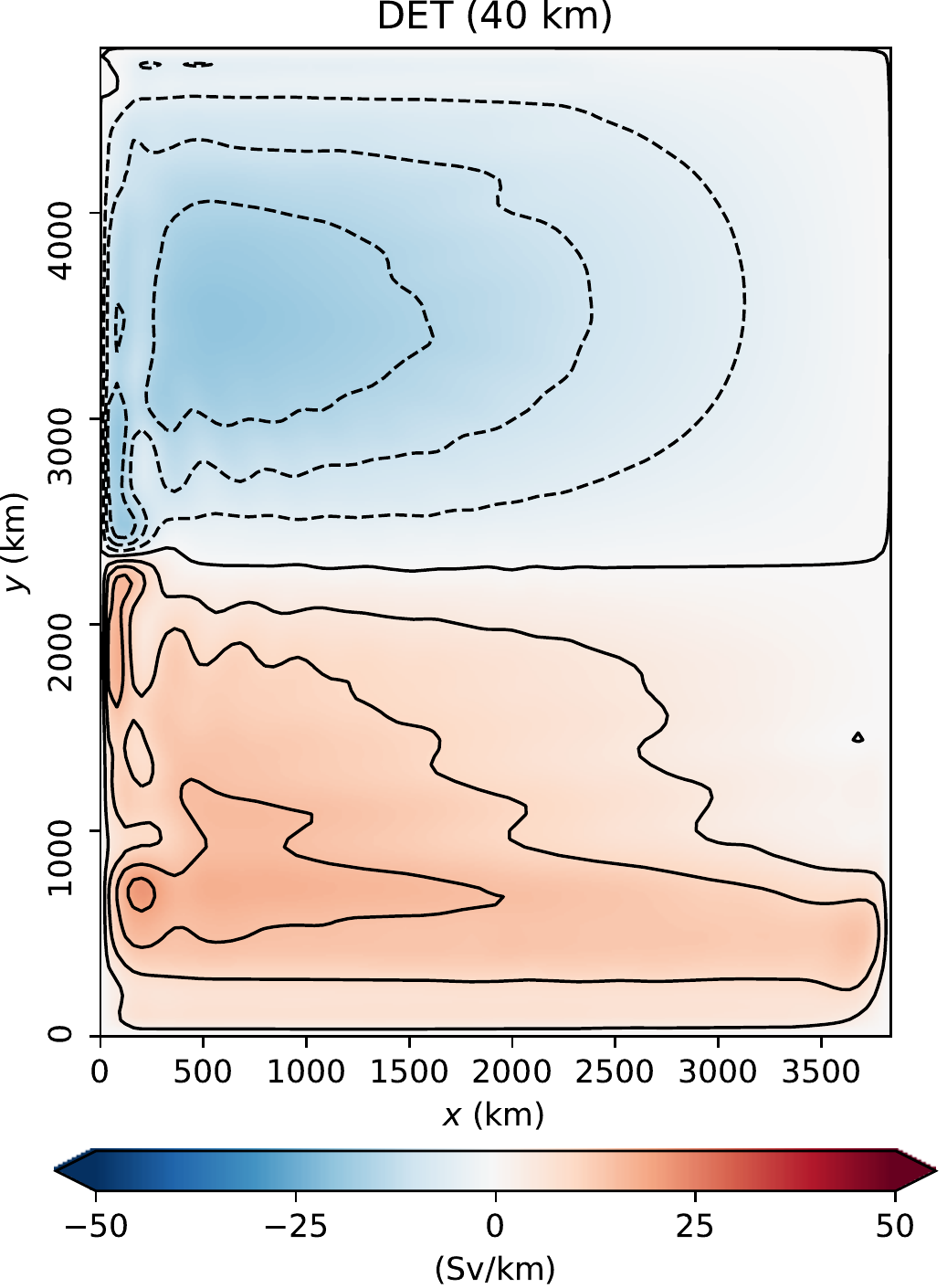} 
\includegraphics[width=3.8cm]{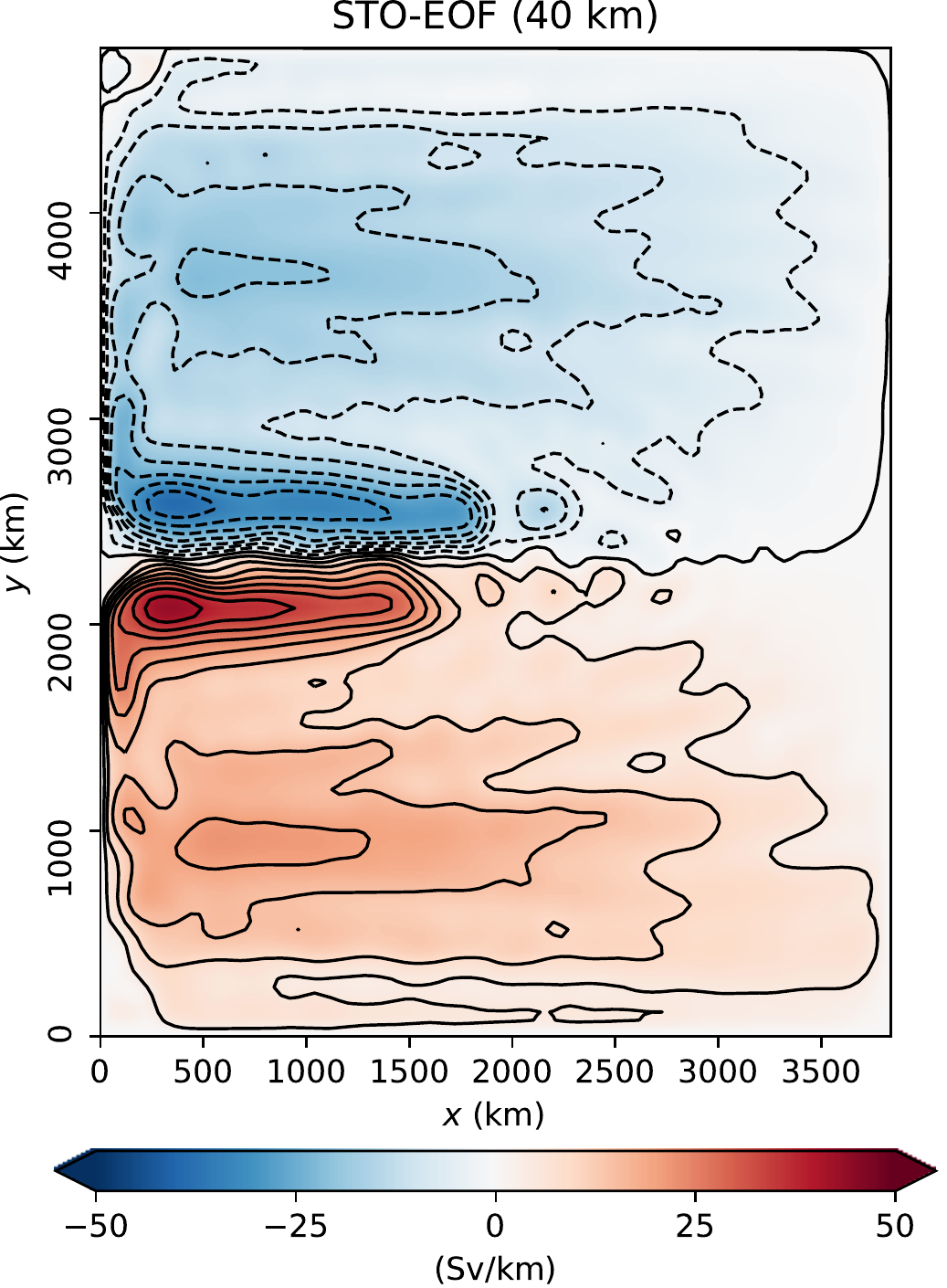}
\includegraphics[width=3.8cm]{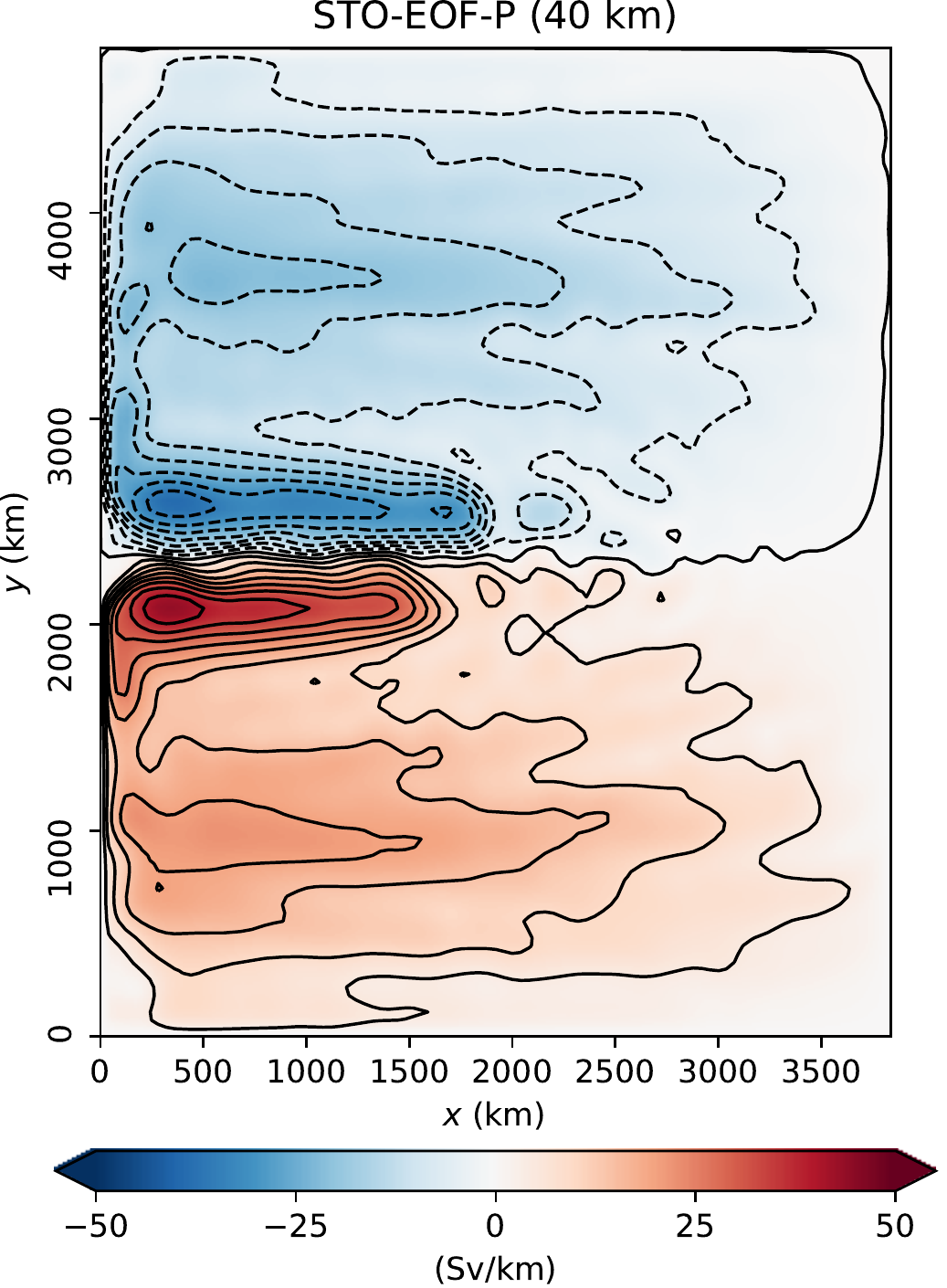}
\end{center}
\caption{Comparison of time-mean contour of barotropic (top row) and first baroclinic (bottom row) streamfunctions for different models (by columns).}
\label{fig:mean-pm40-sst}
\end{figure}

Subsequently, an analogue procedure is performed for all the coarse-resolution models in this configuration. For instance, the results of the time-mean streamfunctions at 40 km are provided in Figures \ref{fig:mean-pm40-sst} and a conclusion consistent with the previous test case is recovered. 

\begin{figure}
\captionsetup{font=footnotesize}
\begin{center}
\includegraphics[width=5cm]{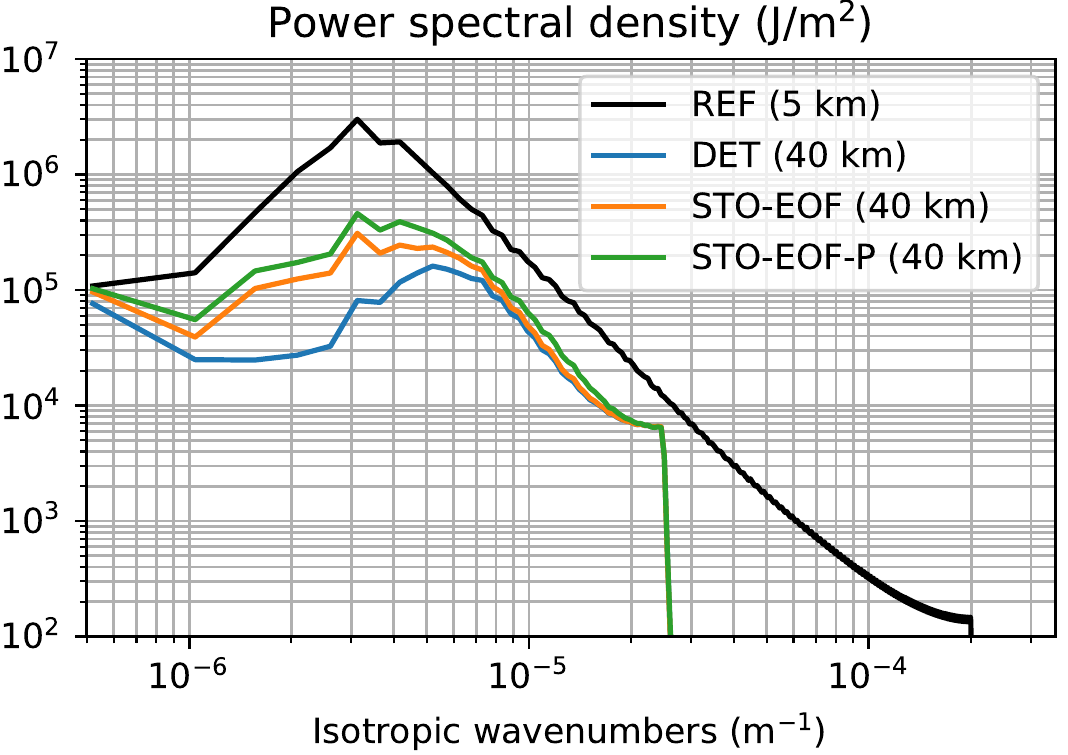}
\includegraphics[width=5cm]{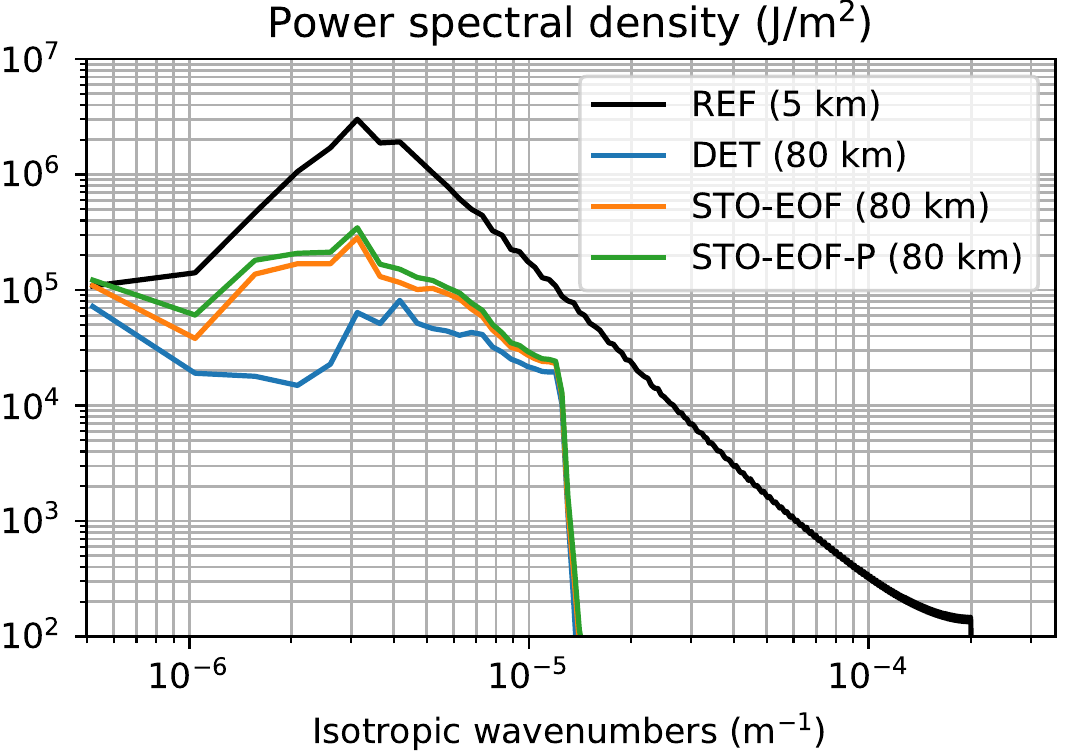} 
\includegraphics[width=5cm]{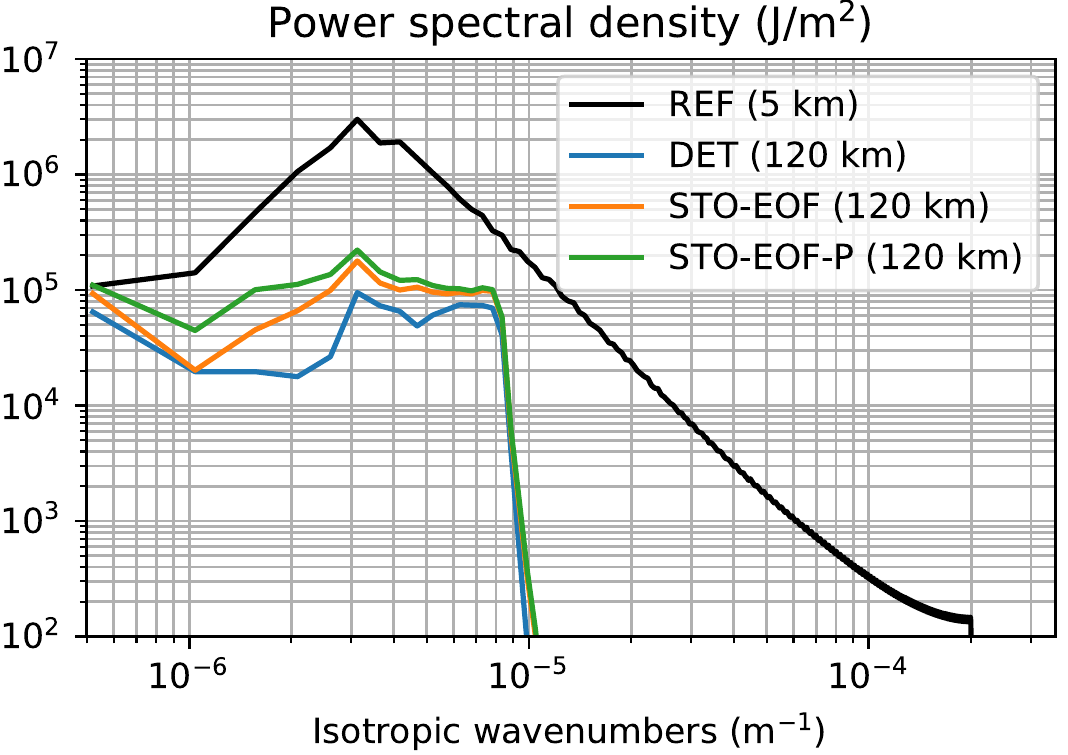} 
\end{center}
\caption{Temporal mean of vertically integrated KE spectra for different models (by colors) at different resolutions (by sub-figures).}
\label{fig:spec-KE-sst}
\end{figure}

As illustrated in Figure \ref{fig:spec-KE-sst}, the \emph{STO-EOF-P} models, with a noise along the iso-surfaces of vertical stratification, involves an additional SST random forcing that brings an higher KE backscattering than the stationary noise model at both resolutions. This seems to highlight the importance of the non-stationary characteristic of the noise brought by the projection. 

\begin{figure}
\captionsetup{font=footnotesize}
\begin{center}
\includegraphics[width=4.8cm]{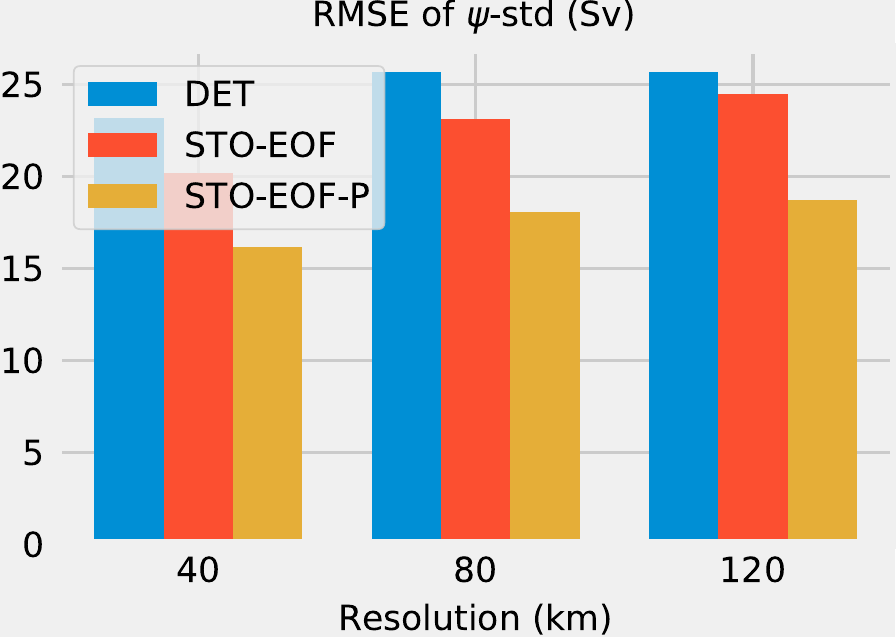} 
\includegraphics[width=4.8cm]{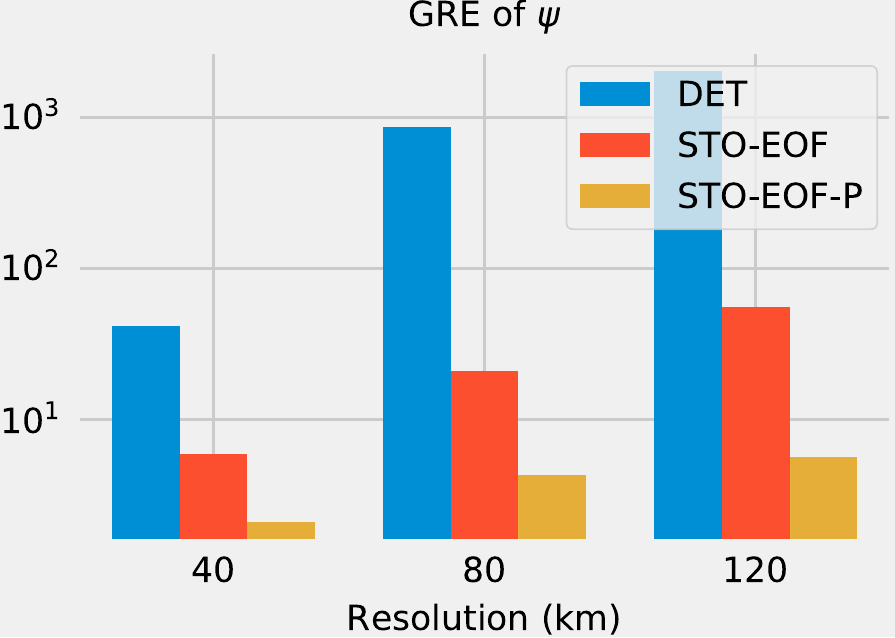}
\includegraphics[width=3.5cm]{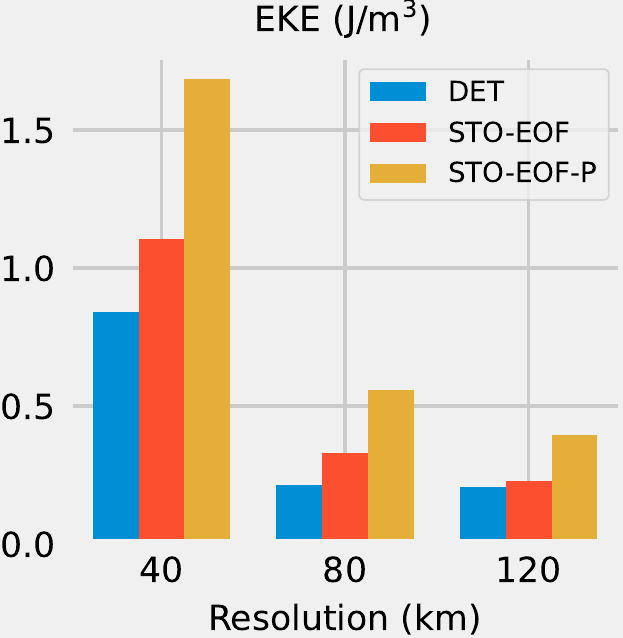} 
\end{center}
\caption{Comparison of variability measures (by sub-figures) for different coarse models (by colors and groups in each sub-figure). The EKE of the \emph{REF} model is 8.26 J/m$^3$.}
\label{fig:stat-sst}
\end{figure}

\begin{figure}
\captionsetup{font=footnotesize}
\begin{center}
\includegraphics[width=3.8cm]{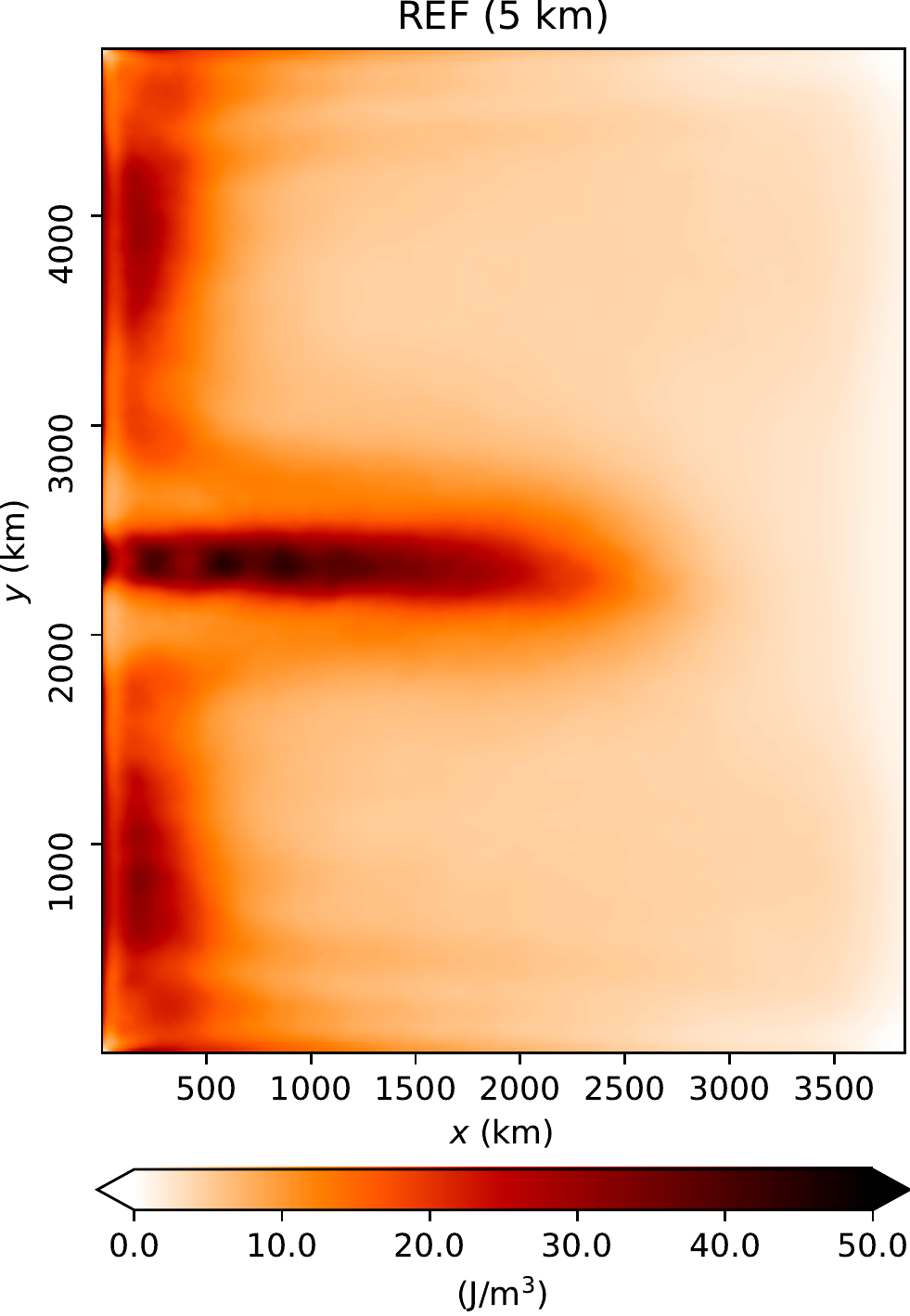}
\includegraphics[width=3.8cm]{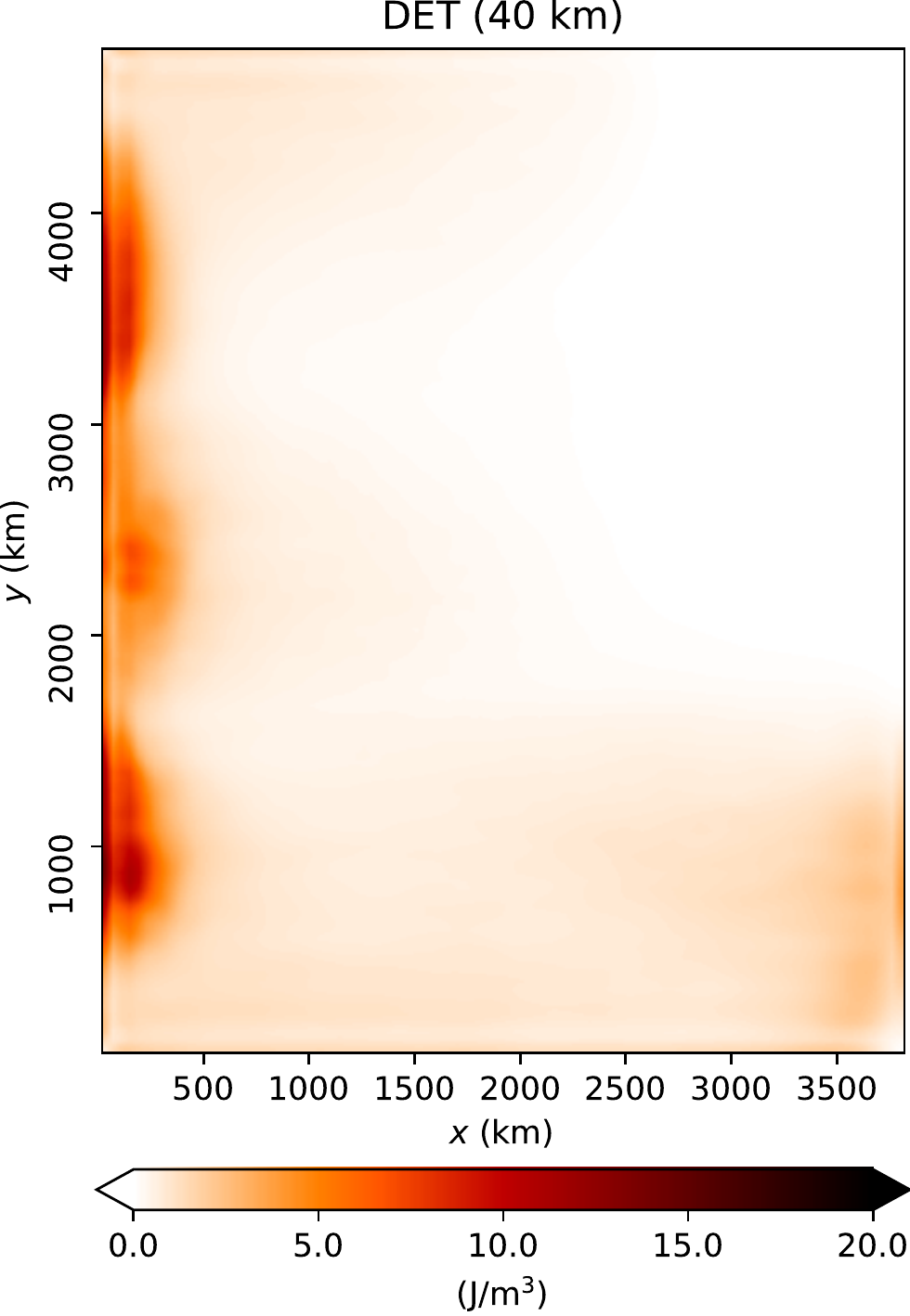}
\includegraphics[width=3.8cm]{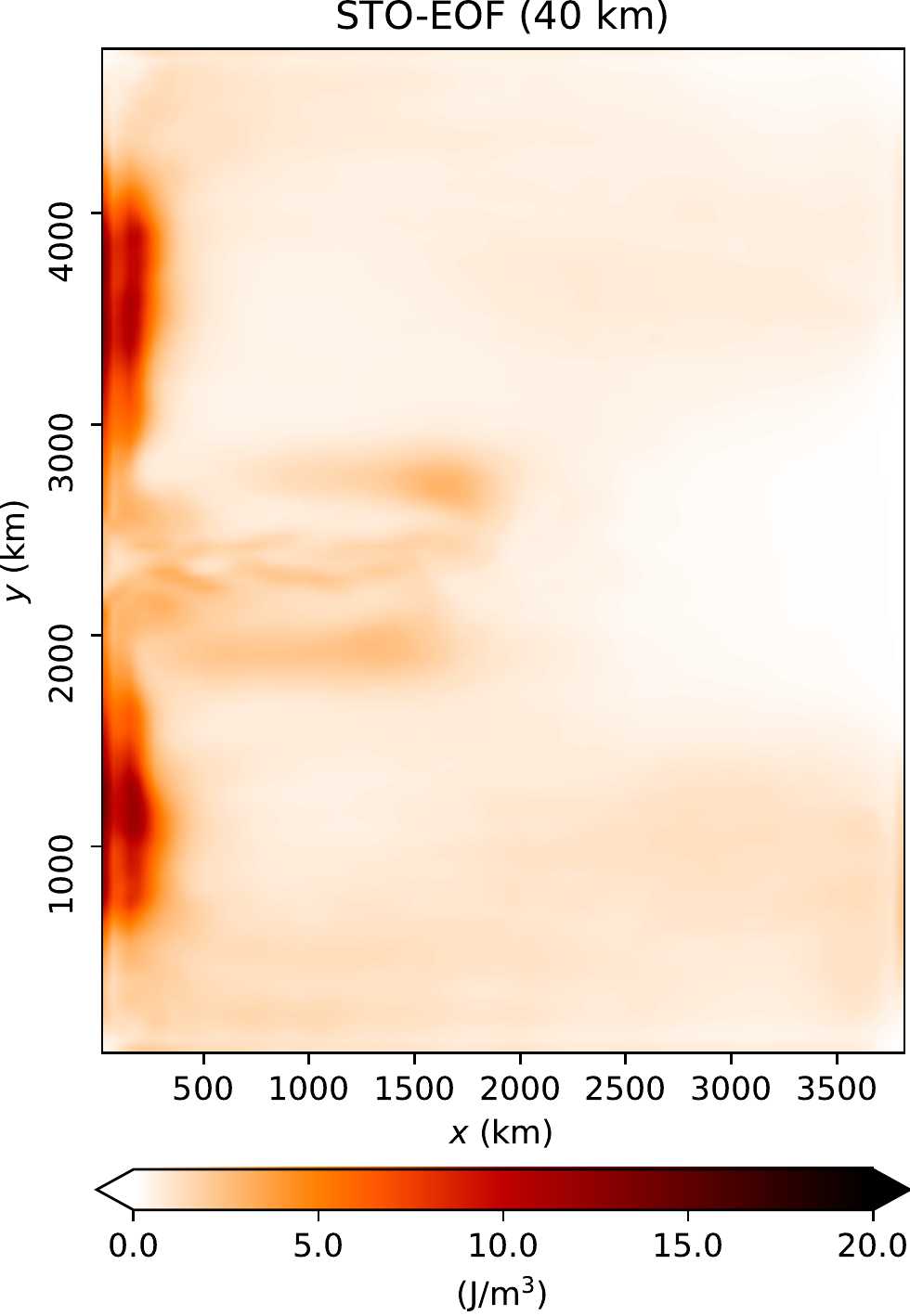}
\includegraphics[width=3.8cm]{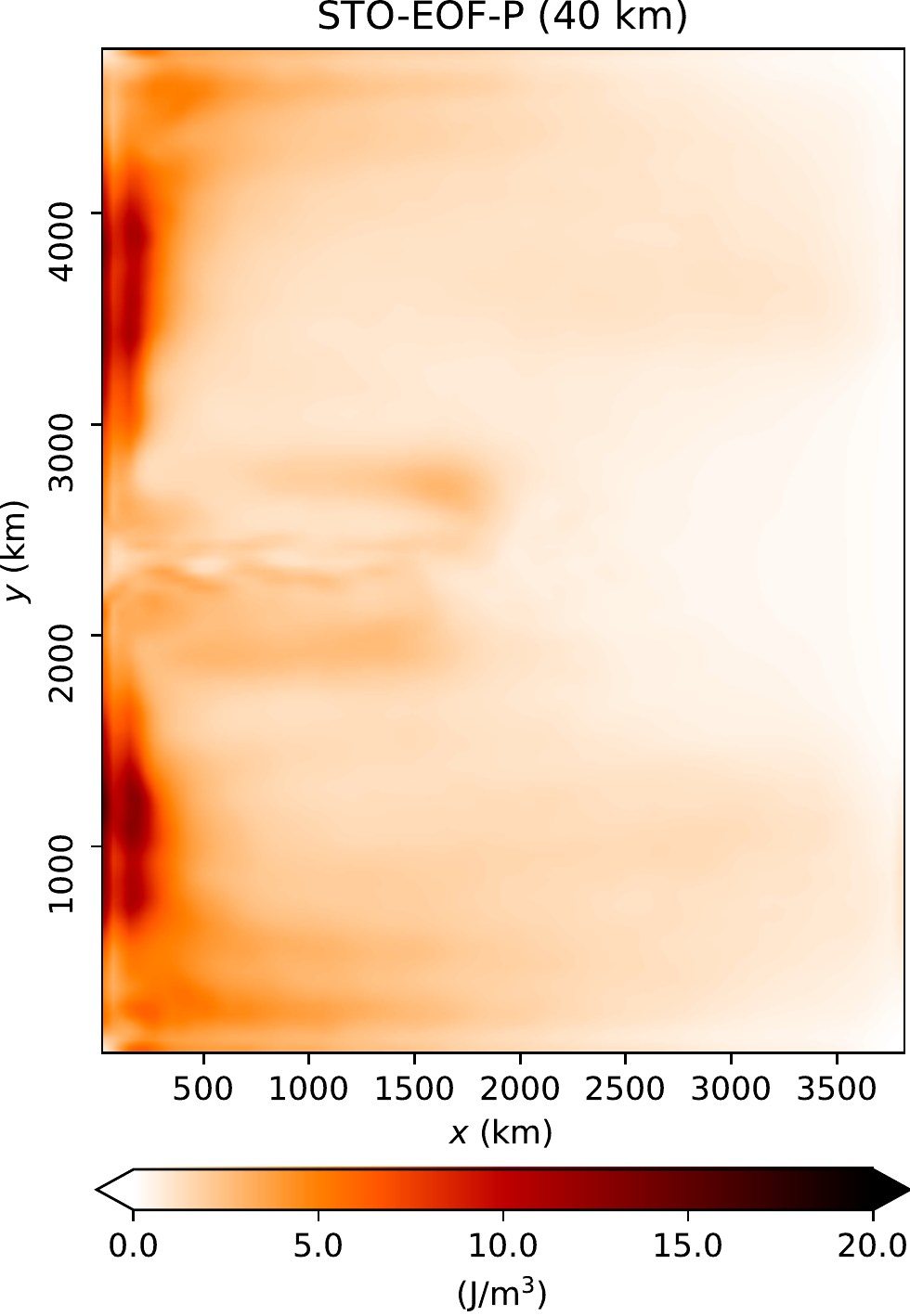} \\
\par\medskip
\includegraphics[width=3.8cm]{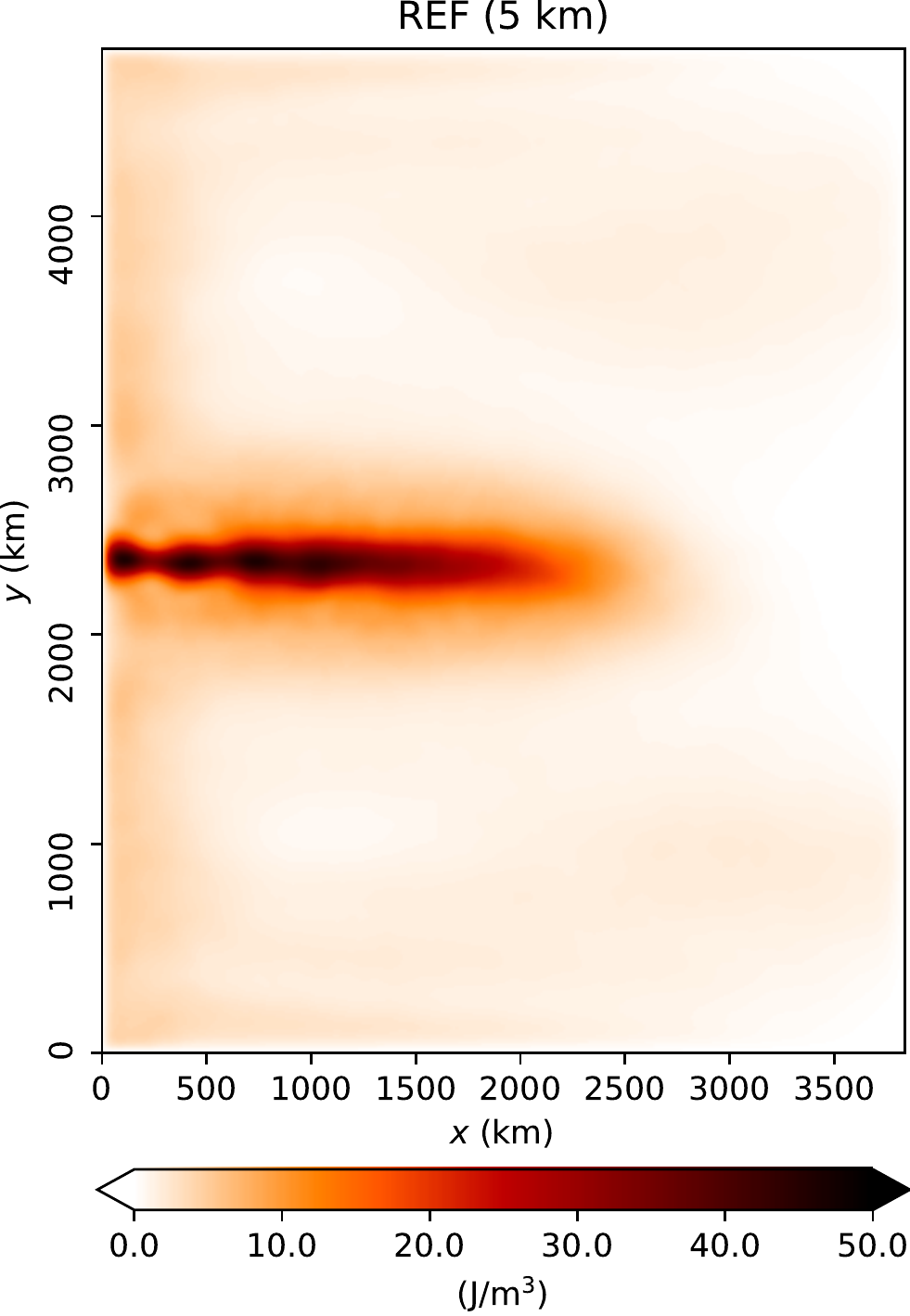}
\includegraphics[width=3.8cm]{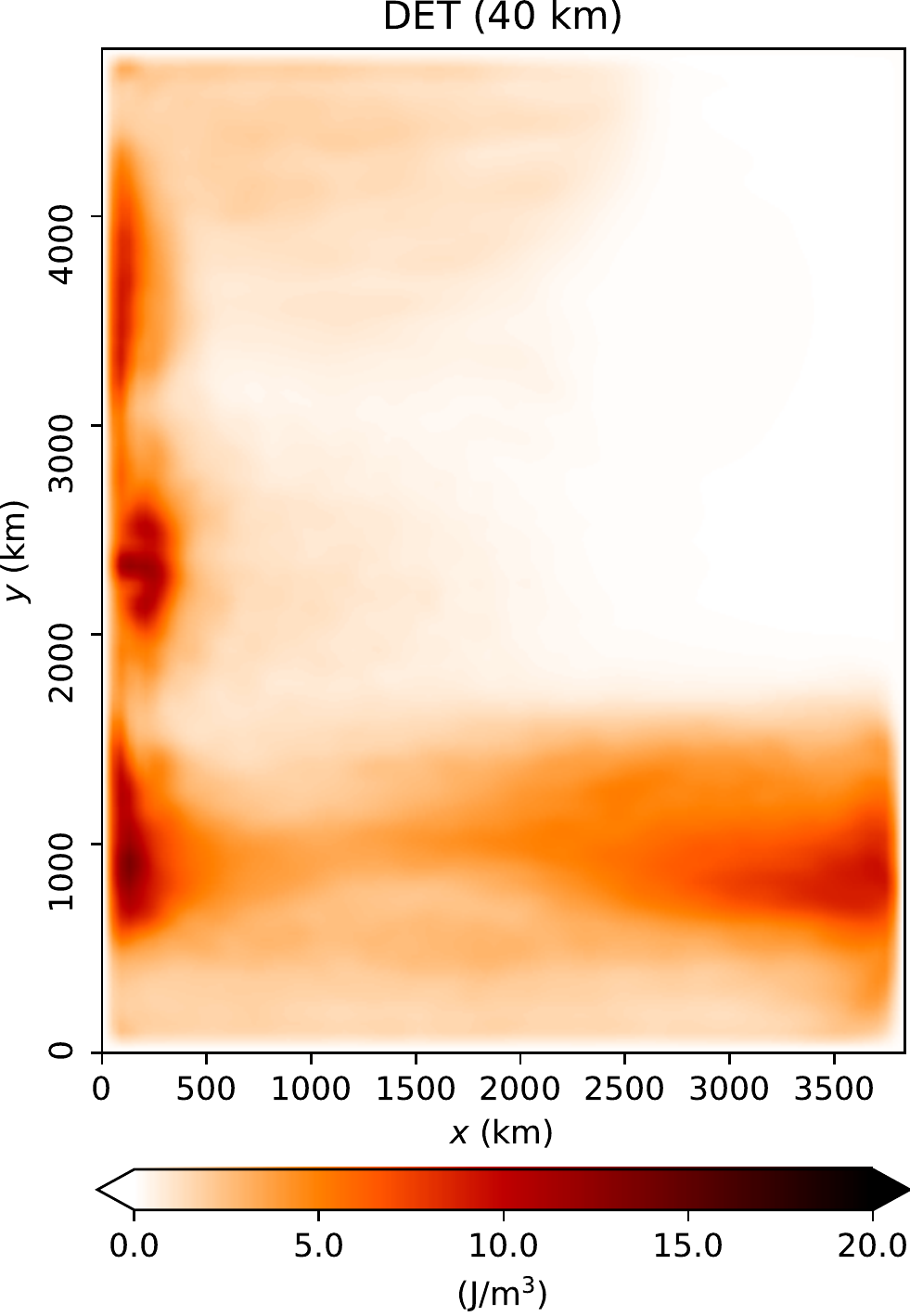}
\includegraphics[width=3.8cm]{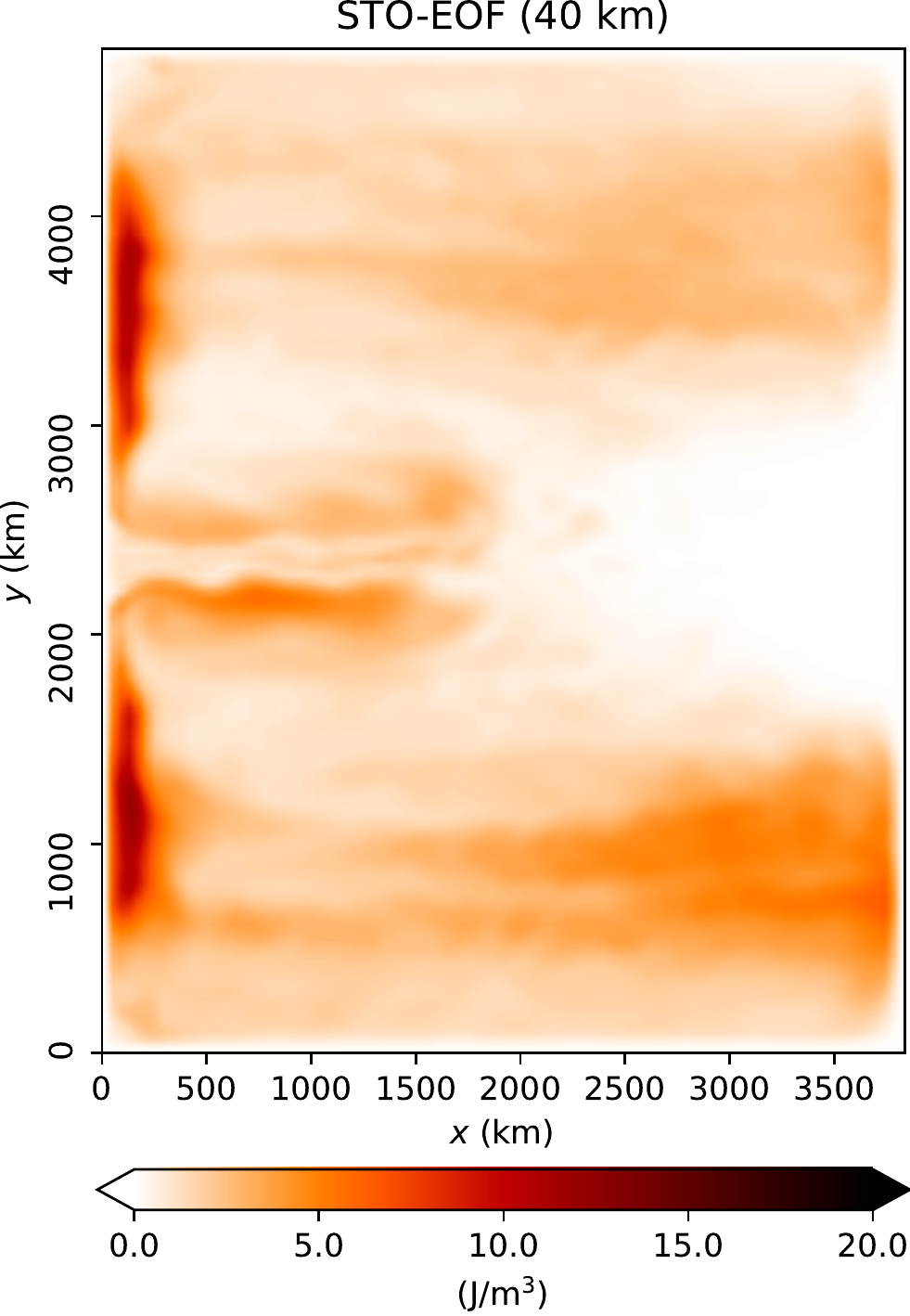}
\includegraphics[width=3.8cm]{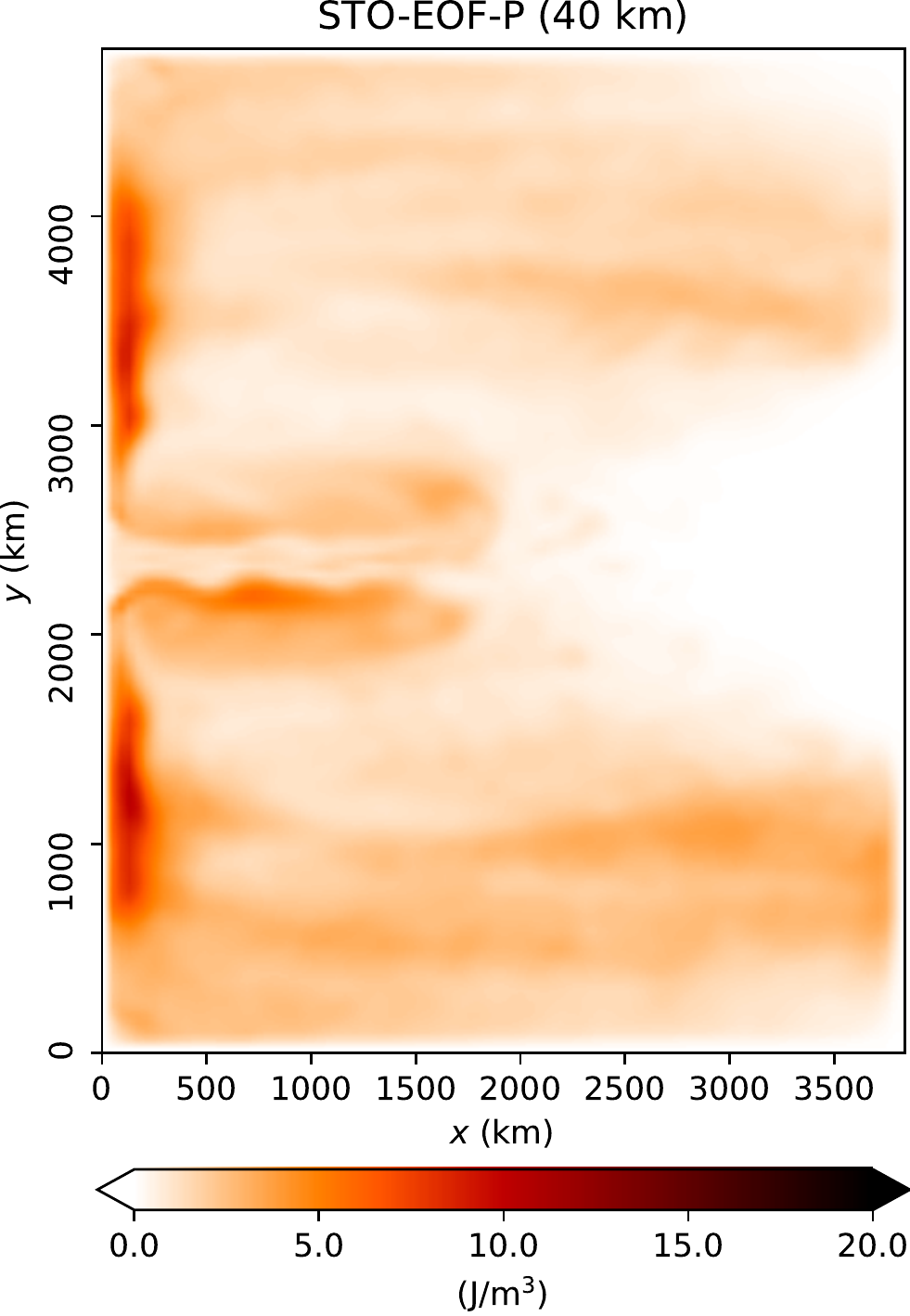}
\end{center}
\caption{Comparison of the vertically-integrated EKE density (top row) and EPE density (bottom row) for different models (by columns).}
\label{fig:denergy-sst}
\end{figure}

As shown in Figure \ref{fig:stat-sst}, the \emph{STO-EOF-P} models provide the best results for all the metrics of variability at different resolutions. The local structures of the eddy energy are also better than those recovered in the other coarse models, which is demonstrated by Figure~\ref{fig:denergy-sst}.

\section{Conclusions}\label{sec:conclu}

A stochastic parameterization of the unresolved eddy field is successfully implemented in a well-known QG model. The empirical spatial correlation of the small-scale noise has been first estimated from eddy-resolving simulation data. An additional correction drift that is fully justified in the stochastic setting considered and ensuing from a change of probability measure has been introduced. This non intuitive term seems quite important in the reproduction of the eastward jet within the wind-driven double-gyre circulation. In addition, a projection method has been proposed to constrain the noise living along the iso-surfaces of the vertical stratification. The resulting non-stationary noise model enables us to improve the intrinsic variability of the large-scale resolved flow. This improvement has been demonstrated through some statistical criterion. We have highlighted that both stationary and non-stationary noise models provide improvements at ocean climatic scale in terms of variability metrics, compared to a corresponding deterministic coarse model. At that resolution, the effects of the mesoscale eddies within the ocean basin at midlatitude (of deformation radius around 30 $\sim$ 40 km) are better represented by the random models, even though the baroclinic instability can not be resolved.

The numerical results presented in this work encourage us to implement the proposed random model on more complex and realistic flow configurations. A following study for a fully coupled model of the  ocean and atmosphere \citep{Hogg2003} is already in progress. Including random forcing into the bottom layer of the atmosphere leads to an unsteady wind stress and to a random diabatic forcing due to Ekman pumping of the sea surface temperature (SST) in both atmospheric and oceanic mixed layers. Thus, it will be interesting to verify the response of the ocean dynamics in terms of variability and energy to these highly variable forcing, and particularly to compare the results to a stationary wind forced ocean dynamics as well as to an unforced dynamics. The stochastic primitive Boussinesq models \eqref{eqs:Boussinesq-LU} could then be tested. However, in that case the random pressure term in the LU model has to be a priori modeled or computed based on proper scaling of hydrostasy.

Another possible research axis focuses on new parameterization methods for the unresolved flow component. On the one hand, techniques relying on the dynamic mode decomposition \citep{Schmid2010, Gugole2020}, or on the spectral analysis of the Koopman operator \citep{Mezic2013, Giannakis2019} could be explored in practice to estimate the dynamics of the noise and the Girsanov drift from high-resolution data. The objective will consist in evaluating their ability to characterize the long-terms impacts of some small-scale events. 

Preliminary very encouraging results for the same stochastic QG model as the one studied in the present work can be found in \citet{Li2022stuod}. On the other hand, it is also interesting to design adequate noises that provide efficient energy conversion \citep{Bachman2019, Jansen2019} based on the analysis presented in this work. The objective is to release the APE through the noise quadratic variation \eqref{eq:QG-LU-buoy} and to backscatter the unresolved KE to the resolved KE through the advection of momentum by the noise \eqref{eq:QG-LU-moment}.

\section*{Data Availability Statement}

The source codes to reproduce the results are available at \url{https://github.com/matlong/qgcm_lu}, all the coarse-resolution data can be found in \url{https://doi.org/10.5281/zenodo.6818751}, and the high-resolution data can be either reproduced by the restart files or accessed by contacting the corresponding author.

\section*{Acknowledgments}

The authors acknowledge the support of the ERC EU project 856408-STUOD. We would like to thank the technical support provided by Charles Deltel and Pranav Chandramouli. 

\section*{Appendix A. Conservation of path-wise energy}\label{sec:proof-energy}

In this section, we show that the stochastic Boussinesq QG system \eqref{eqs:Boussinesq-LU}, without considering the beta effect on the noise, preserves along time the global energy of the resolved geostrophic flow component for each realization. To this end, we first recall the It\^{o}'s integration-by-part formula \cite{Kunita1997}: let $\df_t X = F\, \dt + G\, \df B_t$ and $\df_t Y = F'\, \dt + G'\, \df B_t$ be two S(P)DEs driven by a Brownian motion $B$ satisfying $\Exp [G \df B_t] = \Exp [G' \df B_t] = 0$, then
\begin{align}\label{eq:IPP}
\df_t (X Y) &= X \df_t Y +  Y \df_t X + \df \langle X, Y \rangle_t \nonumber \\
&= X \df_t Y +  Y \df_t X + G G'\, \dt ,
\end{align}
where $\langle X, Y \rangle_t \stackrel{\mb{P}}{=} \lim_{\sub n \rightarrow +\infty} \sum_{i=1}^{p_n} \big( X_{t_i}^n - X_{t_{i-1}}^n \big) \big( Y_{t_i}^n - Y_{t_{i-1}}^n \big)$ stands for the joint quadratic variation between the processes $X$ and $Y$ with $0 = t_0^n < t_1^n < \cdots < t_{p_n}^n = t$ a partition of the interval $[0, t]$, and the limit, if it exists, is defined in the sense of convergence in probability. In particular, we have $\langle B, B \rangle_t = t$.

We now expand equation \eqref{eq:QG-LU-moment} for the resolved geostrophic velocity $\bu = (u,v)\tp$ and equation \eqref{eq:QG-LU-buoy} for the resolved buoyancy $b$,
\beqs\label{eqs:d-u-v-b}
\begin{align}
\df_t u &= - (\us\, \dt + \noi) \adv u + \Big( \alf \bdiv(\ba \grad u) + f_0 v_a + \beta y v - \pd_x p_a \Big)\, \dt \\
\df_t v &= - (\us\, \dt + \noi) \adv v + \Big( \alf \bdiv(\ba \grad v) - f_0 u_a - \beta y u - \pd_y p_a \Big)\, \dt \\
\df_t b &= - (\us\, \dt + \noi) \adv b + \alf \bdiv(\ba \grad b)\, \dt - N^2 w_a\, \dt , 
\end{align}
\eeqs
together with the geostrophic equilibrium \eqref{eq:QG-LU-geost}, the hydrostatic balances \eqref{eq:QG-LU-hydro} and the continuity equation \eqref{eq:QG-LU-cont}. Note that $\ba\, \dt$ corresponds to the quadratic variation of the noise $\noi$, which is $\ba$ defined in \eqref{eq:bracket}. Applying equation \eqref{eq:IPP} for the density of APE, $b^2/(2 N^2)$, and integrating subsequently over the ocean domain $\dom = \area \times [0,-H]$, we have
\begin{align}\label{eq:dPE}
\int_{\dom} \df_t \Big( \frac{b^2}{2 N^2} \Big)\, \dx
&= \int_{\dom} \frac{1}{N^2} \Big( b \df_t b + \alf \df \langle b, b \rangle_t \Big)\, \dx \nonumber\\
&= \int_{\dom} \frac{1}{N^2} \Big( -b (\us\, \dt + \noi) \adv b + \alf b \bdiv(\ba \grad b)\, \dt \nonumber\\
&\:\phantom{= \int_{\dom} \frac{1}{N^2} \Big(}\:
- N^2 w_a b\, \dt + \alf (\grad b) \bdot \ba \grad b\, \dt \Big)\, \dx \nonumber\\
&= - \int_{\dom} \bdiv \Big( (\us\, \dt + \noi) \frac{b^2}{2 N^2}\Big)\, \dx - \int_{\dom} w_a b\, \dx\, \dt \nonumber\\
&= - \int_{\dom} w_a b\, \dx\, \dt ,
\end{align}
where the third equality results from the geostrophic balances \eqref{eq:QG-LU-geost} and the last equality comes from the divergence theorem with ideal boundary conditions, $\bu \bdot \bs{n}\ |_{\pd \area} = \noi \bdot \bs{n}\ |_{\pd \area} = 0$ (which leads to $\us \bdot \bs{n}\ |_{\pd \area} = 0$ from equations \eqref{eq:def-ISD} and \eqref{eq:bracket}).
Similarly to the previous calculation, one can show that
\beqs
\begin{align}
&\int_{\dom} \df_t \Big( \frac{u^2}{2} \Big)\, \dx = \int_{\dom} u (f_0 v_a + \beta y v - \pd_x p_a)\, \dx\, \dt , \\
&\int_{\dom} \df_t \Big( \frac{v^2}{2} \Big)\, \dx = - \int_{\dom} v (f_0 u_a + \beta y u + \pd_y p_a)\, \dx\, \dt .
\end{align}
\eeqs
Summing these two equations, we deduce the evolution of KE:
\begin{align}\label{eq:dKE}
\int_{\dom} \df_t \Big( \frac{|\bu|^2}{2} \Big)\, \dx 
&= -\int_{\dom} \big( (\gradp p) \bdot \bu_a^{\perp} + \bdiv (\bu p_a) \big)\, \dx\, \dt \nonumber \\
&= -\int_{\dom} \big( \pdiv (p \bu_a^{\perp}) - p \bdiv \bu_a \big)\, \dx\, \dt \nonumber \\
&= -\int_{\area} \int_{-H}^{0} p \pd_z w_a\, \dz\, \dA\, \dt \nonumber \\
&= -\int_{\area} \Big( \big[p w_a \big]_{\sub -H}^{0} - \int_{-H}^{0} w_a \pd_z p\, \dz \Big)\, \dA\, \dt \nonumber \\
&= \int_{\dom} w_a b\, \dx\, \dt ,
\end{align}
where the first equality comes from the geostrophic balances \eqref{eq:QG-LU-geost}, the third equality results from both the Stokes theorem with ideal boundary condition of $\bu_a$ and the continuity equation \eqref{eq:QG-LU-cont}, the fourth equality derives from the vertical integration-by-parts formula, and the last equality is based on both the vertical boundary conditions, $w_a (z = 0) = w_a (z = -H) = 0$ and the hydrostatic balances \eqref{eq:QG-LU-hydro}. Finally, by summing equations \eqref{eq:dKE} and \eqref{eq:dPE}, one show that the stochastic system \eqref{eqs:d-u-v-b} conserves the path-wise total energy of the resolved geostrophic flow component, namely
\beq
\df_t \int_{\dom} \alf \Big( |\bu|^2 + \frac{b^2}{N^2} \Big)\, \dx = \int_{\dom} \df_t \Big( \frac{|\bu|^2}{2} \Big)\, \dx + \int_{\dom} \df_t \Big( \frac{b^2}{2 N^2} \Big)\, \dx = 0 .
\eeq
Since this conservation property is path-wise, it holds also for their ensemble-mean.

\section*{Appendix B. Conversion of ensemble energies}\label{sec:conv-energy}

In this section, we show briefly the conversions between the ensemble-variance and the energy of the ensemble-mean. To that end, we first decompose each prognostic variable $\Theta$ (could be $u,v,b$ or $\ba$) of \eqref{eqs:d-u-v-b} into a mean component $\ol{\Theta} := \Exp [\Theta]$ and an eddy component $\Theta' := \Theta - \Exp [\Theta]$. The other time-smooth components $F\, \dt$ apart from the geostrophic transport operator are decomposed in the same way. For instance, $F = - N^2 (\ol{w_a} + w_a')$ in the buoyancy equation. However, the noise is only an eddy component since $\Exp [\noi] = 0$. Substituting such decomposition into equations \eqref{eqs:d-u-v-b} and taking subsequently the ensemble-mean of the expanded equation, we obtain
\beqs
\begin{align}\label{eq:dmean}
\pd_t \ol{\Theta} =  \ol{F} - \ol{\us} \adv \ol{\Theta} + \alf \bdiv \big( \ol{\ba} \grad \ol{\Theta} \big) - \ol{(\us)' \adv \Theta'} + \alf \bdiv (\ol{\ba' \grad \Theta '}) ,
\end{align}
where the second and the third terms on the RHS describe the mean-mean interactions, whereas the last two terms represent the effect of eddy-eddy interactions on the ensemble-mean component. Under natural boundary conditions, on can easily show the time evolution of the global energy for $\ol{\Theta}$:
\begin{align}
\frac{\df}{\dt} \int_{\dom} \frac{\ol{\Theta}^2}{2}\, \dx 
&= \int_{\dom} \ol{F}\, \ol{\Theta}\, \dx + \int_{\dom} \ol{(\us)' \Theta'} \adv \ol{\Theta}\, \dx \nonumber\\ 
&- \alf \int_{\dom} \big( \grad \ol{\Theta} \big) \bdot \big( \ol{\ba} \grad \ol{\Theta} + \ol{\ba' \grad \Theta'} \big)\, \dx .
\end{align}
\eeqs
Applying this result for $\ol{u}$ and $\ol{v}$ (resp. for $\ol{b}$) together with \eqref{eq:dKE} (resp. with \eqref{eq:dPE}), one can deduce the evolution of the MKE (resp. MPE), and the conversion terms are summarized in the diagram in Section \ref{sec:energy}.

Subtracting next \eqref{eq:dmean} from \eqref{eqs:d-u-v-b}, we deduce the evolution of $\Theta'$:
\beqs
\begin{align}
\df_t \Theta' = &- \Big( \ol{\us} \adv \Theta' - \alf \bdiv \big( \ol{\ba} \grad \Theta' \big) \Big)\, \dt \nonumber \\
&- \big( (\us)'\, \dt + \noi \big) \adv \ol{\Theta} - \alf \bdiv \big( \ba' \grad \ol{\Theta} \big)\, \dt + F'\, \dt \nonumber \\
&- \big( (\us)'\, \dt + \noi \big) \adv \Theta' - \alf \bdiv \big( \ba' \grad \Theta' \big)\, \dt \nonumber \\
&+ \Big( \ol{(\us)' \adv \Theta' - \alf \bdiv (\ba' \grad \Theta ')} \Big)\, \dt ,
\end{align}
where the first two lines describe the eddy-mean interactions, whereas the last two lines represent the effect of eddy-eddy interactions on the ensemble-eddy component. To evaluate the ensemble-mean of the energy for $\Theta'$, the It\^{o}'s integration-by-part formula \eqref{eq:IPP} is required again. In this case, one can first show that the mean of the joint quadratic variation process for $\Theta'$ reads
\begin{align}
\ol{\df \langle \Theta', \Theta' \rangle_t} 
&= \ol{(\grad \Theta') \bdot (\ol{\ba} \grad \Theta')} + \ol{(\grad \Theta') \bdot (\ba' \grad \Theta')} + \ol{(\grad \Theta') \bdot (\ba' \grad \ol{\Theta})} \nonumber \\
&+ (\grad \ol{\Theta}) \bdot (\ol{\ba' \grad \Theta'}) + (\grad \ol{\Theta}) \bdot (\ol{\ba} \grad \ol{\Theta}) .
\end{align}
Subsequently, we can deduce the evolution of the global variance for $\Theta$, namely
\begin{align}
\frac{\df}{\dt} \int_{\dom} \ol{\frac{(\Theta')^2}{2}}\, \dx
&= \int_{\dom} \ol{F' \Theta'}\, \dx - \int_{\dom} \ol{(\us)' \Theta'} \adv \ol{\Theta}\, \dx
\nonumber \\
&+ \alf \int_{\dom} \grad (\ol{\Theta}) \bdot (\ol{\ba} \grad \ol{\Theta} + \ol{\ba' \grad \Theta'})\, \dx .  
\end{align}
\eeqs
Applying similarly this result for $\ol{u}$ and $\ol{v}$ (resp. for $\ol{b}$), we can deduce the evolution of the EKE (resp. EPE), and the results are shown in the diagram in Section \ref{sec:energy}.

\bibliography{mybib}

\end{document}